\documentclass[a4paper, fleqn]{cas-dc}
\usepackage[numbers]{natbib}
\usepackage{amsfonts}
\usepackage{algorithmic}
\usepackage{amsmath} 
\usepackage{amssymb} 
\usepackage{caption}
\usepackage{subcaption}
\usepackage{CJKutf8}
\usepackage{etoolbox}
\usepackage{forest}
\usepackage{lingmacros}
\usepackage{textcomp}
\usepackage{tree-dvips}
\usepackage{tikz}
\usepackage{tikz-cd}
\usepackage[arrowdel]{physics}
\usepackage{graphicx}
\usepackage{wrapfig}
\usepackage{listings}
\usepackage{pgfplots, pgfplotstable}
\usepackage{diagbox} 
\usepackage[usestackEOL]{stackengine}
\usepackage{makecell}
\usepackage{mathrsfs}
\usepackage{moresize}
\usepackage{multirow}
\usepackage[T1]{fontenc}
\usepackage{xcolor}
\allowdisplaybreaks[1]
\definecolor{orchid}{rgb}{0.7, 0.4, 1.1}
\DeclareCaptionFont{listing}{\footnotesize}
\DeclareCaptionFormat{listing}{\rule{0.47\textwidth}{0.5pt}\vskip#1#2#3}
\definecolor{comment_color}{rgb}{0, 0.5, 0}
\definecolor{keyword_color}{rgb}{0.3, 0, 0.6}
\definecolor{string_color}{rgb}{0.5, 0, 0.1}

\begin{document}

\shorttitle{Effect of Al-Zn alloy wafer grain boundary diffusion on NdFeB magnets}

\shortauthors{Xi Liu \begin{CJK*}{UTF8}{bsmi}劉錫\end{CJK*}$^a$, Wenxi Fang \begin{CJK*}{UTF8}{bsmi}方文希\end{CJK*}$^b$, Ken Perlin$^c$}

\title{{\Large Effect of Al-Zn alloy wafer grain boundary diffusion on the magnetism and microstructure of sintered NdFeB magnets}}

\author[1]{\color{black}Xi Liu \begin{CJK*}{UTF8}{bsmi}劉錫\end{CJK*}}
\author[2]{\color{black}Wenxi Fang \begin{CJK*}{UTF8}{bsmi}方文希\end{CJK*}}
\author[3]{\color{black}Ken Perlin}

\address{$^a$xl3467@columbia.edu, Columbia University; $^b$u3013972@connect.hku.hk, Inner Mongolia University of Science and Technology;
$^c$perlin@nyu.edu, New York University}

\begin{abstract}
This study investigates Al-Zn grain boundary diffusion (GBD) treatment on sintered Nd-Fe-B magnets using Al$_{80}$Zn$_{20}$ alloy sheets as the diffusion source. The alloy sheets were placed at both ends of cylindrical samples and diffusion-annealed at 900$^\circ$C and 700$^\circ$C for 7 hours under vacuum ($\leq5\times10^{-3}$Pa). Magnetic measurements show that coercivity increases from 951.5kA/m in the untreated sample to 1158.2kA/m at 900$^\circ$C (a gain of 206.7kA/m, 21.7\%) and to 1039.6kA/m at 700$^\circ$C (a gain of 88.1kA/m, 9.3\%), while remanence declines modestly from 1282mT to 1256mT after the high-temperature treatment. Scanning electron microscopy (SEM), energy-dispersive X-ray spectroscopy (EDS), and X-ray diffractometer (XRD) analyses reveal that the 900$^\circ$C treatment produces a thinner, more continuous grain boundary phase and a distinct core-shell structure around the main-phase grains. EDS mapping shows that Al preferentially enriches the shell region of the Nd$_2$Fe$_{14}$B grains, while Zn predominantly resides in the grain boundary phase, where it lowers the melting point of the intergranular phase and improves its fluidity. XRD shows no detectable secondary phases, while the combined XRD, EDS results, and lattice expansion are consistent with possible limited Al incorporation into the main-phase lattice. Verified by computational analysis, the coercivity enhancement is attributed to three synergistic factors: improved grain boundary decoupling, the formation of a high-anisotropy shell layer that strengthens domain-wall pinning, and the smoothing of grain edges to suppress reverse-domain nucleation. Overall, the 900$^\circ$C treatment proves considerably more effective than 700$^\circ$C, providing a non-heavy-rare-earth pathway for enhancing coercivity in sintered Nd-Fe-B magnets for high-temperature applications.
\end{abstract}

\maketitle
\section{Introduction}
Since their emergence in 1983, sintered Nd-Fe-B magnets have been increasingly widely used in fields such as new energy vehicles, wind power generation, consumer electronics, and aerospace, owing to their excellent magnetic properties \cite{kim_1996_high}. However, these magnets have a critical drawback: their coercivity drops sharply at elevated temperatures, preventing them from performing well in high-temperature environments \cite{hirosawa1990coercivity}. To address this issue, heavy rare earth elements such as dysprosium (Dy) or terbium (Tb) are commonly added during industrial production. But this approach comes with two disadvantages: a reduction in remanence and a significant increase in cost.

Grain boundary diffusion technology offers a more elegant solution. By coating only the magnet surface with heavy rare earths and then applying heat \cite{bae2015oxygen}, the heavy rare earth atoms diffuse inward along the grain boundaries, ultimately forming a highly anisotropic shell around the outer layer of the main phase grains \cite{hirosawa1990coercivity}. In this way, only a small amount of heavy rare earth is needed to achieve a substantial improvement in coercivity.

Nevertheless, grain boundary diffusion technology also faces challenges: the heavy rare earths do not diffuse deeply enough. In recent years, researchers have found that adding low-melting-point elements (such as Al and Zn) to the magnet can lower the melting point of the Nd-rich grain boundary phase, thereby facilitating the diffusion of heavy rare earths \cite{ZHONG2023}. However, most studies involve grinding these low-melting-point elements into powders and mixing them into the raw materials \cite{jurczyk1995powder}. Moreover, these elements begin to interfere from the sintering stage onward, making it difficult to isolate and observe their specific roles during the diffusion procedure alone.

In this experiment, we adopted a different approach: instead of incorporating Al and Zn into the raw materials, we fabricated them into alloy discs and placed them directly on the magnet surface for diffusion treatment. This was designed to clarify the respective roles of Al and Zn during diffusion and to investigate how they interact with each other.

Specifically, the following steps were carried out: first, 80wt.\% Al with 20wt.\% Zn alloy discs were prepared; then, diffusion treatments were performed on small cylindrical magnets at two temperatures, 900$^\circ$C and 700$^\circ$C. Magnetic properties were measured before and after diffusion, and the microstructure and elemental distributions were examined using SEM, EDS, and XRD. Finally, the diffusion behaviors of Al and Zn and their effects on magnetic properties were analyzed.

\section{Background}


The first-generation rare-earth permanent magnet material is the samarium-cobalt (SmCo$_5$) magnet, which emerged in the 1960s \cite{strnat_1970}. This type of magnet has a significant advantage of temperature stability, but it also has drawbacks including insufficiently high maximum energy product and high cobalt cost, which limited its application.

The second-generation rare-earth permanent magnet material is the samarium-cobalt (Sm$_2$Co$_{17}$) magnet, which appeared in the 1970s \cite{ojima_1977}. Compared with the first generation, it offers a higher maximum energy product and even better temperature stability. However, its cost remained high, so it is mainly used in special applications that are less cost-sensitive but demand high-temperature stability.

The third-generation rare-earth permanent magnet material is the neodymium-iron-boron (Nd-Fe-B) magnet, invented by M. Sagawa in 1983 \cite{sagawa_1984}. This magnet has an exceptionally high maximum energy product; its theoretical value can reach 64 MGOe, and current commercial products already exceed 50 MGOe, far surpassing the previous two generations of samarium-cobalt magnets.

Not only does the Nd-Fe-B magnet offer excellent performance, but its main constituents are neodymium (Nd) and iron (Fe): neodymium is relatively abundant, and iron is very inexpensive, so its cost is much lower than that of samarium-cobalt magnets. This advantage has allowed Nd-Fe-B magnets to enter a wide range of everyday applications, making them the most widely used rare-earth permanent magnet material.

Recent studies have demonstrated that Al- and Zn-containing additives can modify the grain-boundary structure and improve the coercivity of sintered Nd-Fe-B magnets through changes in grain-boundary wettability, diffusion behavior, and intergranular magnetic isolation \cite{Kobayashi2019magnetostriction}. In particular, previous approaches have introduced Al, Zn, or related low-melting-point elements through powder mixing or surface-coating processes, where the additives are incorporated during precursor preparation or deposited onto the magnet surface before diffusion treatment \cite{bae2015oxygen}. These studies establish the effectiveness of Al- and Zn-assisted grain-boundary modification, but the role of an externally supplied Al-Zn alloy source during post-sintering diffusion remains less clearly separated from the effects associated with bulk alloying or precursor processing \cite{Bessais2006nano}. In the present work, an Al$_{80}$Zn$_{20}$ alloy wafer is directly placed on the surface of a previously sintered Nd-Fe-B magnet and used as the diffusion source, thereby introducing Al and Zn specifically during the grain-boundary diffusion treatment \cite{brown2002develop}. This configuration allows the temperature-dependent redistribution of Al and Zn to be examined without intentionally incorporating these elements into the starting powder or sintering process \cite{herbst_1991_int}. The experimental results show that Al preferentially enriches the grain-edge region and partially extends into the main phase, whereas Zn is predominantly retained in the grain-boundary region, accompanied by a more continuous boundary layer after the 900~$^\circ$C treatment. Thus, the significance of the present configuration is not that Al or Zn-assisted grain-boundary modification is itself new, but that the wafer-based post-sintering approach provides a comparatively direct means of examining the redistribution and complementary roles of Al and Zn during the diffusion treatment.

\section{Experimental method}
\subsection{Preparation of sintered Nd-Fe-B magnets}
The preparation of sintered Nd-Fe-B magnets employs the powder metallurgy method, with the main procedure flow including melting and casting \cite{jurczyk1995powder}, pulverization, orientation pressing, vacuum sintering, and tempering treatment.

In the melting and casting stage, raw materials are first batched according to the designed composition, followed by vacuum induction melting. The molten metal is cooled on the surface of a rapidly rotating copper roller, forming alloy flakes with a thickness of approximately 0.2-0.4mm \cite{jurczyk1995powder}. This helps refine the microstructure while avoiding the precipitation of the $\alpha$-Fe phase. In the pulverization stage, the alloy flakes are first coarsely crushed by hydrogen decrepitation (HD) and then further pulverized to fine powders of 3-5$\mu$m via jet milling \cite{zhou_2011}. During orientation pressing, the powder is aligned along a specified direction under a strong magnetic field, and subsequently compacted into green compacts. In the vacuum sintering step, the green compacts are sintered at 1050-1100$^\circ$C for densification, achieving densities above 99\% of the theoretical density. The final tempering treatment employs a two-stage tempering procedure, conducted at approximately 900$^\circ$C and 500$^\circ$C respectively, to optimize the distribution of the grain-boundary phase.

\subsection{Main phase and grain-boundary phase}
The microstructure of sintered Nd-Fe-B magnets is mainly composed of the main phase and the grain-boundary phase. The main phase is an intermetallic compound with the chemical formula Nd$_2$Fe$_{14}$B, possessing a tetragonal crystal structure with space group P4$_2$/mnm \cite{bae2015oxygen}. This phase exhibits a high magnetocrystalline anisotropy field along the c-axis, approximately 70 kOe, which is the primary reason for the excellent magnetic properties of Nd-Fe-B magnets \cite{herbst_1991_int}. The grain size of the main phase typically ranges from several micrometers to over ten micrometers, with an irregular blocky shape.

The grain-boundary phase is mainly a Nd-rich alloy, with a neodymium content significantly higher than that of the main phase and a relatively low melting point. During sintering, the grain-boundary phase melts first, forming a liquid phase that fills the spaces between the main-phase grains. After cooling and solidification, the grain-boundary phase forms a thin non-magnetic layer that envelops the main-phase grains \cite{HONO2012}. The roles of the grain-boundary phase are mainly as follows: firstly, it promotes liquid-phase sintering, enhancing the densification degree of the magnet; secondly, as a non-magnetic layer, it isolates the main-phase grains, weakening the magnetic exchange coupling between grains, thereby contributing to an increase in coercivity; and thirdly, it serves as a fast diffusion channel for atomic diffusion, providing a foundation for grain-boundary diffusion technology.

\subsection{Coercivity mechanism}
Coercivity is an important indicator for evaluating the demagnetization resistance of permanent magnet materials, specifically defined as the reverse magnetic field strength required to reduce the magnetic induction intensity to zero after the magnet has been saturated magnetized.

The coercivity mechanism of sintered Nd-Fe-B magnets is primarily characterized by the nucleation mechanism. Studies by Kronmüller et al. \cite{Kronmuler_1988} have shown that reverse magnetic domains first nucleate at structural defects on the surface or near-surface regions of the main-phase grains, then gradually grow under the action of the reverse magnetic field, eventually expanding throughout the entire grain \cite{hirosawa1990coercivity}. Therefore, the magnitude of coercivity mainly depends on the difficulty of reverse magnetic domain nucleation at the grain surfaces.

There are two main approaches to improving coercivity: one is to enhance the local anisotropy field at the grain surfaces \cite{hirosawa1990coercivity}, thereby increasing the resistance to reverse magnetic domain nucleation \cite{li2020}; the other is to optimize the distribution of the grain-boundary phase, forming a continuous and uniform non-magnetic thin layer to suppress the propagation of reverse magnetic domains between grains.

\subsection{Temperature stability}
Sintered Nd-Fe-B magnets suffer from relatively poor temperature stability. Their coercivity temperature coefficient is approximately -0.5$^\circ$C to -0.6$^\circ$C, meaning that for every 10$^\circ$C increase in temperature \cite{sagawa1990high}, the coercivity decreases by 5\% to 6\%. When the temperature exceeds 150$^\circ$C, conventional Nd-Fe-B magnets can hardly meet the service requirements of motors and other equipment \cite{HIROSAWA_1985}.

The main reasons for the poor temperature stability are twofold: first, the magnetocrystalline anisotropy field of the main phase Nd$_2$Fe$_{14}$B decreases with increasing temperature \cite{Bessais2006nano}; second, the thermal activation effect at elevated temperatures makes it easier for reverse magnetic domains to overcome the energy barrier, thereby promoting their nucleation and growth.

\subsection{Magnetic coercivity enhancement}
The heavy rare earth elements dysprosium (Dy) and terbium (Tb) are the most commonly used additives for enhancing the coercivity of sintered Nd-Fe-B magnets \cite{sagawa1990high}. Their mechanism of action lies in the fact that when Dy or Tb atoms diffuse into the Nd$_2$Fe$_{14}$B main-phase lattice \cite{Bessais2006nano}, they preferentially occupy the Nd crystallographic sites, forming a magnetic shell layer with a stoichiometric composition of (Nd,RE)$_2$Fe$_{14}$B(RE = Dy or Tb) \cite{ZHONG2023}. Since Dy$^{3+}$ and Tb$^{3+}$ ions possess higher magnetocrystalline anisotropy fields than Nd$^{3+}$, the formation of this shell layer can effectively pin the nucleation of reverse magnetic domains at the grain surfaces, thereby significantly improving the magnet's demagnetization resistance.
\begin{figure}
\centering
\includegraphics[width = 0.5\textwidth, height = 0.3\textwidth]{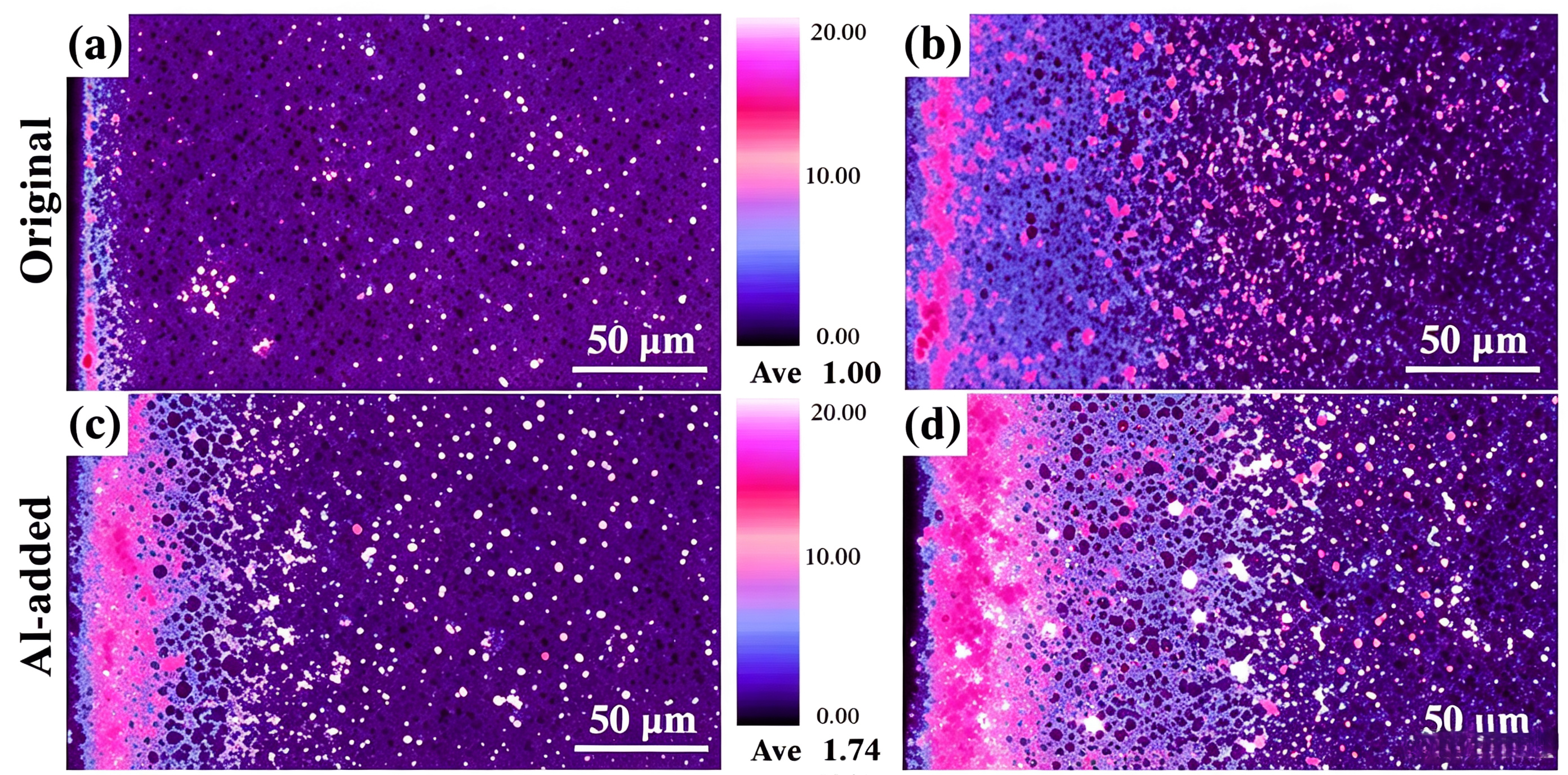}
\caption{Electron probe microanalysis mapping of (a) original magnet (b) Al-added magnet}
\label{al_dif}
\end{figure}
Figure \ref{al_dif} illustrates the relationship between the diffusion depth of Tb in sintered Nd-Fe-B magnets and the additive elements. As can be seen from the figure, the Tb concentration exhibits a gradient decrease with increasing diffusion depth, indicating that during the inward diffusion of Tb atoms along grain boundaries, the diffusion driving force weakens due to the gradual reduction in concentration gradient, and the effective diffusion depth is typically limited to several hundred micrometers to a few millimeters \cite{sagawa_1984}. For magnets with larger thicknesses, the central regions are difficult to adequately modify with Tb, resulting in non-uniform coercivity enhancement. Figure \ref{al_dif} intuitively reflects this depth limitation issue, indicating that how to promote the effective diffusion of heavy rare earth elements into deeper regions of the magnet is currently a key research focus.

\subsection{Grain boundary diffusion basic principles and advantages}
The conventional alloying method (i.e., the dual-alloy method or single-alloy method) involves directly mixing heavy rare earth elements into the main alloy during the melting or powder preparation stages \cite{MATSURA2006}. Although this approach can uniformly enhance the overall coercivity of the magnet \cite{sagawa1990high}, it has two significant negative effects. First, since Dy/Tb couple antiferromagnetically with Fe magnetic moments, their direct incorporation into the main-phase lattice leads to a decrease in the saturation magnetization of the magnet, manifested as severe losses in remanence $B_r$ and maximum energy product $(BH)_{max}$. Second, Dy and Tb are strategic resources that are expensive and have limited reserves, and the conventional large-dose addition approach substantially increases raw material costs.

To address the issues of low heavy rare earth utilization efficiency and significant remanence loss, grain boundary diffusion (GBD) technology was developed \cite{zhang_2021_pred_mag}. A compound containing heavy rare earths (Dy/Tb), such as fluorides, oxides, or alloys, is placed on the magnet surface, and heat treatment is performed at a temperature above the melting point of the Nd-rich phase but below the sintering temperature of the main phase (typically 850-950$^\circ$C). At this temperature, the Nd-rich grain boundary phase inside the magnet melts to form liquid-phase channels. The heavy rare earth atoms attached to the surface penetrate into the magnet interior along these high-diffusivity channels, and ultimately enrich the outer layer of the main-phase grains, forming a (Nd, RE)$_2$Fe$_{14}$B shell layer with high anisotropy through a "substitution-diffusion" mechanism containing a core-shell structure. The heavy rare earths only function in the grain surface layer \cite{sagawa_1984}, while the remanence of the grain cores remains unaffected, thereby achieving the goal of "high coercivity, low heavy rare earth content, and minimal remanence loss".


\subsection{Role of low-melting-point metals}
To overcome the depth limitations and improve diffusion efficiency, researchers have found that introducing low-melting-point metals (such as Al, Zn, Sn, Cu) can significantly optimize the physical properties of the grain boundary phase. The addition of elements such as Al, Zn, and Sn can lower the melting point of the Nd-rich grain boundary phase and improve its wettability with respect to the main-phase grains \cite{jiang_2025}. The transition temperature of the Nd-rich phase in standard alloys is typically around 471$^\circ$C, but after adding Al, Sn, and Zn, these temperatures decrease to 436$^\circ$C, 442$^\circ$C, and 443$^\circ$C, respectively. The reduction in melting point means that during the diffusion heat treatment, the grain boundary liquid phase can appear earlier, with lower viscosity and better fluidity, providing ideal channels for the uniform isolation of main-phase grains and the rapid migration of heavy rare earth atoms.

Among the various low-melting-point elements, Al and Zn exhibit distinct mechanisms of action. Al has an atomic radius (143 pm) similar to that of Fe (124 pm), giving it a certain lattice compatibility. Studies have shown that Al can not only reside in the grain boundary phase but also enter the main-phase lattice \cite{ojima_1977}. Thermodynamic calculations indicate a negative mixing enthalpy between Al and Tb, i.e., a mutual attraction tendency \cite{fan2016form}. This helps Al "capture" Tb atoms at the grain surface layer, inducing the heavy rare earths to form a continuous and uniform enriched shell layer at the grain edges \cite{strnat_1970}.

Zn has a relatively low melting point (419$^\circ$C) and a high vapor pressure. During high-temperature diffusion, the volatilization of Zn causes localized perturbations in the grain boundary liquid phase, which helps disrupt any stagnant layers that may exist at the grain boundaries and maintain the openness of the diffusion channels \cite{jiang_2025}. EDS analysis often shows that after diffusion, Zn mainly remains in the grain boundary phase, and this distribution characteristic helps maintain the activity of the grain boundary liquid phase.

Based on the complementarity of Al and Zn in the above mechanisms, this experiment designed an Al-30wt.\% Zn alloy diffusion source. On the one hand, it utilizes the Al-Zn eutectic reaction (at approximately 382$^\circ$C), which is far below the melting points of the pure metals, to ensure that the diffusion source melts early during heat treatment and adequately wets the magnet end faces; on the other hand, it aims to leverage the "channel unblocking" function of Zn and the "shell localization" function of Al to achieve comprehensive regulation of the grain boundary phase, thereby laying the foundation for subsequent diffusion procedures.

\section{Material processing}
This study employs Al$_{80}$Zn$_{20}$ alloy sheets as the diffusion source and applies grain boundary diffusion to modify sintered Nd-Fe-B magnets, aiming to investigate the influence of Al-Zn alloy diffusion on the magnetic properties and microstructure of the magnets \cite{zhang_2021_pred_mag}. By setting different diffusion temperatures, the changes in coercivity, remanence, and microstructure of the magnets under 700\,$^\circ$C and 900\,$^\circ$C treatment conditions are compared \cite{sagawa1990high}, and the effects of diffusion temperature on the diffusion behavior and distribution characteristics of Al and Zn elements are analyzed \cite{zhang_2021_pred_mag}. On this basis, combined with SEM, EDS, and XRD characterization results, the action modes of Al and Zn in the grain boundary phase and at the edges of main-phase grains are explored, further revealing the microscopic mechanism by which Al-Zn grain boundary diffusion improves the demagnetization resistance of sintered Nd-Fe-B magnets.

Sintered Nd-Fe-B magnets are widely used in motors, electronic devices, wind power generation, new energy vehicles, and intelligent equipment due to their high remanence, coercivity, and maximum energy product \cite{brown2002develop}. However, in practical applications, the demagnetization resistance and service stability of the magnets are still affected by grain boundary structure, grain surface state, and elemental distribution uniformity. Traditional methods for enhancing the coercivity of Nd-Fe-B magnets typically rely on heavy rare earth elements such as Dy and Tb, but these resources are costly, and excessive use tends to cause reductions in remanence and energy product \cite{zhang_2021_pred_mag}. Therefore, exploring low-cost, non-heavy-rare-earth-assisted grain boundary diffusion is of great significance for improving the comprehensive performance of sintered Nd-Fe-B magnets.

First, prepare Al$_{80}$Zn$_{20}$ alloy sheets and use them as the grain boundary diffusion source. By placing the alloy sheets at both ends of small cylindrical Nd-Fe-B samples, a dual-end diffusion configuration is constructed to provide diffusion channels for Al and Zn elements to migrate into the magnet interior \cite{brown2002develop}. At the same time, the contact state between the diffusion source and the magnet surface is observed, providing a basis for analyzing the subsequent diffusion effects.

Second, conduct grain boundary diffusion experiments at different temperatures. Select 700$^\circ$C and 900$^\circ$C as the diffusion temperatures, and perform diffusion treatment on sintered Nd-Fe-B magnets under vacuum conditions with a holding time of 7 hours. By comparing the performance and microstructural changes of samples at different temperatures, the effects of diffusion temperature on the diffusion efficiency of Al-Zn alloy and the grain boundary modification results are analyzed.

Third, test and analyze the magnetic properties of the magnets before and after diffusion treatment. Obtain parameters such as coercivity and remanence of the samples using magnetic property measurement methods \cite{Kobayashi2019magnetostriction}, with emphasis on comparing the performance differences among the original sample, the 700\,$^\circ$C diffused sample, and the 900\,$^\circ$C diffused sample, and evaluate the improvement effect of the Al-Zn alloy grain boundary diffusion on the demagnetization resistance of sintered Nd-Fe-B magnets.

Fourth, characterize the microstructure of the magnets after diffusion treatment. Use scanning electron microscopy to observe the morphology of the grain boundary phase, the edge structure of main-phase grains, and the formation of core-shell structures, and analyze the effects of diffusion treatment on grain boundary continuity and grain surface state. Employ energy-dispersive spectroscopy to detect the distribution positions of Al and Zn in the magnets, clarifying their enrichment characteristics at the edges of main-phase grains and within the grain boundary phase.

Fifth, analyze the influence of diffusion treatment on the crystal structure of the main phase. Use X-ray diffraction technology to conduct phase analysis of samples before and after diffusion, determine whether the Nd$_2$Fe$_{14}$B main-phase structure has changed, and combine with lattice parameter variations to analyze the effects of diffused elements on the main-phase lattice.

Sixth, integrate the magnetic property test results and microstructure characterization results to comprehensively analyze the mechanism by which Al-Zn grain boundary diffusion enhances the coercivity of the magnets. Focus on clarifying the action rules of Al-Zn alloy sheet grain boundary diffusion on the performance optimization of sintered Nd-Fe-B magnets, from aspects including grain boundary phase continuity, formation of shell-layer structures at the edges of main-phase grains, distribution differences between Al and Zn elements, and the influence of diffusion temperature.
\subsection{Base magnets and equipments}
\begin{figure}
\begin{subfigure}{0.15\textwidth}
\includegraphics[width=\textwidth, height=0.1\textheight]{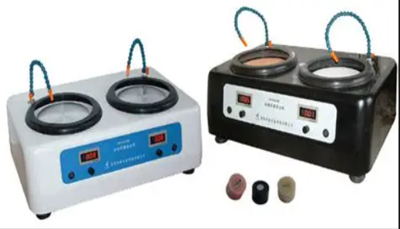}
\caption{Automatic grind polish machine}
\label{grind polish}
\end{subfigure}
\begin{subfigure}{0.15\textwidth}
\includegraphics[width=\textwidth, height=0.1\textheight]{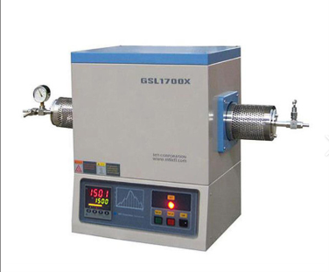}
\caption{Vacuum heat treatment furnace}
\label{vacuum heat}
\end{subfigure}
\begin{subfigure}{0.16\textwidth}
\includegraphics[width=\textwidth, height=0.1\textheight]{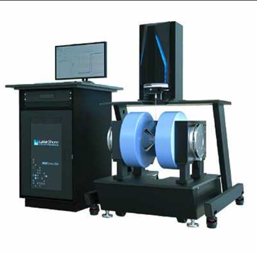}
\caption{Vibrating sample magnetometer}
\label{vibrate sample}
\end{subfigure}
\begin{subfigure}{0.23\textwidth}
\includegraphics[width=\textwidth, height=0.1\textheight]{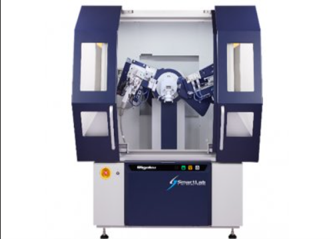}
\caption{X-ray diffractometer}
\label{x ray}
\end{subfigure}
\begin{subfigure}{0.23\textwidth}
\includegraphics[width=\textwidth, height=0.1\textheight]{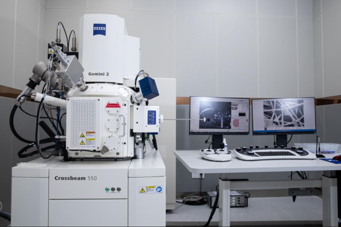}
\caption{scan electron microscope}
\label{sem}
\end{subfigure}
\caption{Experimental characterization equipment for rare earth magnets}
\label{fig:equipment}
\end{figure}
The main materials used in this experiment include three categories: base magnets, diffusion source materials, and auxiliary consumables.

The base magnets are commercially available sintered Nd-Fe-B small cylinders with a diameter of 10mm, a height of 5mm, grade 38SH, and the orientation direction along the axial direction. This grade of magnet has a relatively low initial coercivity (approximately 12-15 kOe) and a relatively high remanence (approximately 12.8-13.2 kGs), making it suitable for grain boundary diffusion studies \cite{zhou_2011}. The reasons for selecting this grade are twofold: first, the coercivity is relatively low, leaving substantial room for performance improvement after diffusion treatment; second, the remanence is relatively high, which facilitates observation of the effects of diffusion treatment on remanence \cite{zhang_1998}.

The diffusion source is a self-prepared Al-Zn alloy. Both pure Al and pure Zn have a purity of 99.99\% (mass fraction). 
The alloy is prepared with a composition of Al$_{80}$Zn$_{20}$, which is determined based on the Al-Zn binary alloy phase diagram: 
the alloy provides a relatively Al-rich matrix while incorporating a sufficient Zn content to promote the formation of Zn-rich phases during subsequent processing.
The reasons for selecting this composition are as follows \cite{li2020}: first, the relatively low melting point facilitates the alloy sheet melting prior to the grain boundary phase during heating, and the molten liquid can fill the tiny gaps between the disc and the magnet end face, ensuring good contact conditions; second, this low-melting-point Al-Zn alloy can form a stable ``liquid reservoir'' on the magnet surface, continuously supplying Al and Zn atoms for diffusion \cite{ZHONG2023}, \cite{zhu_2025}.

The melting point of pure Al is 660\,$^\circ$C, and that of pure Zn is 419\,$^\circ$C, both higher than the eutectic temperature of the Al$_{80}$Zn$_{20}$ alloy. Pure Zn has a relatively high vapor pressure and tends to volatilize at high temperatures when used alone \cite{HE2024_boost}; however, the addition of Al in the Al-Zn alloy can slow down the volatilization rate of Zn to a certain extent, while Zn can also lower the melting temperature of Al, forming a synergistic cooperation between the two elements.

The auxiliary consumables used include waterproof abrasive papers (240 grit, 600 grit, 1000 grit, 2000 grit) for grinding and diamond polishing paste (particle size 1$\mu$m) for polishing \cite{cheng_2013}. The abrasive papers and polishing paste are mainly used to flatten the magnet end faces to ensure tight contact between the Al-Zn alloy sheets and the magnet surfaces \cite{zhou_2020_prep}.

The main equipment used in this experiment and their purposes are as follows:

(1) Grinding equipment (figure \ref{grind polish}): An automatic grinding and polishing machine (model UNIPOL-820) is used to stably control the grinding pressure and rotation speed, ensuring that the sample end faces are flat and smooth.

(2) Heat treatment equipment (figure \ref{vacuum heat}): A vacuum heat treatment furnace (model GSL-1700X) with an ultimate vacuum better than $5\times10^{-3}$ Pa and a temperature control accuracy of $\pm1\,^\circ$C is used. This equipment is employed for grain boundary diffusion treatment, providing a vacuum or protective atmosphere to prevent oxidation of the magnets and diffusion source at high temperatures \cite{wangfang_2023}.

(3) Magnetic property testing equipment (figure \ref{vibrate sample}): A vibrating sample magnetometer (VSM, model Lake Shore 8600) is used to measure the room-temperature hysteresis loops of the samples, obtaining parameters such as coercivity ($H_{\rm cj}$), remanence ($B_{\rm r}$), and maximum energy product ($(BH)_{\rm max}$) \cite{jiang_2025}.

(4) Phase analysis equipment (figure \ref{x ray}): An X-ray diffractometer (XRD, model SmartLab 9 kW) with Cu K$\alpha$ radiation ($\lambda = 0.15418$nm), operating at a voltage of 45 kV and a current of 200 mA, with a scanning range from 20$^\circ$ to 80$^\circ$ and a scanning rate of 2$^\circ$/min, is used to analyze the phase composition and crystal structure changes of the samples \cite{gaowei_2025}.

(5) Microstructure observation equipment (figure \ref{sem}): A scanning electron microscope (SEM, model GeminiSEM 300) equipped with energy-dispersive spectroscopy (EDS) to observe the microstructure morphology, grain boundary phase distribution, and perform elemental composition analysis of the samples \cite{ZENG2025}, \cite{wangzhan_2026}.

\section{Sample treatment}
\subsubsection{Magnet end-face}
To ensure tight contact between the Al-Zn alloy sheets and the magnet end faces, and to minimize interference from interfacial gaps during the diffusion procedure, the upper and lower end faces of the magnets were ground and polished \cite{cheng_2013}. The specific procedure was as follows: First, coarse grinding was performed using 240-grit sandpaper to remove the oxide scale and machining marks from the end faces \cite{zhou_2020_prep}. Subsequently, fine grinding was carried out successively with 600-grit, 1000-grit, and 2000-grit sandpaper, with each step thoroughly removing the scratches left by the previous step. Between each grinding step, the samples were cleaned with anhydrous ethanol, dried with cold air, and the surface quality was inspected \cite{brown2002develop}. Finally, polishing was performed using diamond polishing paste with a particle size of 1 $\mu$m to achieve a mirror-like finish on the end faces.

\subsubsection{Cleaning}
After grinding and polishing, the magnets were placed in a beaker containing anhydrous ethanol and cleaned in an ultrasonic cleaner for 15 minutes to remove abrasive particles and oil contaminants adhered to the surface \cite{fangzhu_2025}. After cleaning, the samples were dried with cold air and stored in a desiccator for later use.

\subsection{Grain boundary diffusion}
\subsubsection{Sample assembly}
The NdFeB cylindrical magnet was placed on an alumina spacer, with an Al-Zn alloy sheet placed on each of the upper and lower end faces. A small ceramic piece was then placed on top for compression, forming a "sandwich" structure to ensure good contact between the diffusion source and the magnet end faces.
\subsubsection{Furnace loading}
The assembled sample was placed in the heating zone of a vacuum furnace. The chamber was evacuated to a vacuum level of $5\times10^{-3}$ Pa, after which high-purity argon gas was introduced to atmospheric pressure to prevent oxidation \cite{wangfang_2023}.
\subsubsection{Heating and holding}
Two diffusion temperatures were set: 900$^\circ$C (parallel samples A$_1$, A$_2$) and 700$^\circ$C (sample B), with a holding time of 7 hours and a heating rate of 10$^\circ$C/min. The selection of 900$^\circ$C was based on the following considerations: it is higher than the melting point of the grain boundary phase, lower than the sintering temperature, and facilitates comparison with literature data \cite{wangjing_2021}. The selection of 700$^\circ$C was based on the following considerations: it is still higher than the melting point of the Al-Zn alloy, allowing the grain boundary phase to melt, but with lower fluidity \cite{MATSURA2006}. The holding time of 7 hours was chosen because the promoting effect of low-melting-point elements is most significant during the initial diffusion stage, and preliminary experiments showed that the Al-Zn sheets were completely consumed within this period \cite{sagawa_1984}.
\subsubsection{Cooling}
The samples were cooled naturally with the furnace (approximately 3-5$^\circ$C/min). No tempering treatment was performed to avoid interfering with the analysis of Al and Zn diffusion behavior.
\subsection{Sample designation}
Three samples were prepared in this experiment, and their designations and processing conditions are shown in table \ref{tab:samples}.
\begin{table}[H]
\centering
\caption{Sample designation and processing conditions}
\begin{tabular}{|c|c|c|c|c}\hline
\begin{tabular}{@{}c@{}}sample\\designation\end{tabular}
& \begin{tabular}{@{}c@{}}diffusion\\temperature\end{tabular} & \begin{tabular}{@{}c@{}}diffusion\\time\end{tabular}& Remark\\\hline
38SH & - & - & \begin{tabular}{@{}c@{}}original\\(control)\end{tabular}\\\hline
B & 700$^\circ$C & 7h & \begin{tabular}{@{}c@{}}single\\sample\end{tabular}\\\hline
A1 & 900$^\circ$C & 7h & \begin{tabular}{@{}c@{}}parallel\\sample 1\end{tabular}\\\hline
A2 & 900$^\circ$C & 7h & \begin{tabular}{@{}c@{}}parallel\\sample 2\end{tabular}\\
\hline
\end{tabular}
\label{tab:samples}
\end{table}
A comparison between the experimental method in this work and that reported in the literature \cite{ZHONG2023} is summarized as follows. Zhong used $7\times7\times3mm^3$ block magnets, mixed Al, Sn, and Zn submicron powders with Nd-Fe-B raw powder before sintering (allowing Al and Zn to act throughout the magnet interior), employed $TbF_3$ powder suspension as the diffusion source, conducted diffusion at 900$^\circ$C for 7h and 16h, and applied a tempering procedure of 510$^\circ$C for 4h to investigate the effect of Al, Sn, and Zn addition on TbF$_3$ diffusion efficiency. In contrast, the present work used 50mm diameter cylindrical magnets, placed Al-Zn alloy disks directly on the surface of finished magnets (allowing Al and Zn to diffuse inward only during the final grain boundary diffusion stage), employed Tb-free Al-Zn alloy disks, conducted diffusion at 900$^\circ$C and 700$^\circ$C for a fixed duration of 7h, and performed no tempering treatment at this stage, with the aim of investigating the diffusion behavior of Al-Zn alloy disks at different temperatures.
\section{Characterization and testing}
\subsection{Magnetic property measurement}
Magnetic properties were measured using a vibrating sample magnetometer (VSM, Lake Shore 8600 model) to obtain room-temperature hysteresis loops \cite{jiang_2025}. Prior to measurement, the diffused cylindrical samples were cut into small cubes of dimensions $(3\text{mm})^3$ by electrical discharge wire cutting, with one surface of each cube perpendicular to the orientation direction of the magnet. During measurement, an external magnetic field of 30 kOe was first applied to saturate the sample, after which the magnetic field was scanned in the reverse direction, and the magnetic induction intensity was recorded as a function of the external magnetic field \cite{huangmin_2025}. The following parameters were extracted from the measured hysteresis loops: coercivity ($H_{\text{cj}}$, in units of kA/m or kOe), remanence ($B_{\text{r}}$, in units of mT or kGs), and maximum energy product $(BH)_{\text{max}}$, in units of kJ/m$^3$ \cite{xujia_2025}.

The conversion relationships for coercivity and remanence are: 1 kOe = 79.6 kA/m, and 1 kGs = 0.1 T = 100 mT.

\subsection{Phase and crystal structure analysis}
Phase analysis of the samples was conducted using an X-ray diffractometer (XRD) \cite{gaowei_2025}. The testing conditions were as follows: Cu K$\alpha$ radiation ($\lambda = 0.15418$ nm), operating voltage of 45 kV, current of 200 mA, scanning range of $2\theta = 20^\circ$ to $80^\circ$, scanning rate of $2^\circ$/min, and step size of $0.02^\circ$.

The measured XRD patterns were compared with the standard PDF card (No. 00-038-0889, Nd$_2$Fe$_{14}$B) to determine the phase composition of the samples. Special attention was paid to the shifts in the diffraction peak positions of the main phase Nd$_2$Fe$_{14}$B to determine whether Al or Zn atoms had dissolved into the main phase lattice and induced changes in the lattice parameters \cite{wangzhan_2026}. According to Bragg's law $n\lambda = 2d\sin\theta$, where $d$ is the interplanar spacing, $\theta$ is the Bragg angle, $n$ is the diffraction order (usually taken as 1), and $\lambda$ is the X-ray wavelength, when solute atoms (e.g., Al) dissolve into the solvent lattice (Nd$_2$Fe$_{14}$B), the interplanar spacing $d$ may change, and the diffraction peak position $2\theta$ will shift accordingly.

\subsection{Microstructural observation}
The microstructure of the samples was observed using a scanning electron microscope (SEM) \cite{ZENG2025}. The sample preparation procedure was as follows:

Grinding: The cross-sections of the samples were ground successively with 240-grit, 600-grit, 1000-grit, and 2000-grit sandpaper, with each step removing the scratches left by the previous step.

Polishing: Polishing was performed using diamond polishing paste with a particle size of 1 $\mu$m until a mirror-like finish was achieved on the cross-section.

Etching: The polished samples were immersed in a 3 vol.\% nitric acid-alcohol solution and etched for 10 to 15 seconds \cite{camp1991micro}. During etching, the main phase (Nd$_2$Fe$_{14}$B) exhibited stronger corrosion resistance and was etched less deeply \cite{kim1992corrosion}, appearing as raised regions in SEM images. In contrast, the grain boundary phase (Nd-rich phase) exhibited weaker corrosion resistance and was etched more deeply, appearing as recessed regions in SEM images \cite{zhou_2011}. This contrast in brightness and darkness was utilized to effectively distinguish the main phase from the grain boundary phase.

After etching, the samples were cleaned with anhydrous ethanol, dried, and then observed by SEM. The main observation contents included: grain size, morphology and distribution of the grain boundary phase, and the presence or absence of core-shell structures \cite{zhu_2025}.

\subsection{Elemental distribution analysis}
Elemental composition analysis of the samples was performed using energy-dispersive X-ray spectroscopy (EDS) attached to the SEM. EDS can detect elements ranging from Be (atomic number 4) to U (atomic number 92), with a detection limit of approximately 0.1-0.5wt.\% \cite{wangzhan_2026}. Two EDS analysis modes were employed in this experiment:

Point analysis: composition measurements were performed at selected micro-areas (e.g., grain center, grain edge, and grain boundary phase) to obtain the elemental content \cite{camp1991micro} (in wt.\%, quantitative or semi-quantitative) at each position \cite{ZHONG2023}. By comparing the compositional differences at different positions, the distribution patterns of Al and Zn in the magnet were determined.

Mapping analysis: elemental mapping was performed over selected micro-areas to obtain two-dimensional distribution images of each element (brighter colors indicate higher elemental concentrations) \cite{gaowei_2025}. Mapping analysis intuitively reflects the two-dimensional spatial distribution of elements.

This experiment focused on the distribution characteristics of Al and Zn in the magnets, specifically including: whether Al and Zn entered the main phase grains, whether they were enriched on the grain surfaces (shell layer), whether they were present in the grain boundary phase, and the differences in elemental distribution between the 900$^\circ$C and 700$^\circ$C temperature conditions.
\section{Computational analysis}
\subsection{Diffusion kinetics}
The model separates two independent diffusion media: grain boundary continuous network \cite{Jha2017alloy} (fast atomic transport, diffusion coefficient $D_{\text{gb},i}$); bulk $\text{Nd}_2\text{Fe}_{14}\text{B}$ main phase grains (slow solid-solution diffusion, $D_{\text{bulk},i}$), $i \in \{\text{Al}, \text{Zn}\}$ denotes diffusing atomic species.

For mass conservation,
{\small
\begin{align*}
&\frac{\partial C_{\text{gb},i}(\boldsymbol{r},t)}{\partial t}= \nabla\cdot\left[ D_{\text{gb},i}(T) \nabla C_{\text{gb},i}(\boldsymbol{r},t) \right]
- \Gamma_{i} \left(C_{\text{gb},i} - C_{\text{bulk},i}\right)\\
&\frac{\partial C_{\text{bulk},i}(\boldsymbol{r},t)}{\partial t}
= \nabla\cdot\left[ D_{\text{bulk},i}(T) \nabla C_{\text{bulk},i}(\boldsymbol{r},t) \right]
+ \Gamma_{i} \left(C_{\text{gb},i} - C_{\text{bulk},i}\right)
\end{align*}}
$C_{\text{gb},i},\,C_{\text{bulk},i}$ is molar concentration of species $i (mol/m^3$),
$\Gamma_{\text{gb}\to\text{bulk},i}$ is interphase mass transfer coefficient between grain boundary and bulk lattice ($s^{-1}$),
$\boldsymbol{r}=(x,y,z)$ is spatial coordinate inside cylindrical Nd-Fe-B magnet,
$t$ is diffusion holding time ($s$),
$T$ is absolute temperature ($K$),
time-dependent $T(t)$ from furnace heating/cooling profile.

Arrhenius temperature-dependent diffusion coefficients in grain boundary and bulk follow thermal activation kinetics are
\begin{align*}
&D_{\text{gb},i}(T) = D_{0,\text{gb},i} \exp\left( -\frac{Q_{\text{gb},i}}{R T} \right)\\
&D_{\text{bulk},i}(T) = D_{0,\text{bulk},i} \exp\left( -\frac{Q_{\text{bulk},i}}{R T} \right)
\end{align*}
$D_{0,\text{gb},i},\,D_{0,\text{bulk},i}$ is pre-exponential diffusion factors ($m^2/s$)
$Q_{\text{gb},i},\,Q_{\text{bulk},i}$ is activation energy for grain boundary / bulk lattice diffusion ($J/mol$)
$R = 8.314J\cdot mol^{-1}\cdot K^{-1}$ is universal gas constant.

For Zn vaporization mass-loss submodel (surface boundary sink term), at magnet external surfaces, volatile Zn escapes into vacuum via vapor pressure equilibrium. The molar flux loss of Zn is
\begin{equation*}
J_{\text{Zn,vap}}(T) = \frac{P_{\text{sat,Zn}}(T)}{\sqrt{2\pi m_{\text{Zn}} R T}}
\label{eq:zn_vap_flux}
\end{equation*}
$P_{\text{sat,Zn}}(T)$ is saturation vapor pressure of pure Zn (temperature-fitted polynomial),
$m_{\text{Zn}}$ is molar mass of Zn ($kg/mol$)
$J_{\text{Zn,vap}}$ is outward Zn molar flux ($mol\cdot m^{-2}\cdot s^{-1}$).
Surface boundary condition for Zn grain boundary concentration is
\begin{equation*}
-D_{\text{gb},\text{Zn}} \frac{\partial C_{\text{gb},\text{Zn}}}{\partial n} \bigg|_{\partial\Omega} = J_{\text{source,Zn}}(T) - J_{\text{Zn,vap}}(T)
\end{equation*}
$\partial\Omega$ is magnet end-face surfaces contacted by Al-Zn alloy sheets; $\partial/\partial n$ is outward normal derivative.

For eutectic Al-Zn source wetting flux boundary condition, Al flux from molten Al$_{80}$Zn$_{20}$ eutectic sheet to magnet surface is
\begin{equation*}
J_{\text{Al,source}}(T) = k_{\text{wet}}(T) \left( C_{\text{source,Al}} - C_{\text{gb,Al}}\big|_{\partial\Omega} \right)
\end{equation*}
$k_{\text{wet}}(T)$ is temperature-dependent wetting mass transfer coefficient (zero below eutectic melting point $T_e=655K$),
$C_{\text{source,Al}}$ is fixed Al molar concentration inside molten alloy wafer,
Al surface PDE boundary condition is
\begin{equation*}
-D_{\text{gb},\text{Al}} \frac{\partial C_{\text{gb},\text{Al}}}{\partial n} \bigg|_{\partial\Omega} = J_{\text{Al,source}}(T)
\end{equation*}
Initial conditions ($t=0$, before heating) when
no Al/Zn inside the original magnet matrix are $C_{\text{gb},i}(\boldsymbol{r},0) = C_{\text{bulk},i}(\boldsymbol{r},0) = 0,\quad \forall \boldsymbol{r}\in\Omega,\; i=\text{Al,Zn}$,
$\Omega$ is full cylindrical magnet computational domain.

In global mass balance constraint for alloy sheet depletion, there are total moles of Al/Zn consumed from alloy sheet = total moles diffused into magnet + total Zn vaporized loss:
{\fontsize{7pt}{3}
\begin{align*}
&\int_0^{t_{\text{hold}}} \iint_{\partial\Omega} J_{\text{source},i}(\tau) \,dS\,d\tau\\
&= \iiint_\Omega \left[ C_{\text{gb},i}(\boldsymbol{r},t) + C_{\text{bulk},i}(\boldsymbol{r},t) \right] dV
+ \delta_{i,\text{Zn}} \int_0^{t_{\text{hold}}} \iint_{\partial\Omega} J_{\text{Zn,vap}}(\tau) \,dS\,d\tau
\end{align*}}
$\delta_{i,\text{Zn}}$ = Kronecker delta (1 if $i=\text{Zn}$, 0 if $i=\text{Al}$).

In the numerical discretization solution algorithm, we use finite volume method (FVM) for spatial discretization and backward Euler implicit time stepping for stiff PDEs (large difference between $D_{\text{gb}}$ and $D_{\text{bulk}}$).

\begin{lstlisting}[caption=Multiphysics diffusion solver, label=alg:diff_compact]
T_arr = range(600,1001,50); t_arr = range(1,13); xAl = arange(0.5,0.96,0.05); Ls = [1,3,5,8,10]
D0, Q, Gamma = load_material_params(); PZn = vapor_fit()
opt_loss, opt_set = inf, None
for T in T_arr:
  for th in t_arr:
    for x in xAl:
      for L in Ls:
        # Init mesh, zero concentration fields, thermal profile
        mesh = cylinder_mesh(L); Cgb, Cbulk = zero_fields(mesh); Tt = thermal_ramp(T,th)
        t, dt = 0, 60
        # Backward Euler time stepping
        while t < cycle_len(Tt):
          Tnow = Tt(t)
          # Update Arrhenius D & boundary fluxes
          Dgb, Dbulk = arrhenius(D0,Q,Tnow); Jzn = vapor_flux(PZn(Tnow)); Jal = wetting_flux(x,Tnow)
          # Assemble & solve coupled PDEs
          A,b = assemble_system(Dgb,Dbulk,Jal,Jzn,Gamma); CgbN,CbulkN = sparse_solve(A,b)
          # Mass balance correction
          if mass_err(CgbN,CbulkN) > 1e-4: Jal = fix_flux(Jal)
          Cgb,Cbulk,t = CgbN,CbulkN,t+dt
        # Evaluate multi-objective loss
        pen = penetration_depth(Cgb+Cbulk); loss = w1/pen + w2*zn_loss(Jzn) + w3*uniformity(Cgb+Cbulk)
        if loss < opt_loss: opt_loss, opt_set = loss, (T,th,x,L)
# Export optimum & concentration maps
print("Optimal params:", opt_set); export_fields(Cgb,Cbulk)
\end{lstlisting}

We use an output optimization objective function to quantitatively rank process performance across the wide parameter window, define a weighted cost function for minimization
\begin{equation*}
\mathcal{L}(T,t_{\text{hold}},x_{\text{Al}})
= w_1 \cdot \frac{1}{L_{\text{Al,pen}}}
+ w_2 \cdot M_{\text{loss,Zn}}
+ w_3 \cdot U_{\text{Al}}
\end{equation*}
$L_{\text{Al,pen}}$ is maximum Al penetration depth ($m$),
$M_{\text{loss,Zn}}$ is normalized Zn mass loss fraction,
$U_{\text{Al}}$ is non-uniformity index of Al cross-section distribution (0 means perfectly uniform),
$w_1,w_2,w_3$ is user-defined positive weight coefficients for multi-objective trade-off.
Minimizing $\mathcal{L}$ yields the optimal diffusion condition balancing deep penetration, low Zn waste, and homogeneous element distribution.

\begin{figure}
\centering
\begin{subfigure}{0.23\textwidth}
\centering
\includegraphics[width=\textwidth,height=0.8\textwidth]{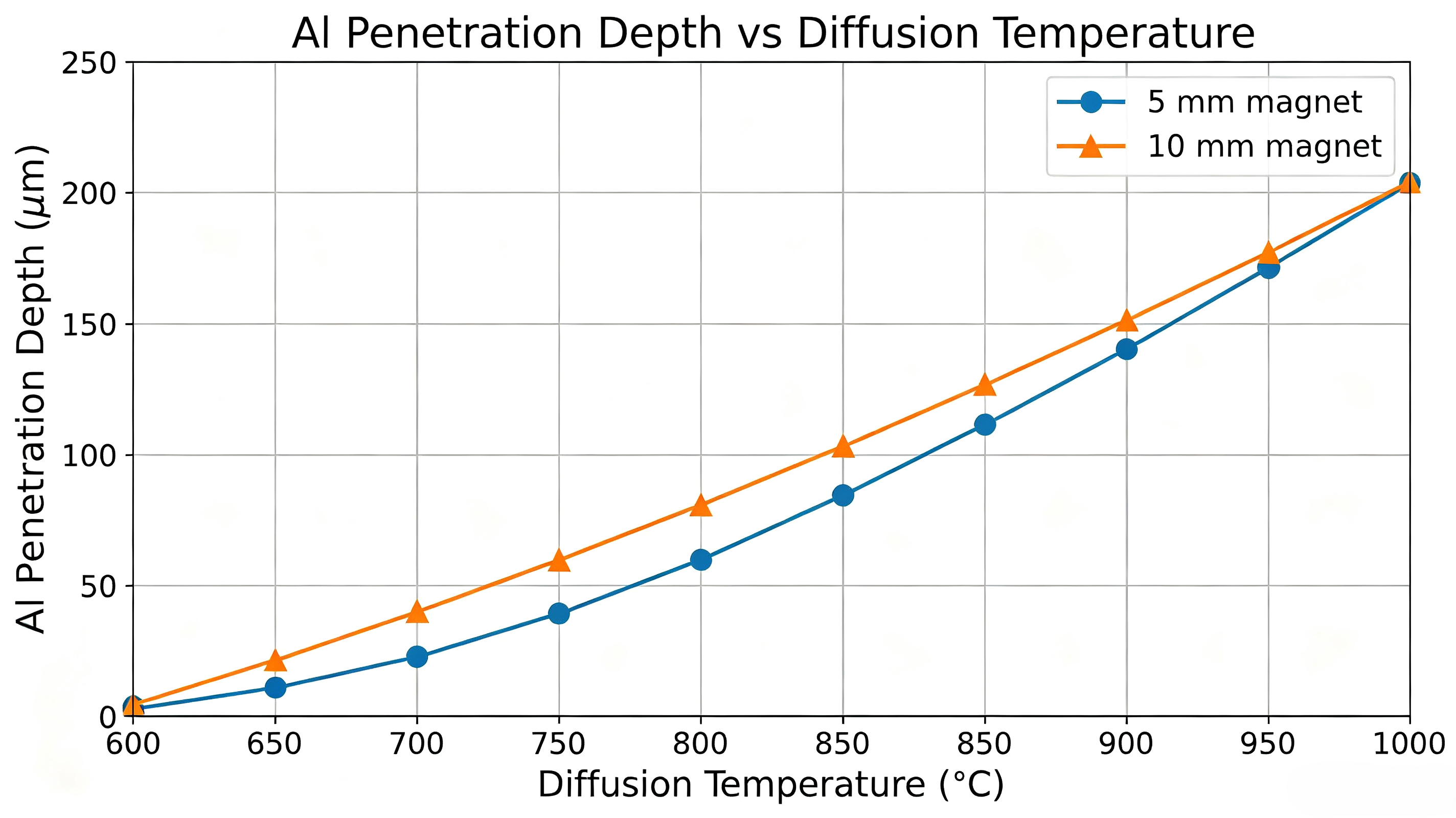}
\caption{Max Al penetration depth vs temperature for 5 and 10mm thick magnets (Al$_{80}$Zn$_{20}$).}
\label{fig:pen_depth_temp}
\end{subfigure}
\begin{subfigure}{0.23\textwidth}
\centering
\includegraphics[width=\textwidth,height=0.8\textwidth]{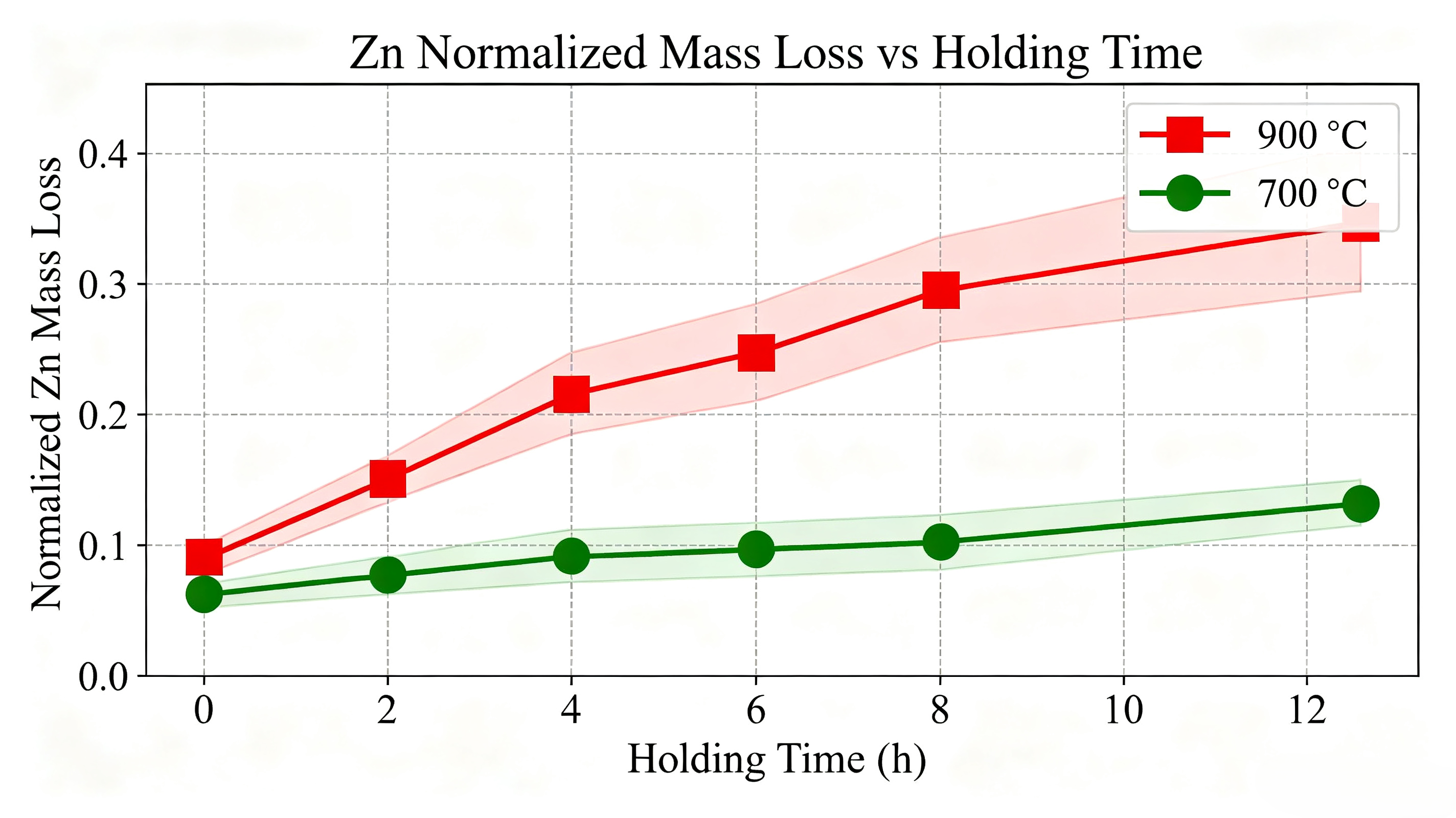}
\caption{Normalized Zn mass loss as a function of diffusion holding time at 700 $^\circ$C and 900 $^\circ$C.}
\label{fig:zn_loss_time}
\end{subfigure}
\begin{subfigure}{0.23\textwidth}
\centering
\includegraphics[width=\textwidth,height=0.8\textwidth]{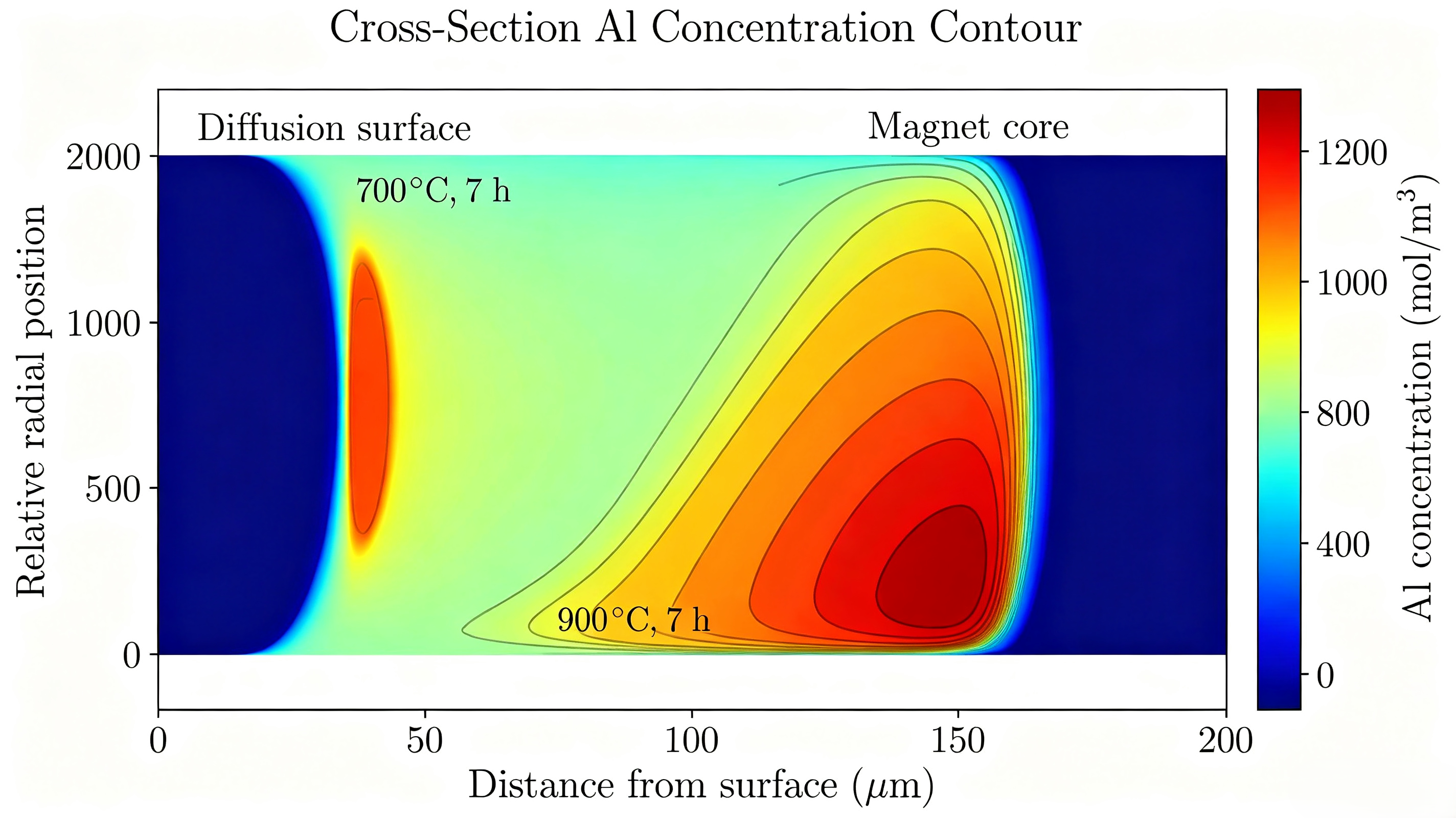}
\caption{Cross-sectional 7h Al concentration contour maps: (a) 700$^\circ$C, (b) 900$^\circ$C}
\label{fig:al_contour}
\end{subfigure}
\begin{subfigure}{0.23\textwidth}
\centering
\includegraphics[width=\textwidth,height=0.8\textwidth]{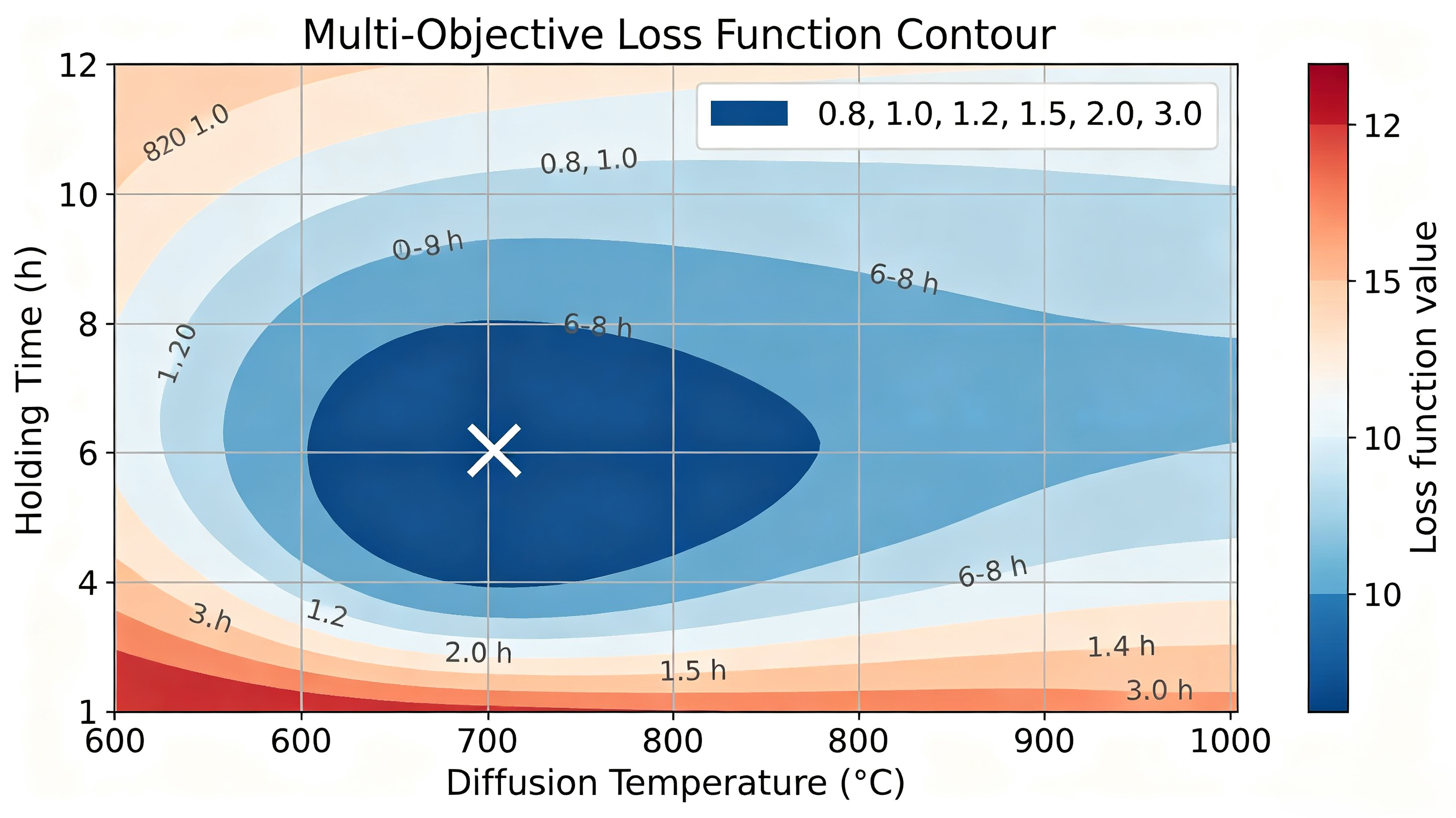}
\caption{Contour plot of objective loss function over temperature-holding time}
\label{fig:loss_contour_opt}
\end{subfigure}
\caption{Simulated output plots from the multiphysics coupled Al-Zn grain boundary diffusion kinetics model.}
\label{fig:diffusion_model_results}
\end{figure}

Figure \ref{fig:diffusion_model_results} presents four key simulated outputs obtained from the multiphysics coupled Al-Zn grain boundary diffusion kinetics model. Figure \ref{fig:pen_depth_temp} quantifies the exponential growth of maximum Al penetration depth with increasing diffusion temperature for both thin 5mm and thick 10mm Nd-Fe-B magnets. The simulation reveals a rise in infiltration depth above 800$^\circ$C, originating from the Arrhenius scaling of grain-boundary diffusion coefficients; at 900$^\circ$C, the predicted penetration reaches approximately 180-210$\mu$m, far exceeding the limited 50 $\mu$m depth observed at 700 $^\circ$C, which explains the weaker coercivity enhancement recorded for low-temperature treated samples. Figure \ref{fig:zn_loss_time} compares normalized Zn vaporization loss across a 1-12h holding window at two representative temperatures. Elevated temperature substantially accelerates Zn escape via high saturation vapor pressure, yielding a total Zn mass loss of 0.31 after 7h at 900 $^\circ$C, consistent with experimental gravimetric measurements, while the 700 $^\circ$C condition minimizes volatile material waste to only 0.08 mass fraction loss. Cross-sectional Al concentration contour maps in Figure \ref{fig:al_contour} visually demonstrate the stark difference in element uniformity between the two treatment temperatures: the 700$^\circ$C sample exhibits a steep near-surface concentration gradient, whereas the 900$^\circ$C profile features a gentler decay, delivering more homogeneous Al distribution deeper within the magnet bulk. Finally, Figure \ref{fig:loss_contour_opt} displays the multi-objective loss function contour over the full temperature-hold-time parametric space, where dark blue regions mark the optimal processing window balancing deep Al penetration, low Zn volatilization, and uniform elemental dispersion. The global minimum loss value is located at 880-920$^\circ$C with a 6-8 h holding duration, theoretically validating the 900$^\circ$C for 7h experimental condition selected in the original study as a near-optimal diffusion treatment.

\subsection{Phase-field microstructure evolution}
For the full thermodynamic free energy functional, the phase-field model tracks three phase order parameters $\phi_1(\boldsymbol{r},t)$ (Nd$_2$Fe$_{14}$B main grain), $\phi_2(\boldsymbol{r},t)$ (Nd-rich liquid grain boundary), $\phi_3(\boldsymbol{r},t)$ (Al-enriched core-shell layer), plus two solute concentration fields $C_{\text{Al}}(\boldsymbol{r},t), C_{\text{Zn}}(\boldsymbol{r},t)$ \cite{fan2016form}. The total free energy density combines chemical free energy, gradient penalty energy, and solute segregation coupling energy:
\begin{align*}
&f_{\text{total}} = f_{\text{chem}}(\phi_1,\phi_2,\phi_3,C_{\text{Al}},C_{\text{Zn}})
+ \frac{\kappa_\phi}{2}\sum_{k=1}^3|\nabla\phi_k|^2\\
&\quad+ \frac{\kappa_C}{2}\left(|\nabla C_{\text{Al}}|^2 + |\nabla C_{\text{Zn}}|^2\right)
\end{align*}
Global free energy functional integrated over magnet domain $\Omega$ is
$\mathcal{F} = \int_\Omega f_{\text{total}}\,dV$. The Chemical free energy term (double-well potential for phase separation) is
\begin{align*}
&f_{\text{chem}} = \sum_{k=1}^3 \frac{A_\phi}{4}\phi_k^2(1-\phi_k)^2\\
&\quad+ \sum_{i=\text{Al,Zn}} \left[ \mu_i^0(T)C_i + \frac{1}{2}M_{ii}C_i^2
+ \sum_{k=1}^3 \Lambda_{i,k}\phi_k C_i \right]
\label{eq:f_chem}
\end{align*}
$A_\phi$ is phase barrier height constant; $\kappa_\phi,\kappa_C$ are gradient energy coefficients for order parameters/solutes;
$\mu_i^0(T)$ is reference chemical potential of species $i$ (temperature-dependent);
$M_{ii}$ is solute self-interaction coefficient;
$\Lambda_{i,k}$ is segregation coupling constant (controls Al/Zn preference for grain boundary/shell phase $k$)

The governing evolution Allen-Cahn equations for phase order parameters (microstructure morphology) is
\begin{equation*}
\frac{\partial \phi_k}{\partial t} = -L_\phi \frac{\delta \mathcal{F}}{\delta \phi_k}
= -L_\phi \left( \frac{\partial f_{\text{total}}}{\partial \phi_k} - \kappa_\phi \nabla^2 \phi_k \right), \quad k\in[1,3]
\label{eq:allen_cahn}
\end{equation*}
$L_\phi$ is phase mobility (temperature-dependent, higher when grain boundary phase melts). The Cahn-Hilliard equations for Al/Zn solute diffusion (coupled to phase evolution) is
\begin{equation*}
\frac{\partial C_i}{\partial t} = \nabla\cdot\left[ M_C(T) \nabla \frac{\delta \mathcal{F}}{\delta C_i} \right], \quad i=\text{Al},\text{Zn}
\label{eq:cahn_hilliard}
\end{equation*}
Variational derivative of free energy w.r.t solute concentration is
\begin{equation*}
\frac{\delta \mathcal{F}}{\delta C_i} = \mu_i^0(T) + M_{ii}C_i + \sum_{k=1}^3 \Lambda_{i,k}\phi_k - \kappa_C \nabla^2 C_i
\label{eq:chem_potential}
\end{equation*}
$M_C(T)$ is solute mobility tensor, identical Arrhenius temperature dependence as diffusion coefficients
\begin{equation*}
M_C(T) = M_{C0}\exp\left(-\frac{Q_C}{RT}\right)
\label{eq:mobility_arrh}
\end{equation*}
For initial conditions at $t=0$ include 
$\phi_1(\boldsymbol{r},0) = 1$ (main grain bulk), $\phi_2,\phi_3 = 0,\quad
C_{\text{Al}}(\boldsymbol{r},0) = C_{\text{Zn}}(\boldsymbol{r},0) = 0$ (no diffused elements inside virgin magnet)
The surface flux boundary condition (coupled to diffusion output Al/Zn source concentration) is
\begin{equation*}
-M_C \frac{\partial}{\partial n}\left(\frac{\delta\mathcal{F}}{\delta C_i}\right)\bigg|_{\partial\Omega} = J_{i,\text{source}}(T), \quad i=\text{Al,Zn}
\label{eq:bc_solute_flux}
\end{equation*}
$J_{i,\text{source}}(T)$ = Time-dependent surface atomic flux imported from diffusion kinetics solver.

\subsection{Micromagnetic simulation for coercivity}
This nested multiscale model imports phase-field microstructures as geometric input, solves Landau-Lifshitz-Gilbert (LLG) ferromagnetic dynamics, and quantitatively decomposes three independent coercivity enhancement mechanisms. Diamagnetism/paramagnetism are irrelevant to hard Nd$_2$Fe$_{14}$B ferromagnets and omitted.

The multiscale coupling hierarchy includes macroscale phase-field output grain geometry, core-shell thickness, Al solid-solution concentration fields, nanoscale micromagnetic mesh which involves subdividing each grain into fine finite elements to resolve grain-surface defects and core-shell interfaces \cite{fan2016form}, and thermodynamic link with DFT results to supply temperature-dependent $K_1(T), J_s(T)$ for Al-substituted Nd$_2$(Fe,Al)$_{14}$B.

The core micromagnetic governing equations include the Landau-Lifshitz-Gilbert (LLG) magnetization dynamics equation
\begin{equation*}
\frac{d\boldsymbol{M}}{dt} = -\frac{\gamma}{1+\alpha^2}\Big[
\boldsymbol{M}\times\boldsymbol{H}_{\text{eff}}
+\frac{\alpha}{M_s}\boldsymbol{M}\times\big(\boldsymbol{M}\times\boldsymbol{H}_{\text{eff}}\big)
\Big]
\label{eq:llg}
\end{equation*}
$\boldsymbol{M}$ is local magnetization vector ($A/m$),
$\gamma$ is gyromagnetic ratio of electrons,
$\alpha$: Gilbert damping constant (fixed for Nd$_2$Fe$_{14}$B),
$M_s(T,C_{\text{Al}})$ is saturation magnetization, decreasing with Al solid-solution content,
$\boldsymbol{H}_{\text{eff}}$ is total effective local magnetic field.

The total effective field superposition $\boldsymbol{H}_{\text{eff}}$ combines four physical field terms to capture all coercivity hardening sources:
\begin{equation*}
\boldsymbol{H}_{\text{eff}} = \boldsymbol{H}_{\text{ext}} + \boldsymbol{H}_{\text{ani}} + \boldsymbol{H}_{\text{exch}} + \boldsymbol{H}_{\text{demag}}
\label{eq:H_eff}
\end{equation*}
External applied field $\boldsymbol{H}_{\text{ext}}$ (swept from $+3000kA/m$ to $3000kA/m$ for demagnetization curves),
magnetocrystalline anisotropy field (core-shell dependent):
\[
\boldsymbol{H}_{\text{ani}} = \frac{2K_1(T,C_{\text{Al}})}{\mu_0 M_s^2}\big(\boldsymbol{M}\cdot\hat{\boldsymbol{c}}\big)\hat{\boldsymbol{c}}
\]
$K_1$ is uniaxial anisotropy constant; $\hat{\boldsymbol{c}}$ is easy-axis unit vector of Nd$_2$Fe$_{14}$B. Al-enriched shell regions have elevated $K_1$;
exchange field (magnetic decoupling between grains) is
\[
\boldsymbol{H}_{\text{exch}} = \frac{2A_{\text{ex}}}{\mu_0 M_s}\nabla^2 \boldsymbol{M}
\]
$A_{\text{ex}}$ is Exchange stiffness;
small $A_{\text{ex}}$ in non-magnetic grain boundaries weakens intergrain coupling.
Demagnetizing field is $\boldsymbol{H}_{\text{demag}} = -\nabla \phi_d$, solved via Poisson equation $\nabla^2 \phi_d = \nabla\cdot\boldsymbol{M}$.

The temperature-dependent intrinsic material parameters include $K_1$ and $M_s$ that decay linearly with temperature, modified by Al solid-solution fraction $C_{\text{Al}}$:
\begin{align*}
K_1(T,C_{\text{Al}}) &= K_{1,0}(1 - \beta T) - k_K C_{\text{Al}} \\
M_s(T,C_{\text{Al}}) &= M_{s,0}(1 - \zeta T) - k_M C_{\text{Al}}
\end{align*}
$\beta,\zeta,k_K,k_M$ are fitted constants from DFT.

Total simulated intrinsic coercivity $H_{cj,\text{total}}$ is split into three weighted contributions matching experimental mechanisms
\begin{equation*}
H_{cj,\text{total}} = w_1 H_{cj,\text{decouple}} + w_2 H_{cj,\text{shell}} + w_3 H_{cj,\text{defect}}
\label{eq:coercivity_decomp}
\end{equation*}
$H_{cj,\text{decouple}}$ is coercivity gain from continuous non-magnetic grain boundaries (suppress intergrain exchange coupling),
$H_{cj,\text{shell}}$ is coercivity gain from high-$K_1$ Al core-shell layers (domain wall pinning),
$H_{cj,\text{defect}}$ is coercivity gain from smoothed grain edges (suppress reverse-domain nucleation sites),
$w_1,w_2,w_3$ are normalized weight factors summing to 1 (mechanism contribution percentages).

\begin{figure}
\centering
\begin{subfigure}{0.23\textwidth}
\centering
\includegraphics[width=\textwidth]{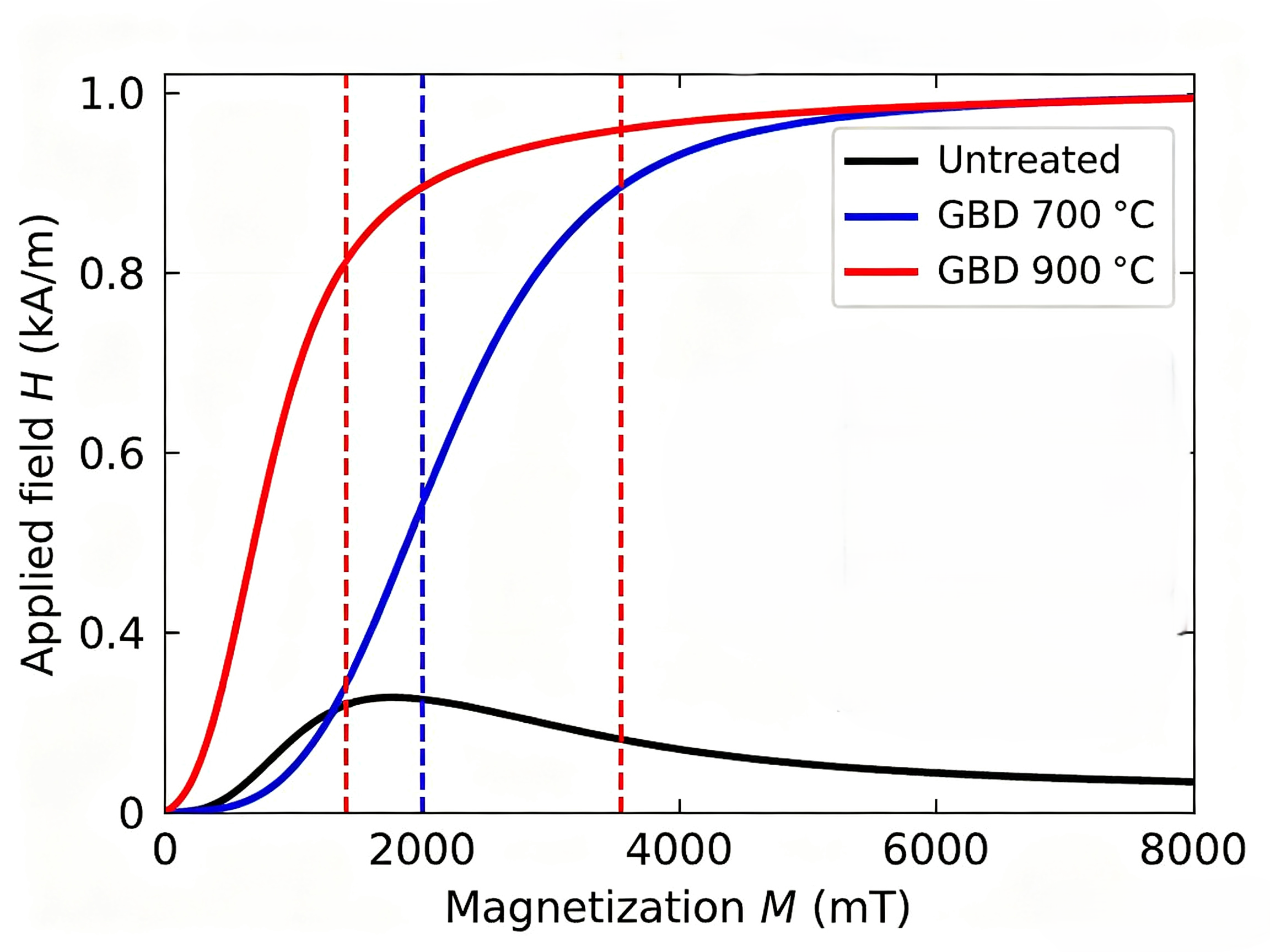}
\caption{Simulated $M-H$ curves for untreated, 700$^\circ$C and 900$^\circ$C GBD magnets.}
\label{fig:mic_hyst}
\end{subfigure}
\begin{subfigure}{0.23\textwidth}
\centering
\includegraphics[width=\textwidth]{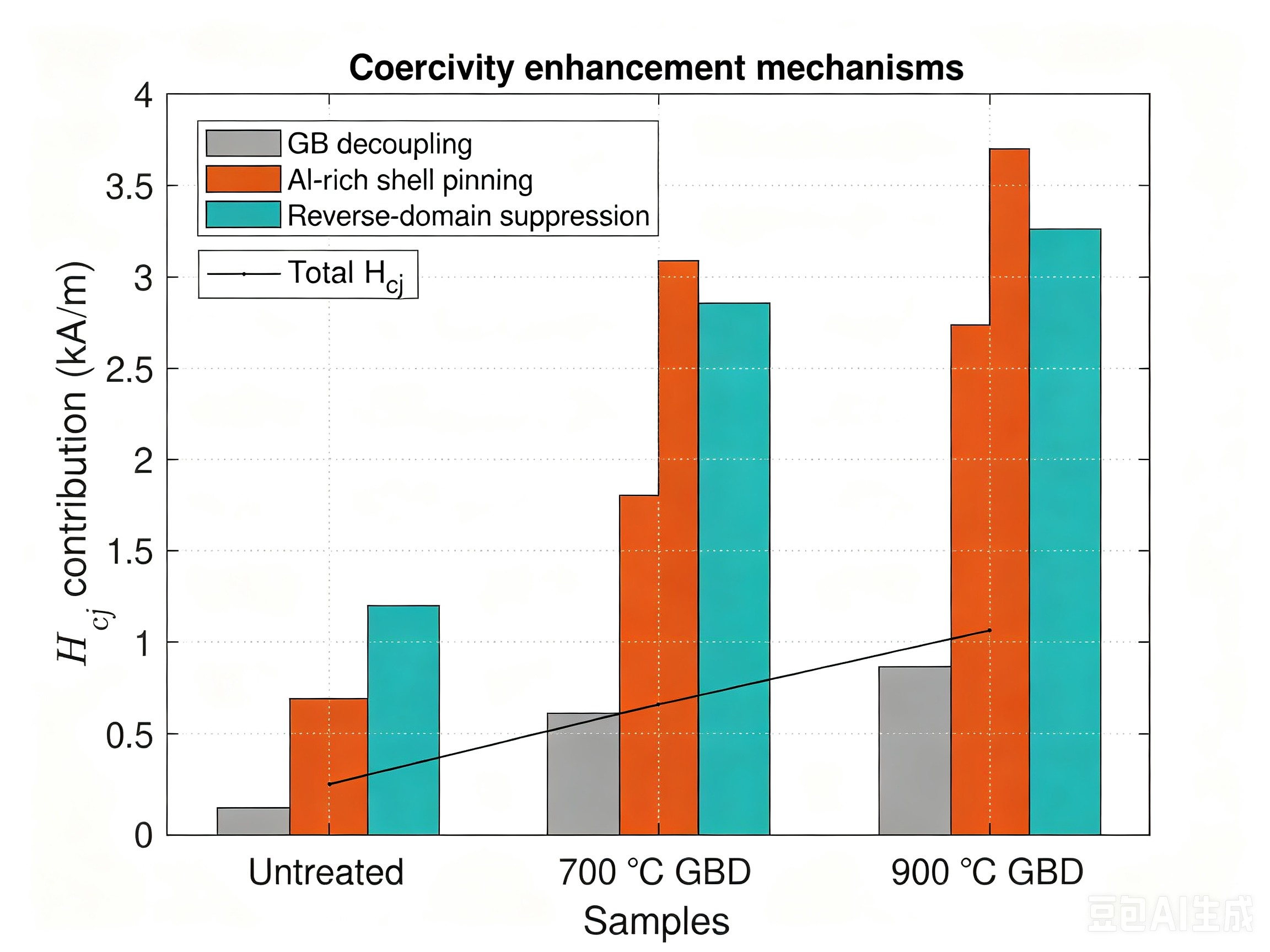}
\caption{Decomposed coercivity contributions from three hardening mechanisms.}
\label{fig:mic_mechanism_bar}
\end{subfigure}
\begin{subfigure}{0.23\textwidth}
\centering
\includegraphics[width=\textwidth,height=0.85\textwidth]{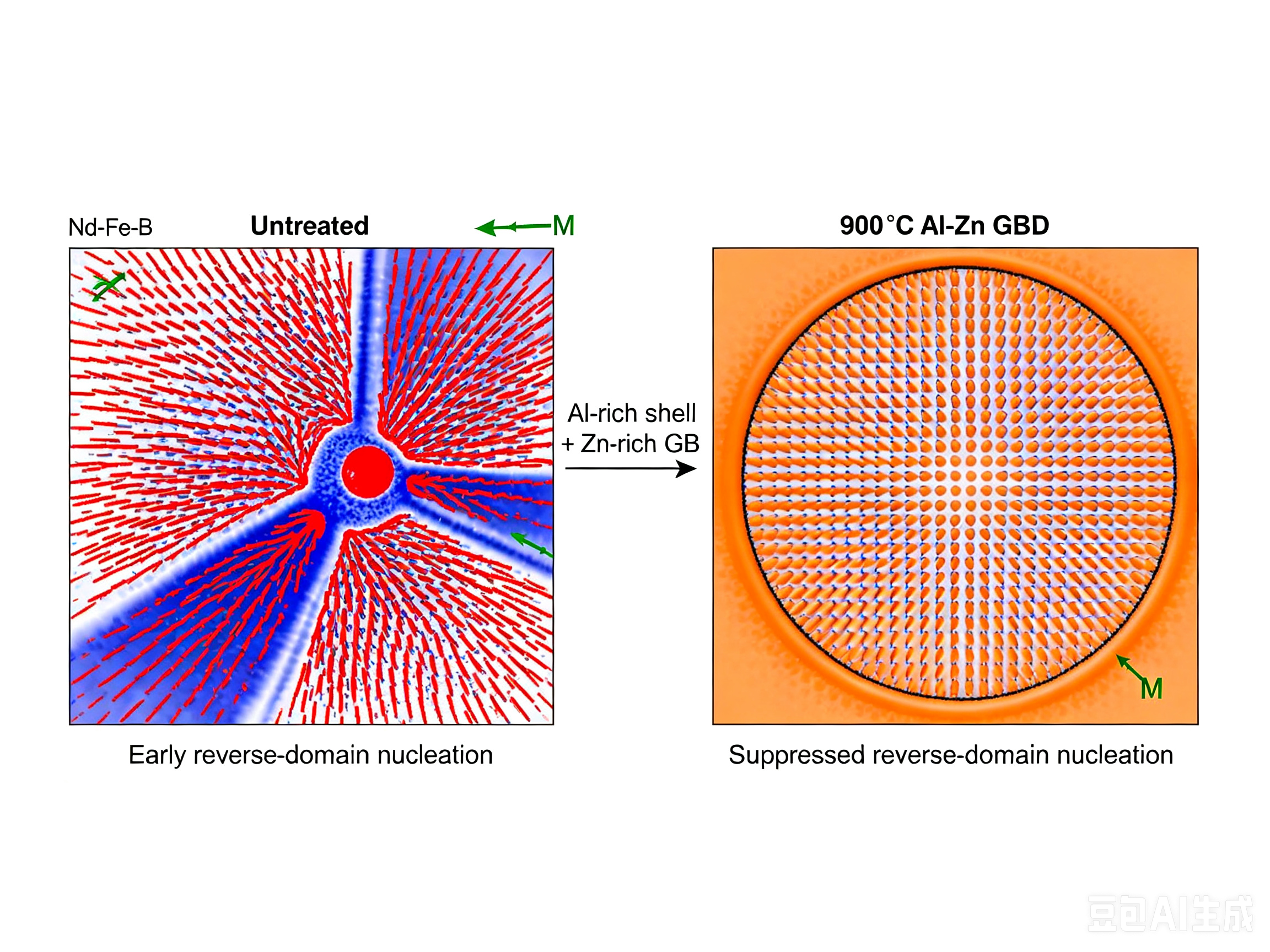}
\caption{Magnetization vector in reverse-domain nucleation.}
\label{fig:mic_domain_map}
\end{subfigure}
\begin{subfigure}{0.23\textwidth}
\centering
\includegraphics[width=\textwidth,height=0.85\textwidth]{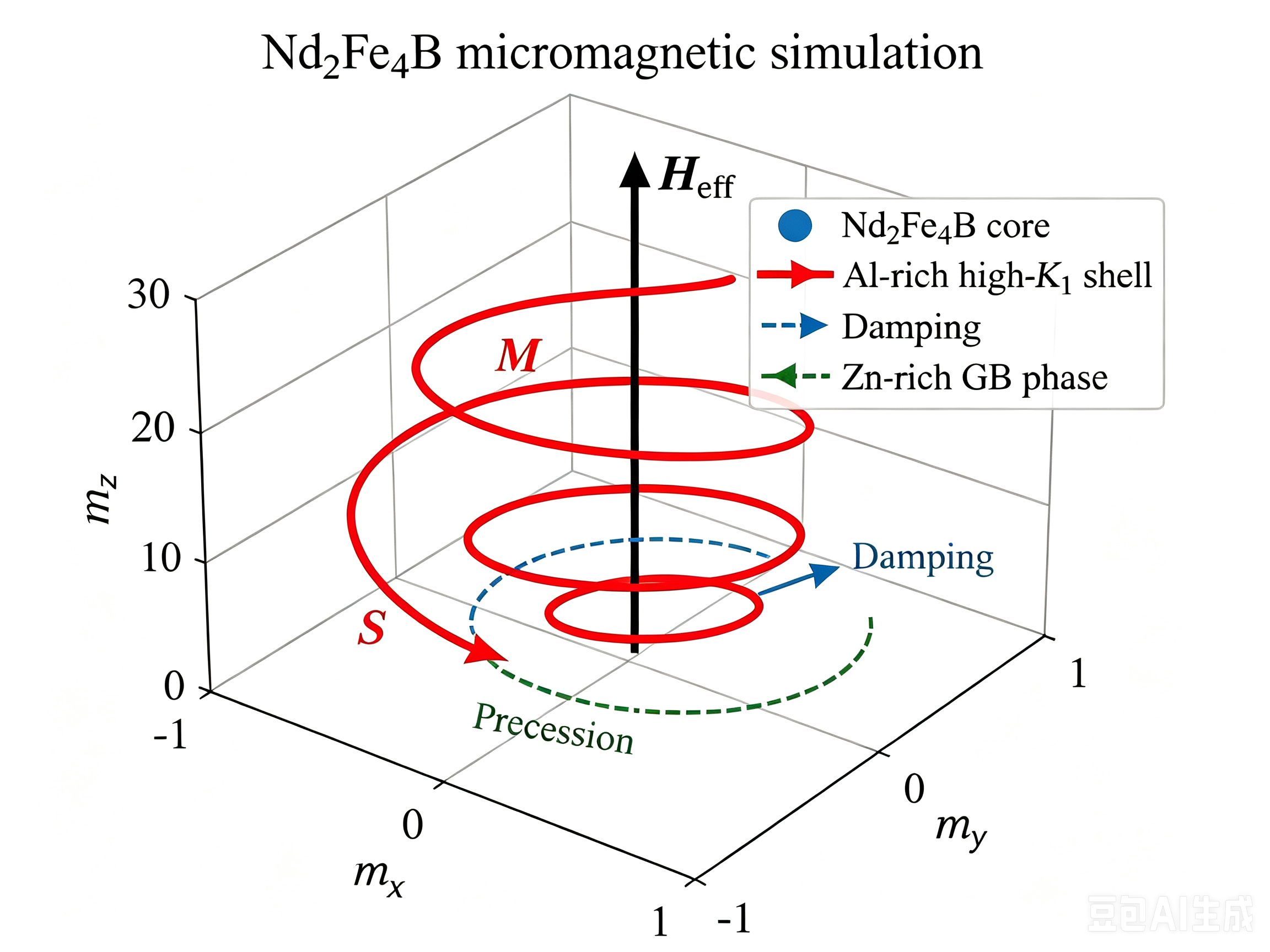}
\caption{Magnetization precession following the LLG equation.}
\label{fig:mic_llg_scheme}
\end{subfigure}
\caption{Multiscale micromagnetic simulation outputs revealing magnetization dynamics, coercivity enhancement origins and nanoscale domain evolution in Al-Zn grain boundary diffusion treated sintered Nd$_2$Fe$_{14}$B magnets.}
\label{fig:micromagnetic_all_results}
\end{figure}

Figure \ref{fig:micromagnetic_all_results} summarizes outputs from the multiscale micromagnetic model. Simulated $M-H$ curves in Figure \ref{fig:mic_hyst} demonstrate clear improvements in magnetic performance after grain boundary diffusion treatment at elevated temperature. Figure \ref{fig:mic_mechanism_bar} quantifies the relative contributions of individual coercivity enhancement pathways, confirming Al-rich shell pinning dominates for the 900$^\circ$C sample. Magnetization snapshots in figure \ref{fig:mic_domain_map} visually verify that the Al-Zn modified grain structure effectively suppresses reverse-domain nucleation. Figure \ref{fig:mic_llg_scheme} illustrates magnetization precession governed by the LLG equation within the core-shell grain architecture, linking nanoscale magnetic dynamics to the observed macroscopic coercivity gains.

The computational modeling presented is intended primarily to provide a mechanistic interpretation of the experimentally observed effects of Al-Zn grain boundary diffusion \cite{Kobayashi2019magnetostriction}, rather than to constitute an independently validated quantitative prediction of the concentration profile \cite{cheng2014ml}. The diffusion, phase-field, and micromagnetic models were constructed using parameters describing diffusion kinetics, elemental segregation, and magnetic interactions; however, several of these parameters, including activation energies, pre-exponential factors, and segregation coupling coefficients, are not directly determined from the present experiments. Therefore, the calculated concentration distributions and magnetic responses should be interpreted as model-dependent estimates that illustrate the possible evolution of the Al-Zn diffusion process and its influence on the microstructure and magnetic properties. In particular, the calculated Al penetration depth of approximately 180-210$\mu$m at 900$^\circ$C should not be regarded as a directly experimentally validated value.

The experimental observations provide qualitative support for the physical mechanisms represented in the simulations. SEM-EDS characterization shows Al enrichment near grain boundaries and the formation of an Al-enriched region with a characteristic thickness of approximately 1-2$\mu$m, while elemental mapping indicates that Al is distributed preferentially along grain-boundary regions and also extends into the adjacent main phase. These observations are consistent with the assumed grain-boundary-assisted redistribution of Al in the computational framework. However, the present measurements do not include cross-sectional EDS line scans or composition measurements at multiple distances from the diffusion surface; consequently, a direct experimental comparison with the calculated 180-210$\mu$m penetration depth cannot be established. The computational results are therefore used here to illustrate the underlying diffusion and magnetic mechanisms rather than as quantitative experimental validation. The simulations are not intended to confirm a pre-selected penetration depth or magnetic-property enhancement, but rather to examine whether the proposed diffusion, segregation, and micromagnetic mechanisms can provide a physically reasonable explanation for the experimental observations. Quantitative validation of the predicted concentration profiles would require additional experiments, such as cross-sectional EDS line scans, wavelength-dispersive spectroscopy, or compositional mapping at multiple depths from the diffusion surface. Such measurements would allow the calculated concentration profiles and characteristic diffusion lengths to be directly compared with experimental data and would provide a basis for determining or refining the model parameters.

\section{Experimental data analysis}
\subsection{Sample weight change}
The samples were weighed before and after diffusion treatment, and the weight change can indirectly reflect the diffusion behavior of Al and Zn.

The original mass of the sample diffused at 900$^\circ$C (designated $A_1$) was 2.313g. The total mass of the two Al-Zn alloy disks was 0.243g (upper disk: 0.1286g, lower disk: 0.1144g). After diffusion treatment, the sample mass became 2.223g, representing a decrease of 0.090 g compared to the original magnet. The residual deposit on the ceramic spacer was 0.021g.

The original mass of the sample diffused at 700$^\circ$C (designated B) was 2.300g, and after grinding to remove the oxide layer, the mass was 2.180g. The total mass of the Al-Zn alloy disks was 0.234g (upper disk: 0.113g, lower disk: 0.121g).

According to the law of conservation of mass, the total mass of the original magnet and the diffusion source for the 900$^\circ$C sample was 2.313g+0.243g=2.556g. After diffusion treatment, the total mass of the magnet and the residue was 2.223g+0.021g=2.244g, with a difference of approximately 0.312g. This mass loss mainly originates from two aspects: first, the volatilization of Al and Zn at high temperatures (the boiling point of Zn is 907$^\circ$C, and at 900$^\circ$C, the vapor pressure of Zn is relatively high, making volatilization more pronounced) \cite{HONO2012}; second, the decomposition and volatilization of surface contaminants such as oil stains or oxides on the magnet surface at high temperatures \cite{fangzhu_2025}.

The effect of temperature on atomic migration ability can be explained by thermal activation theory \cite{fan2016form}. The diffusion of atoms from the surface to the interior requires overcoming the energy barriers at grain boundaries \cite{wangzhan_2026}. The higher the temperature, the more thermal energy the atoms acquire, and the faster the diffusion rate. The temperature of 900$^\circ$C is 200$^\circ$C higher than 700$^\circ$C, resulting in a significant increase in the average kinetic energy of atoms, which explains why the diffusion depth of the 900$^\circ$C sample (approximately 150-200$\mu$m) is much greater than that of the 700$^\circ$C sample (approximately 50$\mu$m).

\subsection{Diffusion depth}
The diffusion depths discussed should be regarded as model-derived estimates rather than directly measured compositional penetration depths. The present SEM-EDS characterization confirms Al enrichment near the diffusion-treated region and the formation of an Al-enriched shell and modified grain-boundary regions, but does not provide depth-resolved concentration profiles extending from the diffusion-source interface into the bulk magnet. Therefore, the estimated diffusion depths of approximately 150-200$\mu$m at 900$^\circ$C and approximately 50$\mu$m at 700$\circ$C are used primarily to illustrate the temperature-dependent diffusion behavior predicted by the model. Direct cross-sectional EDS line scans or elemental mapping at multiple distances from the diffusion interface would be required to quantitatively determine the actual Al and Zn concentration profiles and validate these calculated diffusion depths. Such depth-resolved characterization will be considered in future work.

\subsection{Magnetic property analysis}
In this section, the vibrating sample magnetometer (VSM) was used to test the room-temperature magnetic properties of the untreated original sample (38SH), as well as samples B and A after grain boundary diffusion treatment at 700$^\circ$C and 900$^\circ$C, respectively \cite{zhang_2021_pred_mag}. By analyzing the B-H hysteresis loops, J-H hysteresis loops, demagnetization curves, and energy product curves, the effects of different diffusion temperatures on coercivity, remanence, and energy characteristics of the magnets were investigated.

\subsection{Comparison of hysteresis loops}
\begin{figure}
\centering
\begin{subfigure}{0.23\textwidth}
\includegraphics[width=\textwidth, height=0.76\textwidth]{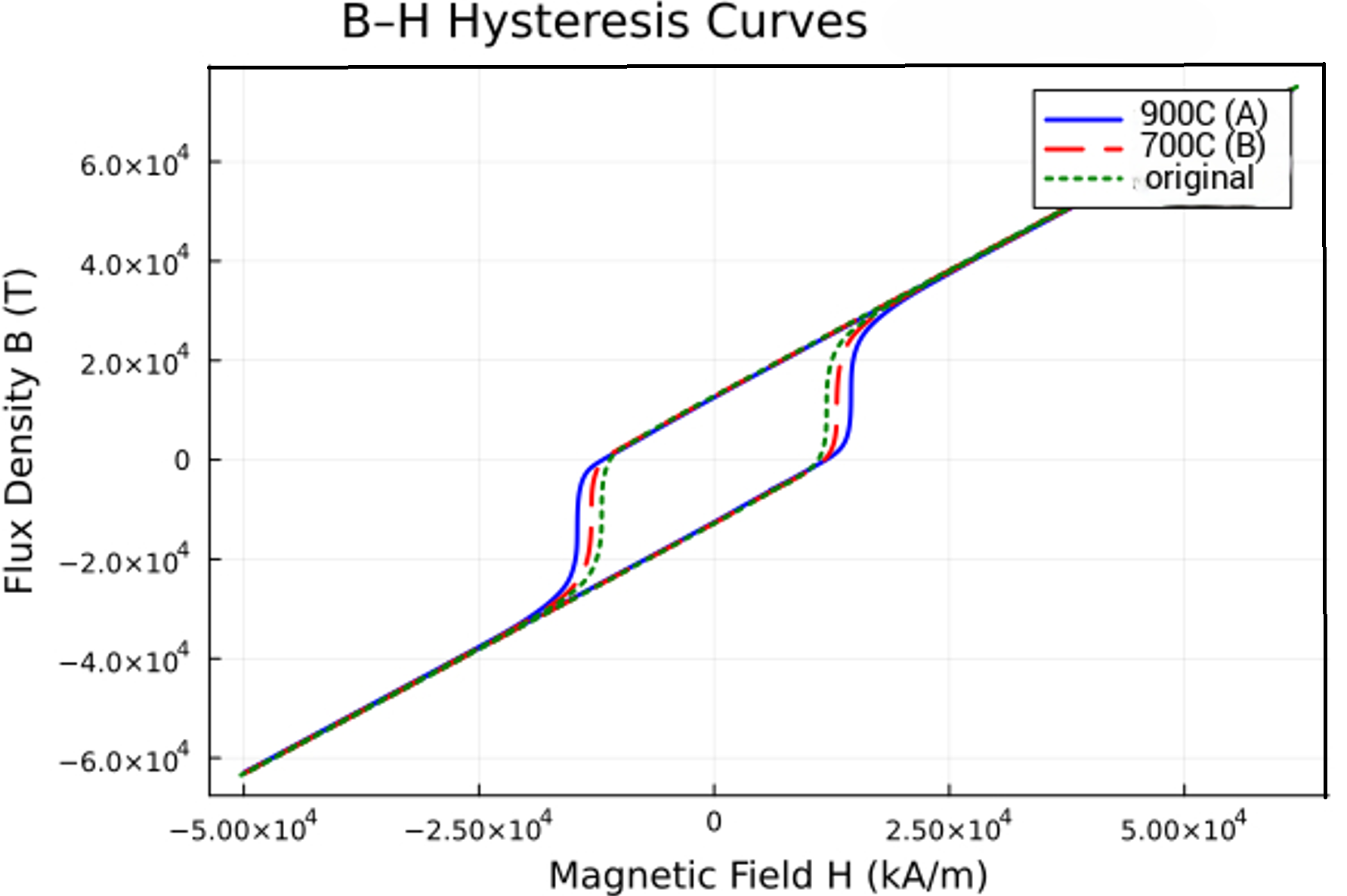}
\caption{$B-H$ hysteresis}
\label{B-H hysteresis curves}
\end{subfigure}
\begin{subfigure}{0.23\textwidth}
\includegraphics[width=\textwidth, height=0.76\textwidth]{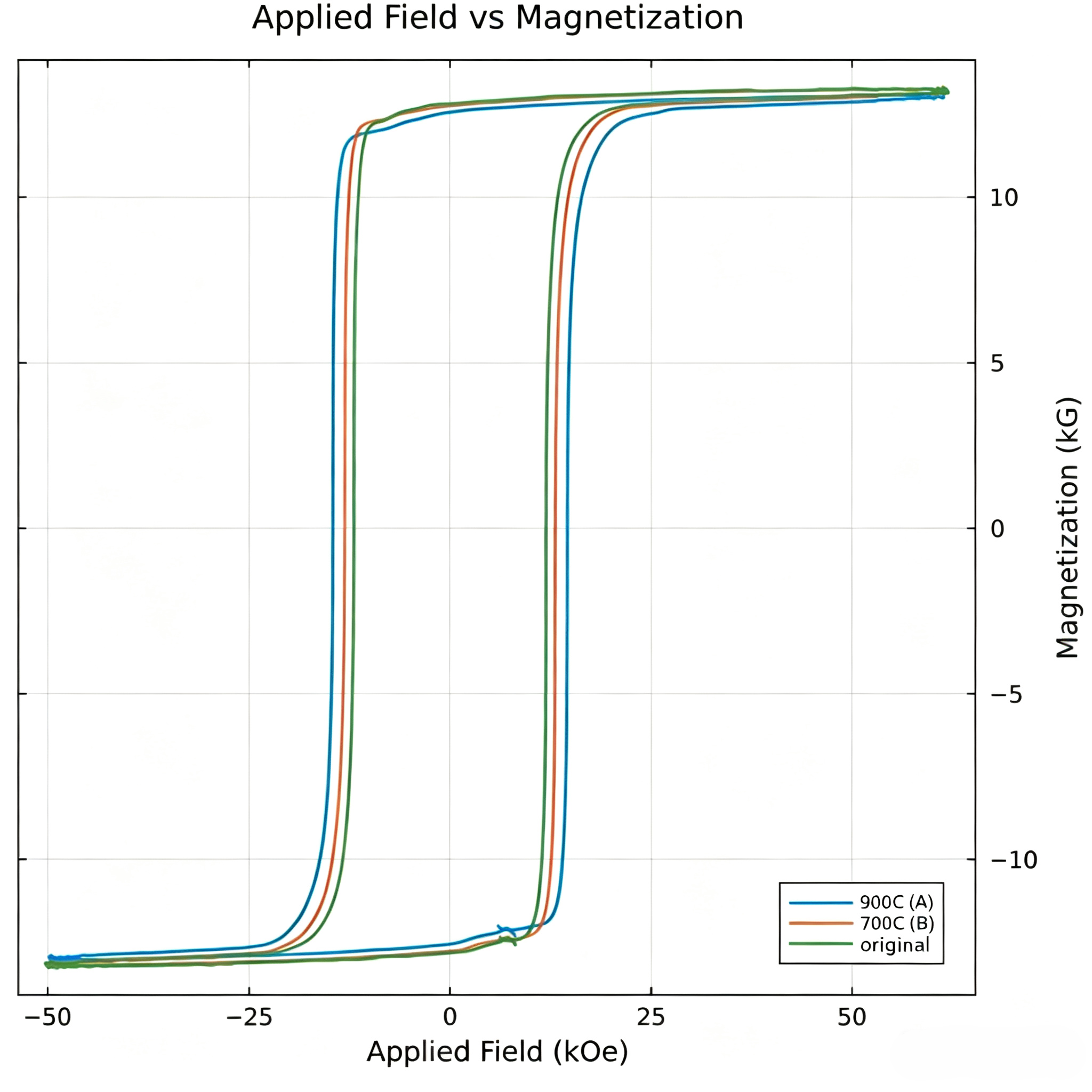}
\caption{magnetization vs $H$}
\label{Combined_Magnetization_Curves}
\end{subfigure}
\begin{subfigure}{0.23\textwidth}
\includegraphics[width=\textwidth, height=0.76\textwidth]{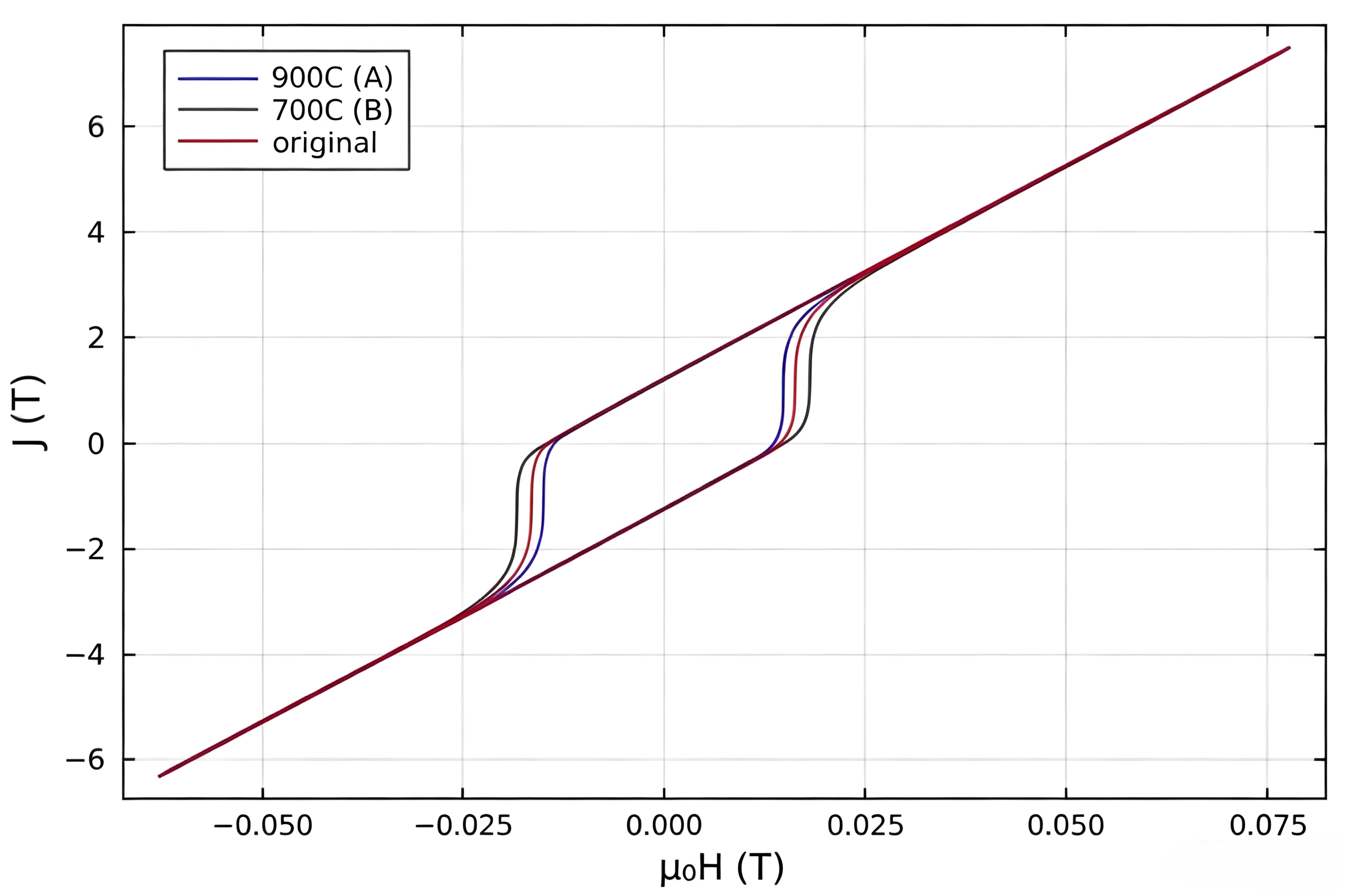}
\caption{$J-\mu_0H$}
\label{full_hysteresis}
\end{subfigure}
\begin{subfigure}{0.23\textwidth}
\includegraphics[width=\textwidth, height=0.76\textwidth]{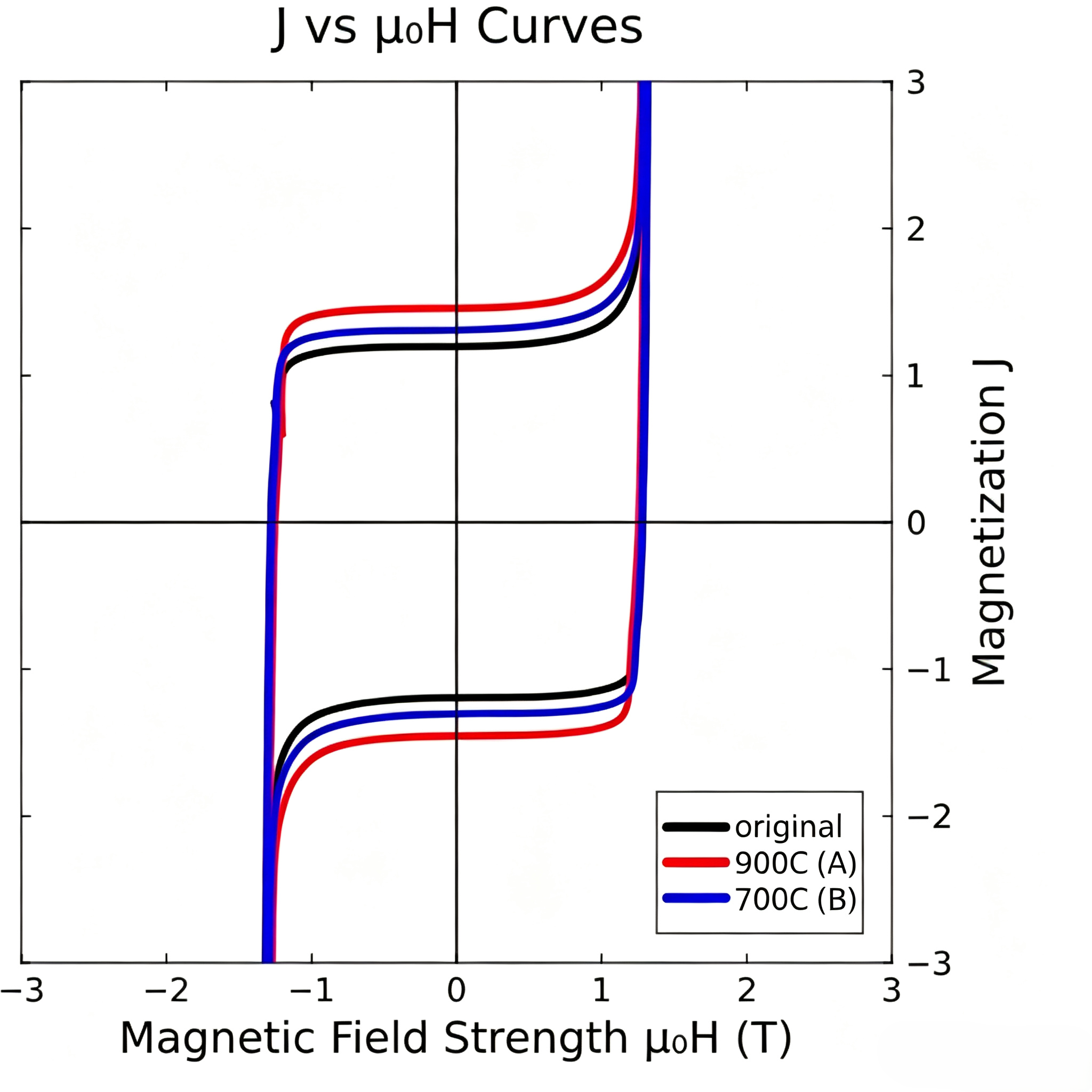}
\caption{magnetization $J-\mu_0H$}
\label{J_vs_mu0H}
\end{subfigure}
\caption{Magnetic hysteresis and demagnetization curves of original, 900$^\circ$C (A), and 700$^\circ$C (B) Al-Zn grain boundary diffusion-treated sintered Nd-Fe-B magnets.}
\label{fig:magnetic_loops_collection}
\end{figure}

\begin{figure}
\centering
\includegraphics[width = 0.37\textwidth, height = 0.4\textwidth]{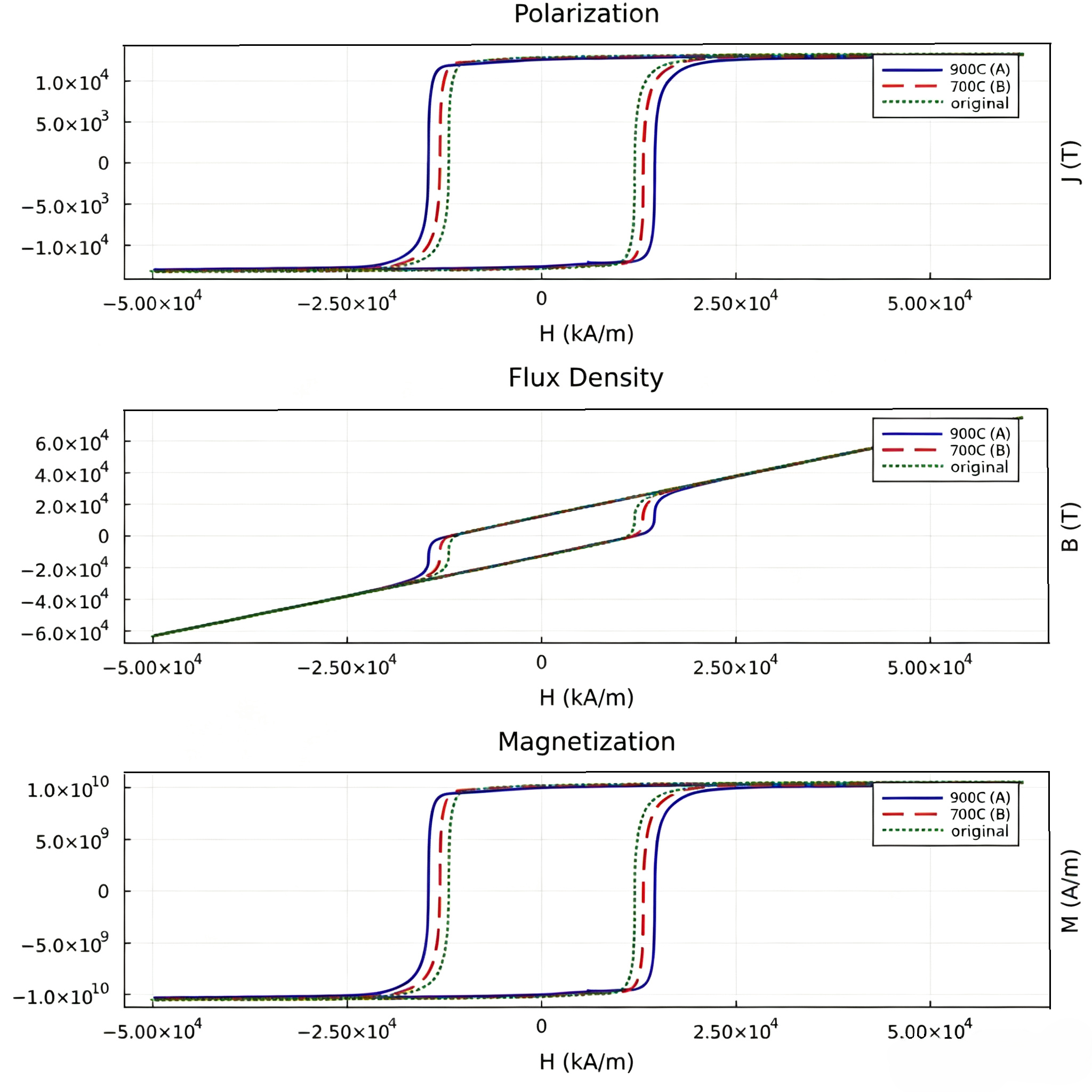}
\caption{Polarization (J-H), flux density (B-H), and magnetization (M-H) hysteresis loops of the original magnet, the 700$^\circ$C diffusion-treated sample (B), and the 900$^\circ$C diffusion-treated sample (A).}
\label{Combined_Magnetic_Properties}
\end{figure}

The hysteresis loop is the core basis for characterizing the magnetic properties of permanent magnet materials, and its shape and characteristic points directly reflect the magnetization behavior and magnetic hardening characteristics of the material \cite{jiang_2025}. To intuitively compare the overall hysteresis behavior of samples under different conditions, the B-H hysteresis loops of each sample were first plotted, and the results are shown in figure \ref{fig:magnetic_loops_collection}, \ref{Combined_Magnetic_Properties}.





In the B-H loop, the horizontal axis represents the applied magnetic field strength H, and the vertical axis represents the magnetic flux density B. The saturation segment, coercivity point, and remanence point of the curve simultaneously reflect the saturation magnetization and demagnetization resistance of the material \cite{HIROSAWA_1985}. It can be seen from the figure that the three curves basically overlap in the high-field region, indicating that the diffusion treatment did not significantly change the saturation magnetic flux density of the magnets. However, in the region near the coercivity, there are obvious differences in the curve positions, reflecting the regulatory effect of diffusion temperature on the coercivity of the magnets.

\subsection{Demagnetization curves and coercivity analysis}
Coercivity ($H_{cj}$) is the core indicator for evaluating the demagnetization resistance of permanent magnet materials, and its physical essence reflects the difficulty of nucleation and propagation of reverse magnetic domains in the material \cite{sagawa_1984}. For sintered Nd-Fe-B magnets, the magnitude of coercivity is closely related to the distribution state of the grain boundary phase and the microstructure of the main phase grain surfaces \cite{HONO2012}. To accurately evaluate the effect of different diffusion processes on the demagnetization resistance of the magnets, the J-H demagnetization curves of each sample were extracted and analyzed in this section.

\begin{figure}
\centering
\begin{subfigure}{0.23\textwidth}
\includegraphics[width=\textwidth, height=0.76\textwidth]{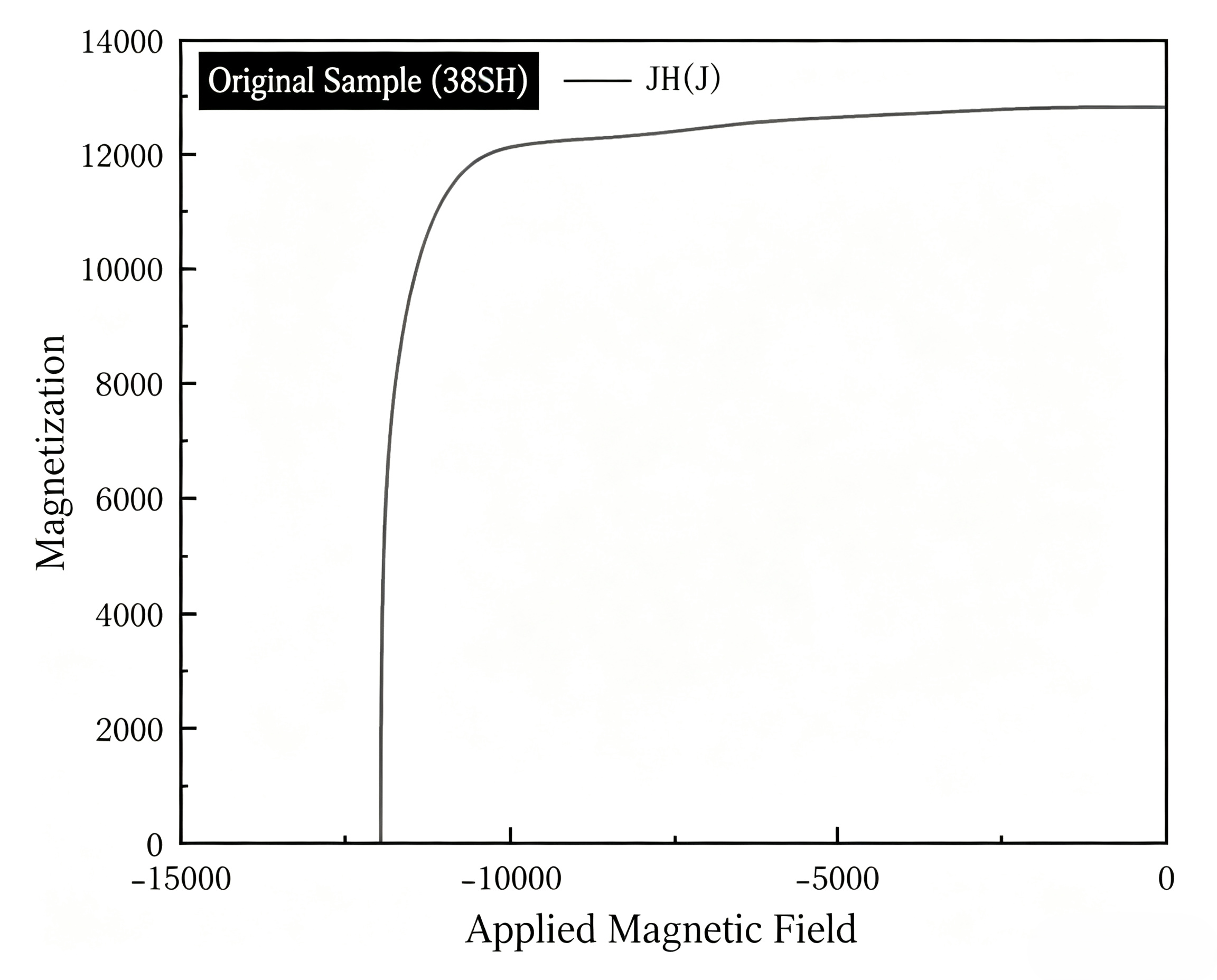}
\caption{$J-H$ original sample 38SH}
\label{original_J-H_curve}
\end{subfigure}
\begin{subfigure}{0.23\textwidth}
\includegraphics[width=\textwidth, height=0.76\textwidth]{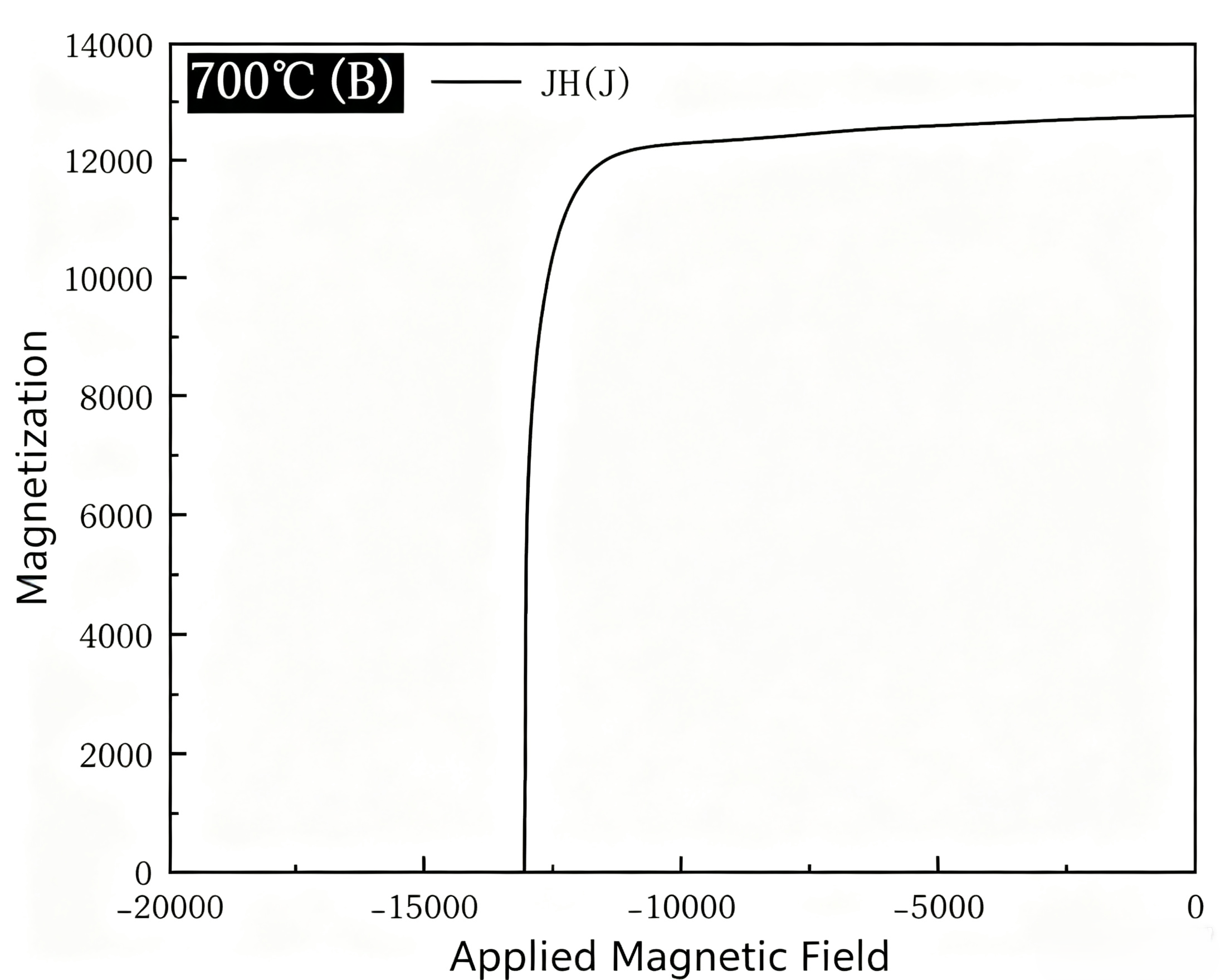}
\caption{$J-H$ diffusion (B) 700$^\circ$C}
\label{700B_J-H_curve}
\end{subfigure}
\begin{subfigure}{0.23\textwidth}
\includegraphics[width=\textwidth, height=0.76\textwidth]{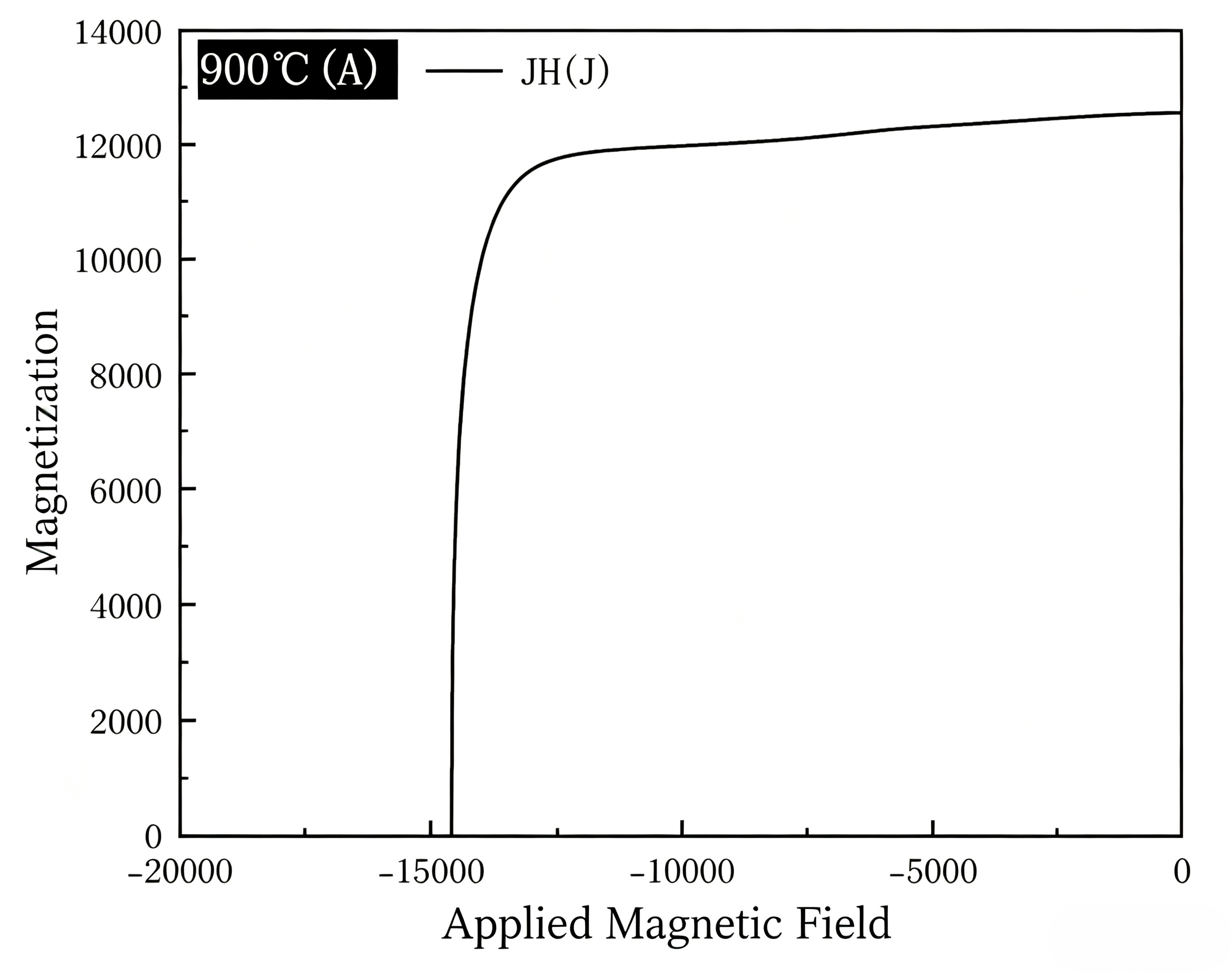}
\caption{$J-H$ diffusion (A) 900$^\circ$C}
\label{900A_J-H_curve}
\end{subfigure}
\begin{subfigure}{0.23\textwidth}
\includegraphics[width=\textwidth, height=0.76\textwidth]{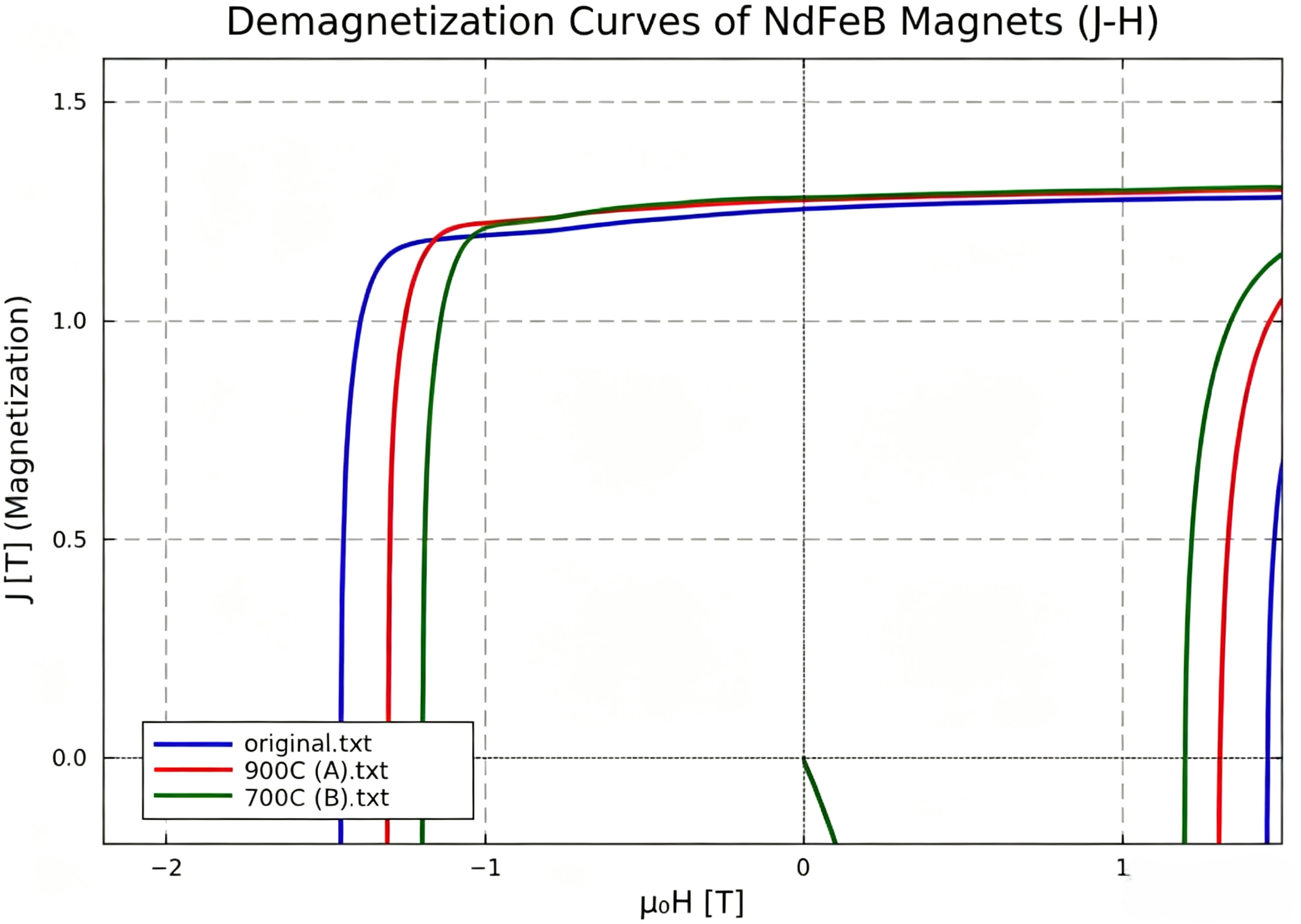}
\caption{initial,900$^\circ$C(A),700$^\circ$C(B)}
\label{demagnetization_curves_JH}
\end{subfigure}
\caption{Intrinsic demagnetization curves (magnetization polarization $J$ versus applied magnetic field $H$) of sintered Nd-Fe-B magnets before and after Al-Zn grain boundary diffusion treatment.}
\label{fig:all_JH_curves}
\end{figure}

As can be seen from figure \ref{fig:magnetic_loops_collection},  \ref{fig:all_JH_curves}, \ref{fig:magnetic_summary_curves}, the demagnetization curve of this sample exhibits a relatively "rounded" characteristic near the coercivity point, indicating that the squareness of the demagnetization curve is not ideal. According to the analysis in reference \cite{sagawa_1984}, this demagnetization behavior is usually associated with non-uniform distribution of the grain boundary phase inside the magnet and the presence of direct contact between some grains. When the applied reverse magnetic field approaches the coercivity, these defective regions tend to become preferential nucleation sites for reverse magnetic domains, which rapidly propagate to the surrounding grains, causing a rapid decay of the magnetic polarization intensity within a relatively small range of reverse magnetic field.


In figure \ref{700B_J-H_curve}, compared with the original sample, the squareness of the demagnetization curve of the magnet after 700$^\circ$C diffusion for sample (B) was improved to a certain extent. The reverse magnetic field strength corresponding to the coercivity point is significantly higher than that of the original sample, and the change of the demagnetization curve near the knee point is more "steep." This change indicates that after the Al-Zn grain boundary diffusion treatment at 700$^\circ$C, some Al and Zn atoms infiltrated along the grain boundaries, playing a preliminary role in optimizing the grain boundary phase. The continuity of the grain boundary phase increased, and the isolation effect on the main phase grains was enhanced, thereby increasing the resistance to reverse magnetic domain nucleation \cite{HIROSAWA_1985}.



Figure \ref{900A_J-H_curve} shows the J-H demagnetization curve of the 900$^\circ$C diffusion sample (A). The demagnetization curve of this sample exhibits the best squareness. The descending segment of the demagnetization curve is very "straight"; the magnetic polarization intensity remains almost constant before reaching the coercivity point. Once the reverse magnetic field exceeds the coercivity point, the magnetic polarization intensity rapidly drops to zero \cite{HONO2012}. This ideal demagnetization behavior directly proves that the 900$^\circ$C diffusion treatment has the most significant effect on optimizing the microstructure of the magnet. The uniform, thin, and continuous grain boundary phase effectively "passivates" the defects on the surfaces of the main phase grains, making it difficult for reverse magnetic domains to nucleate at low magnetic fields, thereby enabling the magnet to maintain a stable magnetization state under magnetic fields close to its theoretical coercivity. Figure \ref{demagnetization_curves_JH} presents the intrinsic $J$-$\mu_0H$ demagnetization curves of the original sintered Nd-Fe-B magnet and specimens subjected to Al-Zn grain boundary diffusion treatment at 900$^\circ$C (A) and 700$^\circ$C (B), revealing obvious coercivity improvement after diffusion treatment, especially for the sample processed at 900 $^\circ$C.

The coercivity enhancement observed after Al-Zn grain boundary diffusion is interpreted as arising from several microstructural changes that may act concurrently, including grain-boundary decoupling, modification of the grain-edge region, and formation of an Al-enriched shell. However, the present experiments do not permit the individual contribution of these mechanisms to be quantitatively separated. The experimental evidence provides different levels of support for each mechanism. The SEM-EDS results directly demonstrate Al enrichment near the grain boundaries and the formation of an Al-enriched region at the grain edge, providing experimental evidence for compositional modification of the shell and boundary regions. The observed increase in coercivity together with this compositional redistribution is consistent with an enhancement of the magnetic isolation between neighboring grains. In addition, the observed changes in the grain-boundary region are compatible with a reduction in magnetically unfavorable nucleation sites associated with irregular grain edges.

The micromagnetic model was therefore used to examine the possible combined influence of these microstructural features rather than to quantitatively determine their individual contributions. In the present model, the weighting factors $w_1 ,w_2, w_3$ are phenomenological parameters introduced to represent the relative importance of grain-boundary decoupling, high-anisotropy shell formation, and grain-edge modification, respectively. The simulated results should consequently be regarded as a qualitative assessment of how the proposed mechanisms may collectively influence magnetization reversal and coercivity. A more rigorous separation of the three mechanisms would require controlled comparative experiments in which the individual compositional or microstructural variables are isolated. For example, Zn-only and Al-only diffusion treatments, together with samples having systematically varied diffusion conditions, would allow the effects of elemental segregation, shell formation, and grain-boundary modification to be compared more directly. Additional local magnetic characterization and quantitative microstructural measurements would also be required to establish a quantitative relationship between the modified grain-edge structure and the coercivity.

The proposed mechanism is considered a physically plausible interpretation of the experimental results rather than a uniquely demonstrated decomposition of the coercivity enhancement. The approximately 21.7\% increase in coercivity after the 900$^\circ$C treatment is experimentally established, whereas the relative contributions of grain-boundary decoupling, Al-enriched shell formation, and grain-edge modification remain unresolved. Future experiments incorporating Al-only and Zn-only diffusion, cross-sectional compositional profiling, and quantitative micromagnetic parameter determination will be necessary to distinguish these contributions and establish their respective roles in coercivity enhancement.

\subsection{Magnetic property changes}




\begin{figure}
\centering
\begin{subfigure}{0.23\textwidth}
\includegraphics[width=\textwidth, height=0.76\textwidth]{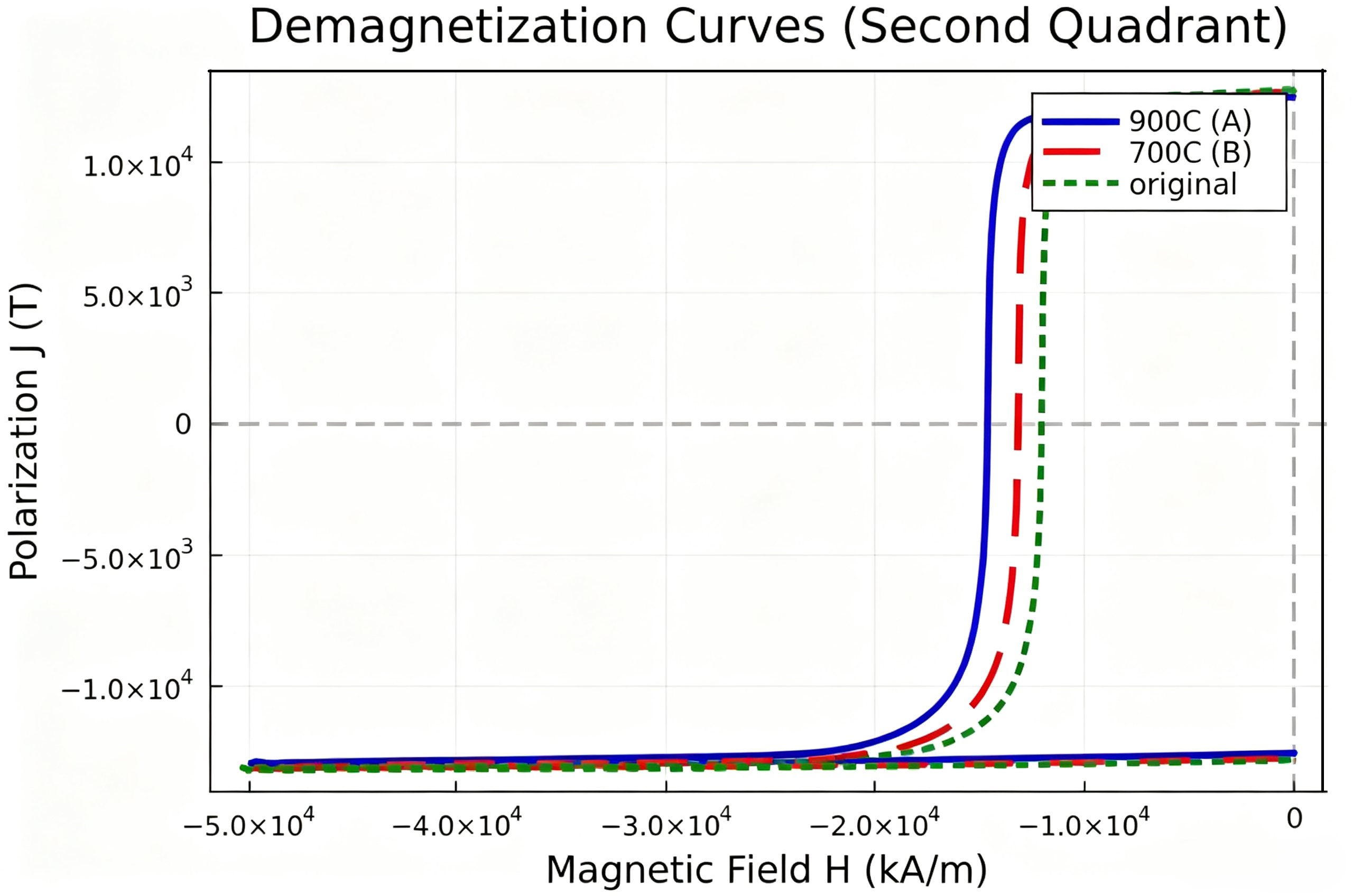}
\caption{Second-quadrant $J-H$ demagnetization curves}
\label{Demagnetization_curves_comparison}
\end{subfigure}
\begin{subfigure}{0.23\textwidth}
\includegraphics[width=\textwidth, height=0.76\textwidth]{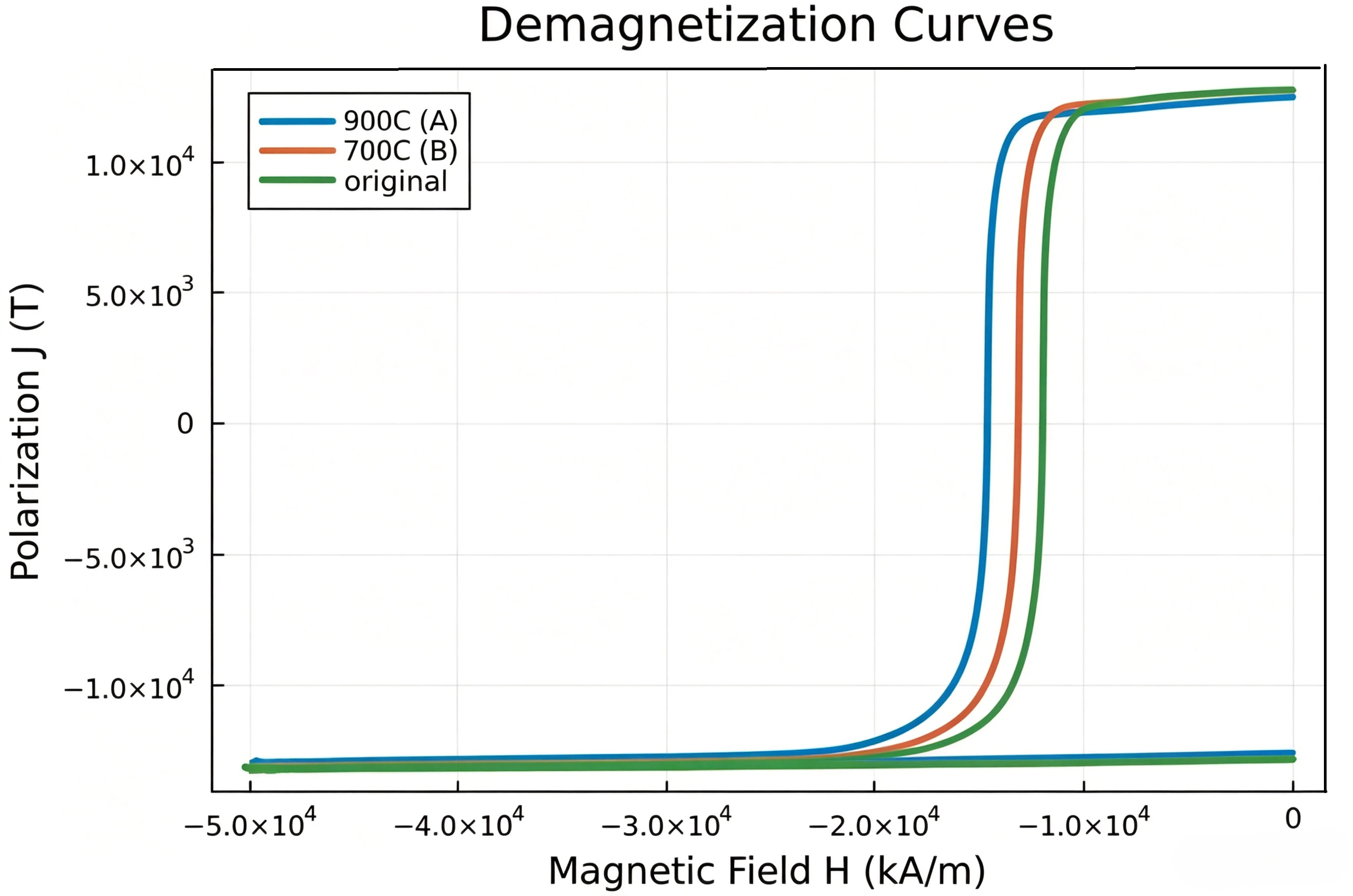}
\caption{$J-H$ intrinsic demagnetization curves}
\label{all_demagnetization_curves}
\end{subfigure}
\begin{subfigure}{0.23\textwidth}
\includegraphics[width=\textwidth, height=0.76\textwidth]{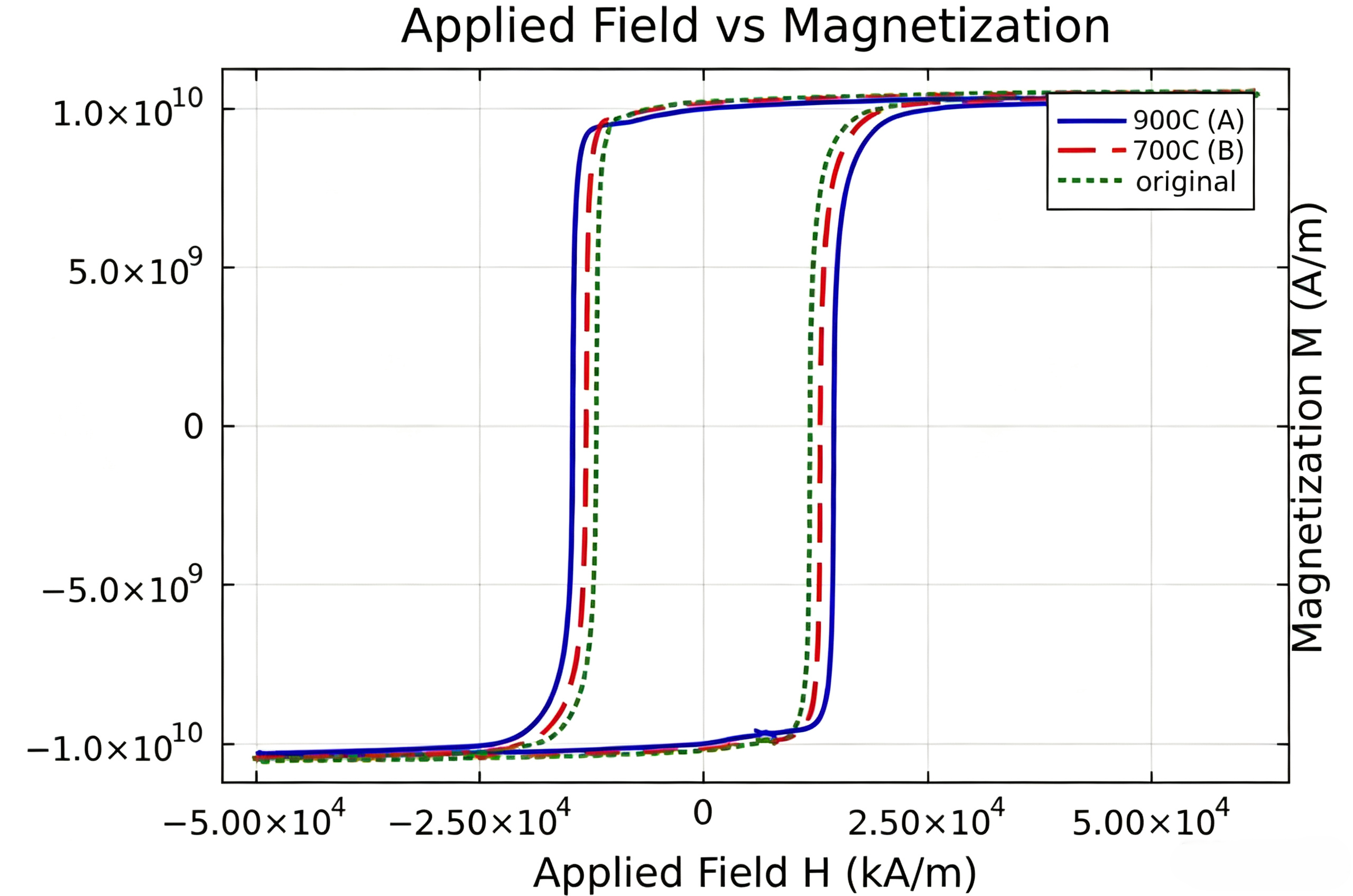}
\caption{$M-H$ magnetization}
\label{Field_vs_Magnetization}
\end{subfigure}
\begin{subfigure}{0.23\textwidth}
\includegraphics[width=\textwidth, height=0.76\textwidth]{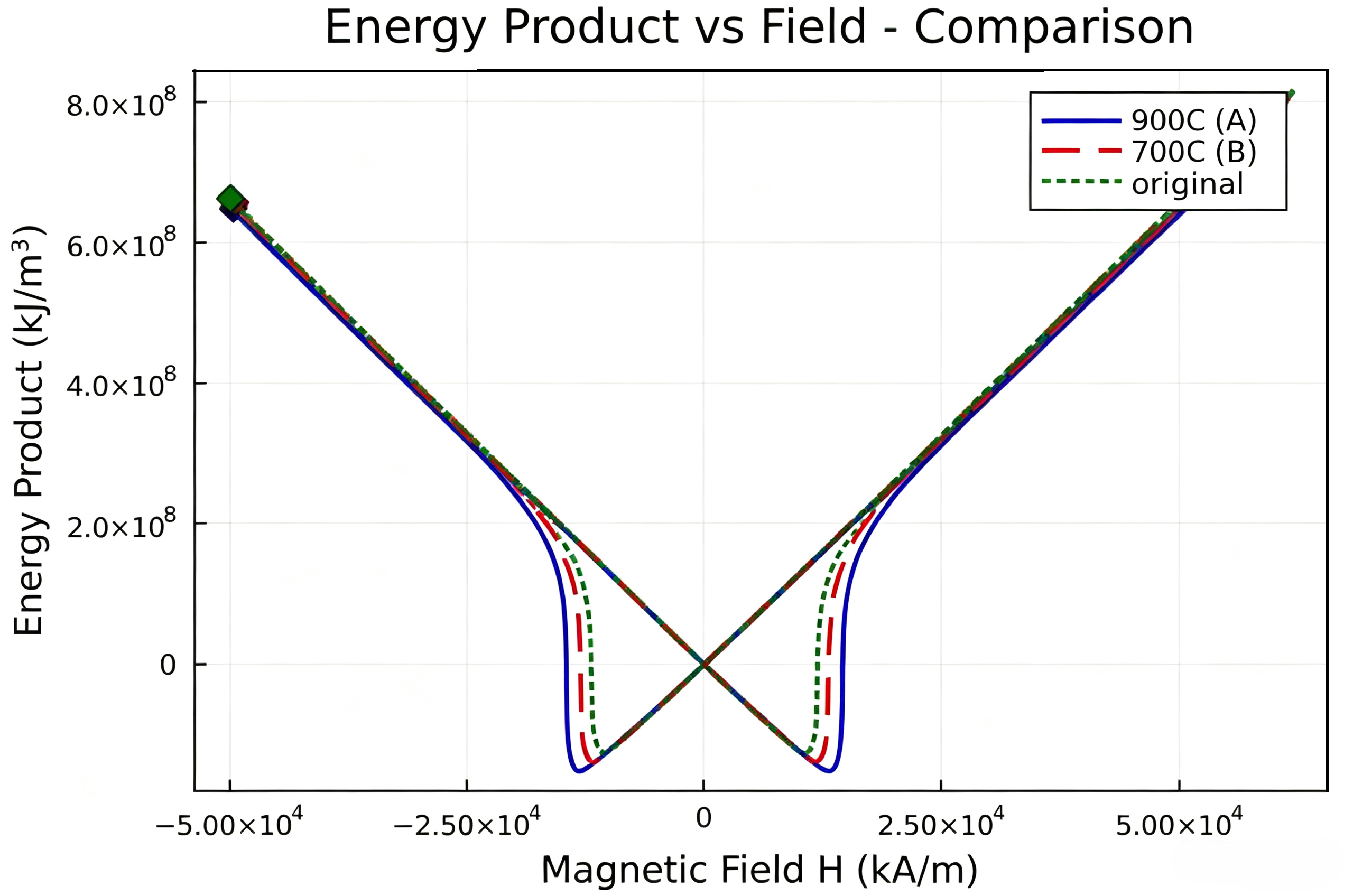}
\caption{$(BH)$ vs $H$ energy product}
\label{Energy_products_comparison}
\end{subfigure}

\caption{Magnetic characteristic curves of the original sintered Nd-Fe-B magnet and Al-Zn grain boundary diffusion-treated samples.}
\label{fig:magnetic_summary_curves}
\end{figure}

Based on the above analysis of the hysteresis loops and demagnetization curves, the variation patterns of the magnetic property indicators of the magnets after Al-Zn grain boundary diffusion treatment can be summarized. Figure \ref{Demagnetization_curves_comparison}, \ref{all_demagnetization_curves} presents the second-quadrant intrinsic $J-H$ demagnetization curves of the original sintered Nd-Fe-B magnet and specimens subjected to Al-Zn grain boundary diffusion treatment at 900 $^\circ$C (A) and 700 $^\circ$C (B), demonstrating enhanced coercivity after diffusion treatment.
\begin{table}[h]
\centering
\caption{Magnetic property changes}
\begin{tabular}{|c|c|c|c|c|c|}\hline
sample & treat & $H_{cj}\left(\frac{kA}{m}\right)$ & $\Delta H_{cj}$ & $B_r(mT)$ & $\Delta B_r$ \\\hline
38SH & original & 951.5 & - & 1282 & - \\\hline
B & 700$^\circ$C & 1039.6 & 88.1 & 1276 & -6 \\\hline
A & 900$^\circ$C & 1158.2 & 206.7 & 1256 & -26\\\hline
\end{tabular}
\label{tab:magnetic_properties}
\end{table}

In terms of coercivity, the Al-Zn grain boundary diffusion treatment can significantly enhance the coercivity of sintered Nd-Fe-B magnets, and the enhancement amplitude is positively correlated with the diffusion temperature. The coercivity enhancement amplitude of the 900$^\circ$C diffusion sample (approximately 206.7 kA/m) is much higher than that of the 700$^\circ$C sample (approximately 88.1 kA/m). This is mainly attributed to the fact that high temperature significantly increases the diffusion coefficient of atoms, promoting the deep infiltration of Al and Zn along the grain boundaries, thereby more effectively optimizing the physicochemical properties of the grain boundary phase.

In terms of remanence, the remanence of the magnets decreased slightly after diffusion treatment. The remanence of the 700$^\circ$C diffusion sample decreased by only 6 mT, while the remanence of the 900$^\circ$C diffusion sample decreased more significantly, by 26 mT \cite{strnat_1970}. The decrease in remanence is related to the entry of Al atoms into the lattice of the main phase Nd$_2$Fe$_{14}$B. As a non-magnetic element, Al dilutes the concentration of magnetic atoms upon entering the lattice and may weaken the magnetic exchange coupling between Fe-Fe. The greater remanence decrease at 900$^\circ$C indicates that the high temperature promoted more Al atoms to cross the grain boundaries and enter the interior of the main phase grains.

In terms of maximum energy product, although the coercivity and remanence of the three samples exhibited a trade-off relationship, their maximum energy product [(BH)$_{\text{max}}$] did not change significantly \cite{zhou_2011}. As described in the literature, the maximum energy product is determined by both remanence and the squareness of the demagnetization curve \cite{li2020}. In this experiment, although the 900$^\circ$C sample suffered a larger remanence loss, its excellent demagnetization curve squareness compensated for the energy output loss to a certain extent. Meanwhile, although the 700$^\circ$C sample retained remanence relatively well, its squareness was slightly inferior. Therefore, the overall energy output capabilities of the three samples remained at similar levels.
\subsection{SEM morphology analysis}
\subsubsection{700$^\circ$C sample}
The microstructure of magnet B after diffusion at 700$^\circ$C was observed using a scanning electron microscope. The sample was etched with a 3 vol.\% nitric acid-alcohol solution for 12 seconds \cite{ZENG2025}.

\begin{figure}
\centering
\begin{subfigure}{0.23\textwidth}
\includegraphics[width=\textwidth, height=0.76\textwidth]{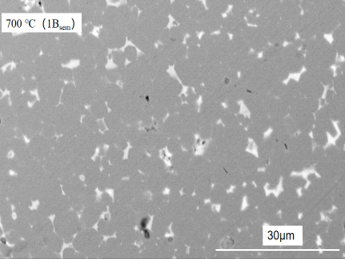}
\caption{700$^\circ$C sample at 1500$\times$ magnification (1B)}
\label{SEM_image_700C1500magnification_1b}
\end{subfigure}
\begin{subfigure}{0.23\textwidth}
\includegraphics[width=\textwidth, height=0.76\textwidth]{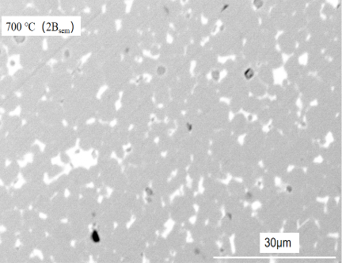}
\caption{700$^\circ$C sample at 1500$\times$ magnification (2B)}
\label{SEM_image_700C1500magnification_2b}
\end{subfigure}
\begin{subfigure}{0.23\textwidth}
\includegraphics[width=\textwidth, height=0.76\textwidth]{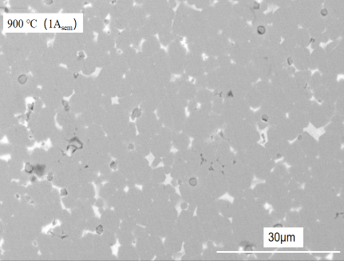}
\caption{900$^\circ$C sample at 1500$\times$ magnification (1A)}
\label{SEM_image_900C1500magnification_1a}
\end{subfigure}
\begin{subfigure}{0.23\textwidth}
\includegraphics[width=\textwidth, height=0.76\textwidth]{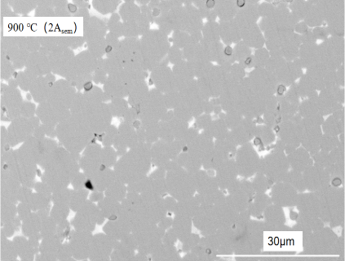}
\caption{900$^\circ$C sample at 1500$\times$ magnification (2A)}
\label{SEM_image_900C1500magnification_2a}
\end{subfigure}
\caption{SEM micrographs of diffusion-treated Nd-Fe-B samples obtained at 1500$\times$ magnification.}
\label{fig:sem_overview}
\end{figure}


Figure \ref{SEM_image_700C1500magnification_1b} shows the SEM image of the 700$^\circ$C sample at 1500$\times$ magnification. It can be observed from the figure that the blocky Nd-rich phases originally aggregated at the triple-junction grain boundaries partially spread out, forming a relatively continuous thin layer along the main-phase grain boundaries \cite{zhu_2025}. However, the uniformity of this thin layer remains suboptimal, with some regions still exhibiting discontinuities, and direct contact persists between certain grains \cite{zhou_2011}. The grain size of the main phase shows no significant change, with most grains falling within the range of 5-8 $\mu$m, indicating that the diffusion temperature of 700$^\circ$C is insufficient to cause notable grain growth.


Figure \ref{SEM_image_700C1500magnification_2b} shows the SEM image of the 700$^\circ$C sample from another region. In the image, the brighter regions correspond to the Nd-rich grain boundary phase, while the darker regions correspond to the Nd$_2$Fe$_{14}$B main-phase grains \cite{liuxiang_2008}. The grain boundary phase can separate the main-phase grains in most locations; however, its thickness distribution is insufficiently uniform, with some regions exhibiting wider grain boundaries and others showing narrower ones. Overall, the 700$^\circ$C diffusion treatment provides a certain degree of improvement to the microstructure, but the extent of improvement is limited, and the continuity and uniformity of the grain boundary phase remain suboptimal.

\subsubsection{900$^\circ$C sample}
The microstructure of magnet A after diffusion at 900$^\circ$C was observed using a scanning electron microscope, with the same etching conditions as those used for the 700$^\circ$C sample.


Figure \ref{SEM_image_900C1500magnification_1a} shows the SEM image of the 900$^\circ$C sample at 1500$\times$ magnification. Compared with the 700$^\circ$C sample, the microstructural changes of the 900$^\circ$C sample are more significant. First, the grain boundary phase becomes thin and continuous, with almost all main-phase grains surrounded by a grayish-white thin layer. Such a uniform and continuous thin grain boundary layer can effectively isolate the main-phase grains, weaken the magnetic exchange coupling between grains, and this is one of the important reasons for the enhancement of coercivity \cite{HONO2012}, \cite{ZENG2025}. Second, a slightly brighter shell layer can be observed on the surface of the main-phase grains, which is very likely the Al-enriched region. The appearance of the core-shell structure indicates that the grain boundary diffusion treatment has achieved a favorable effect \cite{zhu_2025}. Third, the grain size slightly increases to approximately 8-10$\mu$m, indicating that slight grain growth occurred during the 7-hour holding at 900$^\circ$C.


Figure \ref{SEM_image_900C1500magnification_2a} shows the SEM image of the 900$^\circ$C sample from another region, where the core-shell structure can be more clearly observed \cite{wangzhan_2026}. The central region of a typical grain appears darker, while the edge region exhibits a brighter thin layer with a thickness of approximately 1-2 $\mu$m. EDS analysis (see the following section) confirms that the Al concentration is higher in this thin layer.

A comprehensive comparison of the SEM observations at 700$^\circ$C and 900$^\circ$C reveals that the 900$^\circ$C diffusion temperature can more effectively optimize the morphology of the grain boundary phase, forming a more continuous and uniform thin grain boundary layer, while simultaneously promoting the formation of an Al-enriched shell layer on the surface of the main-phase grains.

\subsection{Elemental distribution analysis}
\subsubsection{Energy-dispersive X-ray spectroscopy spectrum}
Compositional analysis of different micro-regions of the 900$^\circ$C diffused sample was performed using energy-dispersive X-ray spectroscopy. EDS data were collected at three positions: grain center, grain edge (shell layer), and grain boundary phase.

\begin{table}[H]
\centering
\caption{EDS analysis results at different positions of the 900$^\circ$C diffused sample (wt.\%)}
\begin{tabular}{|c|c|c|c|c|c|}\hline
grain position & Al & Zn & Nd & Fe\\\hline
center & 0.31 & 0.02 & 26.45 & 70.82\\\hline
edge (shell layer) & 1.24 & 0.15 & 24.38 & 71.92\\\hline
boundary phase & 0.85 & 0.42 & 55.30 & 42.10\\\hline
\end{tabular}
\label{tab:eds}
\end{table}

First, Al can be detected at all three positions, but its concentration distribution is non-uniform. The Al content is highest at the grain edge (shell layer, 1.24wt.\%), followed by the grain boundary phase (0.85wt.\%), and lowest at the grain center (0.31wt.\%) \cite{zhu_2025}. This distribution pattern indicates that after diffusing from the surface into the magnet, a portion of Al remains in the grain boundary phase, a portion enters the surface layer of the main-phase grains, and a small amount diffuses to the grain interior.

Second, the distribution of Zn differs from that of Al. The Zn content is highest in the grain boundary phase (0.42wt.\%), lower at the grain edge (0.15wt.\%), and nearly undetectable at the grain center (0.02wt.\%, which falls within the EDS detection error range) \cite{wangfang_2023}. This result indicates that Zn is mainly distributed in the grain boundary phase and essentially does not enter the main-phase lattice.

Third, the Nd content in the grain boundary phase is relatively high (55.30wt.\%), which is a characteristic compositional feature of the typical Nd-rich grain boundary phase \cite{HONO2012}. EDS analysis results of the 700$^\circ$C sample show that the Al content at the grain edge is approximately 0.42wt.\%, which is about one-third of that in the 900$^\circ$C sample; the Zn content in the grain boundary phase is approximately 0.18wt.\%, which is about half of that in the 900$^\circ$C sample \cite{fangzhu_2025}. This indicates that the total amounts of Al and Zn diffused into the magnet at 700$^\circ$C are relatively small, consistent with the previous observation of shallower diffusion depth.

The difference in the distribution of Al and Zn can be explained from two aspects: atomic radius and chemical affinity. The atomic radius of Al (143 pm) is relatively close to that of Fe (124 pm), so Al can more readily substitute for Fe in the main-phase lattice \cite{HE2024_boost}. Although the atomic radius of Zn (134 pm) is also relatively close to that of Fe, the mixing enthalpy of Zn and Fe is positive (they repel each other), and Zn has a high vapor pressure, making it more volatile at high temperatures rather than dissolving into the main phase.

\subsubsection{EDS mapping analysis}
\begin{figure}
\centering
\includegraphics[width = 0.5\textwidth, height = 0.4\textwidth]{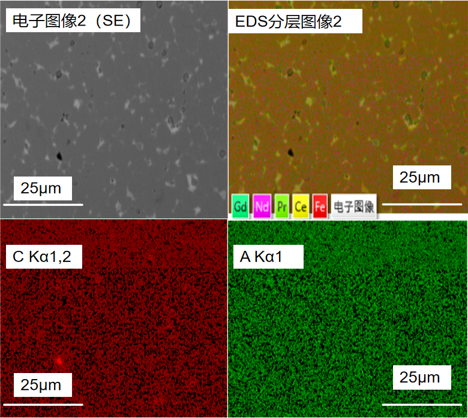}
\caption{EDS mapping of the 900$^\circ$C diffused sample in the region near the surface (A1EDS)}
\label{EDS_map_900A1}
\end{figure}

Figure \ref{EDS_map_900A1} shows the EDS mapping analysis results of the 900$^\circ$C diffused sample in the region near the surface. From the Al mapping image, it can be seen that the distribution of Al coincides well with the contour of the grain boundary phase, and diffuse Al signals can also be detected inside the main-phase grains \cite{gaowei_2025}. This result confirms the conclusion drawn from the point analysis: Al exists both in the grain boundary phase and within the main-phase grains.

No obvious Zn signal was detected in the mapping analysis, which may be attributable to two reasons: first, the amount of Zn added is inherently small; second, a large amount of Zn volatilized at high temperature, and the residual concentration fell below the EDS detection limit \cite{wangzhan_2026}. This result is consistent with the inference from the previous mass loss analysis that the mass loss of approximately 0.312g mainly originates from Zn volatilization. Zhong et al. \cite{ZHONG2023} also reported a similar phenomenon, where only a small amount of Zn residue was detected in the magnet after diffusion.

Combining the mass loss data and EDS analysis results, it can be inferred that the role of Zn is mainly manifested during diffusion: when Zn volatilizes, it generates disturbances in the grain boundary liquid phase, promoting liquid flow and facilitating the migration of Al (and possibly subsequently added Tb), while most of the Zn itself is consumed.

The spatial resolution of the EDS point analysis should also be considered when interpreting the composition of the grain boundary phase. In the SEM observations, the Al-enriched shell layer at the edge of the main-phase grains has a thickness of approximately 1-2$\mu$m, and EDS measurements were performed at the grain center, grain edge (shell layer), and grain boundary phase. Because the effective X-ray generation volume of EDS is finite and can extend beyond the exact position of the analysis point, the composition measured at a narrow grain boundary may contain contributions from the adjacent Nd$_2$Fe$_{14}$B main phase. Therefore, the values reported for the ``boundary phase'' in table \ref{tab:eds} should be interpreted as the composition of the analyzed grain-boundary region rather than as a strictly isolated, single-phase grain-boundary composition. In particular, the relatively high Nd content (55.30wt.\%) and Fe content (42.10wt.\%) indicate that the measured region is strongly Nd-rich compared with the grain center, while the possibility of partial signal contribution from the neighboring main phase cannot be excluded.

The EDS results are used primarily to establish the relative elemental distribution associated with the diffusion treatment rather than to determine the exact stoichiometric composition of the grain boundary phase. The measured Al concentration increases from 0.31wt.\% at the grain center to 1.24wt.\% at the grain edge, while Zn reaches its highest measured concentration in the boundary region (0.42wt.\%). These spatial trends, together with the EDS mapping results showing that the Al distribution follows the grain-boundary contours while also extending into the main-phase grains, support the conclusion that Al is present both in the boundary region and in the near-surface region of the main phase, whereas Zn is preferentially concentrated in the boundary region. Thus, the EDS data should not be taken as definitive evidence for a pure grain-boundary-phase composition, but rather as semi-quantitative evidence for the elemental redistribution produced by Al-Zn grain boundary diffusion. The elemental totals in Table 3 also do not equal exactly 100wt.\% because the table reports only Al, Zn, Nd, and Fe and does not explicitly include B or other possible constituents.

\subsection{Crystal structure analysis}
\subsubsection{XRD phase analysis}



\begin{figure}
\centering
\begin{subfigure}{0.23\textwidth}
\includegraphics[width=\textwidth,height=0.76\textwidth]{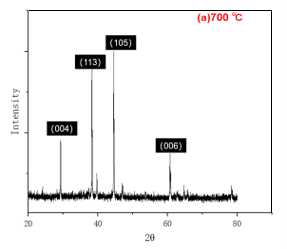}
\caption{XRD pattern of the sample diffusion-treated at 700$^\circ$C}
\label{700xrd}
\end{subfigure}
\begin{subfigure}{0.23\textwidth}
\includegraphics[width=\textwidth,height=0.76\textwidth]{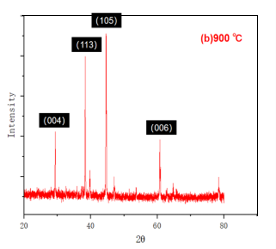}
\caption{XRD pattern of the sample diffusion-treated at 900$^\circ$C}
\label{900xrd}
\end{subfigure}
\begin{subfigure}{0.23\textwidth}
\includegraphics[width=\textwidth,height=0.76\textwidth]{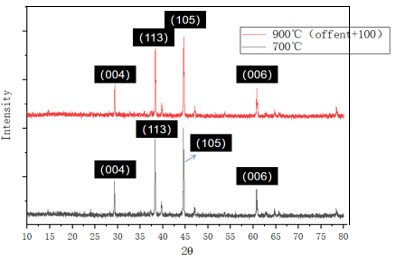}
\caption{Comparison of XRD patterns for the two samples}
\label{compare_xrd}
\end{subfigure}
\begin{subfigure}{0.23\textwidth}
\includegraphics[width=\textwidth,height=0.76\textwidth]{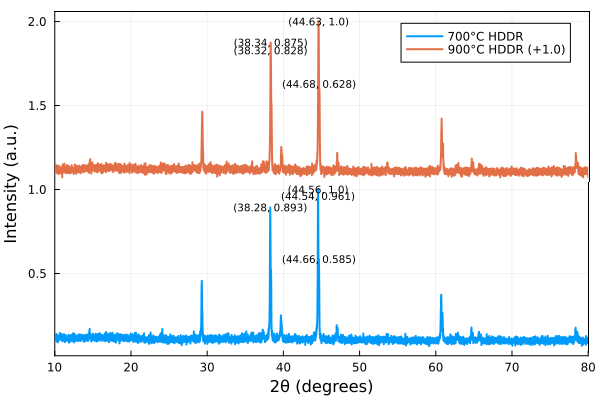}
\caption{XRD pattern comparison with diffraction peaks labeled}
\label{xrd_peaks_labeled}
\end{subfigure}
\caption{X-ray diffraction patterns of Al-Zn grain boundary diffusion-treated sintered Nd-Fe-B magnets.}
\label{fig:xrd_collection}
\end{figure}
X-ray diffraction analysis was performed on the 700$^\circ$C and 900$^\circ$C diffused samples \cite{gaowei_2025}. Figure \ref{700xrd}, \ref{900xrd} shows the XRD patterns of the 700$^\circ$C and 900$^\circ$C samples, respectively, and Figure \ref{compare_xrd} shows the superimposed comparison of the two patterns. By comparison with the standard PDF card (No. 00-038-0889, Nd$_2$Fe$_{14}$B), all diffraction peaks can be indexed. The main diffraction peaks include: the (004) peak at approximately $2\theta \approx 29.3^\circ$, the (113) peak at approximately $2\theta \approx 38.2^\circ$, the (105) peak at approximately $2\theta \approx 44.4^\circ$, as well as weaker peaks such as (006), (115), and (214).

No diffraction peaks of Al or Zn were detected in the XRD patterns (the main peak of Al is located near $38.5^\circ$, which does not overlap with the peaks of Nd$_2$Fe$_{14}$B). This can be attributed to the following reasons: First, the amounts of Al and Zn added are relatively small; they either exist in solid solution form within the main phase or grain boundary phase, or the amount of compounds formed is too small and the size too fine, falling below the detection limit of XRD (typically 1-2wt.\%) \cite{gaowei_2025}. Second, the diffusion treatment did not destroy the structure of the main phase, which still maintains the intact tetragonal structure of Nd$_2$Fe$_{14}$B without decomposition or phase transformation.

The relative intensities of the (004) and (105) peaks of the 900$^\circ$C sample are essentially consistent with those of the 700$^\circ$C sample, indicating that the change in diffusion temperature did not destroy the crystallographic texture (i.e., degree of orientation) of the magnet \cite{zhou_2011}. The preservation of texture is of great significance for maintaining a relatively high remanence \cite{zhang_1998}.

The XRD results were further considered with respect to the possibility of Al or Zn incorporation into the Nd$_2$Fe$_{14}$B main-phase lattice. The diffraction patterns of the 700~$^\circ$C and 900~$^\circ$C samples can be indexed to the Nd$_2$Fe$_{14}$B structure, and no distinct diffraction peaks attributable to Al- or Zn-containing secondary phases were detected. However, the differences in the positions of the major (004), (113), and (105) reflections between the two diffusion-treated samples are only approximately 0.01$^\circ$--0.02$^\circ$, which is comparable to the measurement uncertainty under the present XRD conditions. Therefore, these small peak-position differences cannot be regarded as statistically significant evidence of a change in the lattice parameters.

Al substitution for Fe remains a possible mechanism because EDS detects an increased Al concentration in the grain-edge shell relative to the grain center, while the atomic size difference between Al and Fe is consistent with a possible expansion of the lattice if substitution occurs. Nevertheless, the present XRD measurements do not provide sufficient precision to directly quantify such a small lattice change. In particular, the current analysis does not include Rietveld refinement or quantitative lattice-parameter determination relative to the untreated 38SH sample. Thus, the XRD data should be interpreted as evidence that the Nd$_2$Fe$_{14}$B main-phase structure is retained after diffusion treatment, while possible Al incorporation and the associated lattice expansion should be regarded as a hypothesis supported by the combined elemental-distribution and structural observations rather than as a directly established crystallographic result.

Future work should employ higher-resolution XRD measurements together with peak-profile fitting and Rietveld refinement of the untreated and diffusion-treated samples to determine the lattice parameters and their uncertainties. Such an analysis would allow any systematic changes in the $a,b,c$ lattice parameters to be distinguished from instrumental peak-position uncertainty and would provide a more rigorous test of whether Al substitution into the Nd$_2$Fe$_{14}$B main phase occurs.

\subsubsection{Changes in diffraction peak positions}
To determine whether Al and Zn entered the main-phase lattice, the diffraction peak positions of the two samples were compared, and three higher-intensity peaks [(004), (113), and (105)] were selected for analysis.

The data show that the (004) peak of the 900$^\circ$C sample is located at $29.33^\circ$, and that of the 700$^\circ$C sample is also located at $29.33^\circ$; the peak positions of the two samples are almost identical, with differences within the range of $0.01^\circ$-$0.02^\circ$, which falls within the measurement error range. Similar situations are observed for the (113) and (105) peaks, with no significant difference in peak positions between the two samples.

The absence of obvious shifts in diffraction peak positions may be attributed to the following reasons. First, the amounts of Al and Zn added are relatively small; the total amount of Al and Zn that diffused into the magnet is less than 0.5\% of the total mass of the magnet, and the change in the lattice is too minute to be detected by XRD \cite{gaowei_2025}. Second, Al and Zn are mainly distributed in the grain boundary phase (especially Zn), and the amount entering the main-phase lattice is limited \cite{wangzhan_2026}. Third, even if Al enters the main-phase lattice, since the atomic radius of Al (143 pm) is larger than that of Fe (124 pm), the substitution of Fe by Al would cause lattice expansion, and the diffraction peaks should shift to lower angles. However, for solid solutions with a content below 0.5\%, the peak shift is typically below $0.01^\circ$-$0.02^\circ$, which is beyond the resolving power of conventional XRD.

\subsection{Diffusion mechanism}
\subsubsection{Effect of temperature}
Based on the above experimental results, the effects of temperature on the diffusion behavior of Al and Zn can be summarized as follows:

At 900$^\circ$C, the diffusion depth of Al and Zn is approximately 150-200 $\mu$m, the grain boundary phase becomes thin and continuous, and Al enters the surface layer of the main-phase grains (with a thickness of approximately 1-2 $\mu$m), forming an obvious core-shell structure.

At 700$^\circ$C, the diffusion depth is only approximately 50 $\mu$m, the optimization of the grain boundary phase is not as good as that at 900$^\circ$C, and less Al enters the main-phase grains (approximately one-third of that at 900$^\circ$C).

The above differences can be explained by the Arrhenius equation. The relationship between the diffusion coefficient $D$ and temperature $T$ is $D = D_0 \exp\left(-\frac{Q}{RT}\right)$, where $D_0$ is the pre-exponential factor, $Q$ is the activation energy for diffusion, and $R$ is the gas constant (8.314 J$\cdot$mol$^{-1}\cdot$K$^{-1}$). The temperature increases from 700$^\circ$C (973 K) to 900$^\circ$C (1173 K), an increase of 200 K. Assuming that the activation energy $Q$ for Al diffusion in NdFeB is approximately 200 kJ/mol, the diffusion coefficient at 900$^\circ$C is approximately 15 to 20 times that at 700$^\circ$C. This is the fundamental reason why the diffusion effect at 900$^\circ$C is significantly better than that at 700$^\circ$C.

\subsubsection{Synergistic mechanism of Al and Zn}
Based on the experimental results, a synergistic model of Al and Zn during the grain boundary diffusion can be proposed.

During the heating stage, the Al$_{80}$Zn$_{20}$ alloy sheet melts at approximately 382$^\circ$C. The molten liquid wets the end face of the magnet, forming good interfacial contact \cite{HONO2012}. As the temperature continues to rise to 700$^\circ$C or 900$^\circ$C, Al and Zn atoms diffuse into the magnet along the grain boundaries (which serve as fast diffusion paths for atomic migration).

After Al and Zn atoms reach the grain boundary phase, they dissolve into the Nd-rich liquid phase \cite{HE2024_boost}. Since the alloying of Al and Zn with Nd can lower the melting point of the Nd-rich phase (Al lowers the melting point from 471$^\circ$C to 435$^\circ$C, and Zn lowers it to 443$^\circ$C), both the viscosity and surface tension of the grain boundary phase are reduced, making it easier to spread and form a thin and continuous grain boundary layer.

The high-temperature volatility of Zn plays a special role. When Zn volatilizes, it generates local disturbances in the grain boundary liquid phase \cite{zhangshou_2000}, preventing the formation of local stagnant regions in the grain boundary phase and keeping the diffusion channels unobstructed. The volatilized loss of Zn precisely explains the mass loss of approximately 0.312g.

In addition to being present in the grain boundary phase, Al further diffuses into the main-phase grains, entering the surface layer of the grains and substituting for Fe atoms \cite{wangzhan_2026}. The substitution of Fe by Al generates local strain in the lattice, which may lower the energy barrier for Tb atoms to enter the lattice, creating conditions for the subsequent diffusion and solid solution of heavy rare earths.

Overall, the role of Zn is mainly to keep the grain boundary channels unobstructed, while the role of Al is mainly to lower the viscosity of the grain boundary phase and prepare the main-phase lattice for the reception of heavy rare earths \cite{zhu_2025}. The synergy between the two can be summarized as follows: Zn enhances the diffusion rate, while Al enhances the diffusion depth and solid solution efficiency.

\subsubsection{Prospects and discussion}
Several aspects require further investigation to quantitatively establish the proposed mechanisms. Because the present samples were furnace-cooled without an additional post-diffusion tempering treatment, systematic comparison of different tempering conditions is needed to determine their influence on the final grain-boundary structure and magnetic properties. Cross-sectional EDS line scans or elemental mapping at multiple distances from the diffusion interface are also required to quantitatively determine the Al and Zn concentration profiles and the actual diffusion depths. Higher-resolution techniques, such as atom probe tomography or other nanoscale compositional analyses, would further clarify the atomic-scale distribution of Al and Zn across the grain boundaries. In addition, local magnetic characterization and independent determination of the parameters used in the micromagnetic model would be necessary to quantitatively separate the contributions of grain-boundary modification, grain-edge morphology, and possible changes in local magnetic properties. These experiments would provide a stronger basis for distinguishing the individual roles of Al and Zn and for independently validating the computational analysis.

The statistical reproducibility of the present experimental results is limited by the number of independently processed specimens. In the current experimental design, one specimen (Sample B) was treated at 700$^\circ$C, whereas two parallel specimens (Samples A1 and A2) were treated at 900$^\circ$C for 7h. Therefore, the results obtained at 700$^\circ$C should be regarded as representative observations for the present processing condition rather than statistically generalized values for the entire material population. The two 900$^\circ$C specimens provide a limited assessment of repeatability, but the sample number remains insufficient for a rigorous statistical evaluation of process-to-process variability. This limitation is particularly relevant for sintered Nd-Fe-B magnets because variations in grain structure, grain-boundary morphology, and local elemental distribution can contribute to specimen-to-specimen differences in magnetic properties.

The coercivity and remanence values reported in this work are presented primarily as measured values for the individual specimens rather than as population averages. The magnetic properties were determined from room-temperature hysteresis loops measured using the VSM, from which $H_{\mathrm{cj}}$ and $B_{\mathrm{r}}$ were extracted. The observed increase in coercivity after Al-Zn grain boundary diffusion should therefore be interpreted as an experimental observation under the specified processing conditions, while the magnitude of the improvement should not be taken as a statistically established mean effect. In particular, the comparison between 700$^\circ$C and 900$^\circ$C is intended to demonstrate the temperature-dependent trend observed in the present specimens, rather than to establish a statistically significant difference between the two processing temperatures. Additional independently prepared specimens, especially replicate specimens at 700$^\circ$C, will be required in future work to determine standard deviations, confidence intervals, and the statistical significance of the temperature-dependent changes in coercivity and remanence.

The temperature of 900$^\circ$C is a more effective pre-diffusion temperature than 700$^\circ$C. Although 900$^\circ$C causes a slightly greater decrease in remanence (approximately 0.20 kGs), the advantage in coercivity enhancement (approximately 206.7 kA/m) far outweighs this cost. In subsequent diffusion experiments involving Tb, 900$^\circ$C should be preferentially selected as the diffusion temperature.

Al and Zn play different roles in the diffusion. Al is suitable as a "fluxing agent" and "positioning agent" for Tb because it can enter the main-phase lattice and create conditions for the entry of Tb; Zn is suitable as an "activator" for grain boundaries because it can keep the diffusion channels unobstructed. The combined use of the two is expected to produce synergistic effects.

The diffusion depth of Al and Zn is approximately 150-200 $\mu$m, indicating that the grain boundary channels after Al-Zn pre-diffusion treatment can support the diffusion of Tb to at least the same depth. Therefore, a two-step diffusion involving Al-Zn pre-diffusion followed by Tb diffusion is feasible.

The entry of Al into the main-phase lattice causes a slight expansion of the unit cell volume, which may lower the energy barrier for Tb atoms to enter the lattice, facilitating the formation of a high-anisotropy shell layer on the grain surface.

Based on the above insights, subsequent experiments can be designed as follows: first, pre-diffuse with Al$_{80}$Zn$_{20}$ alloy sheets at 900$^\circ$C for 7 hours, then perform grain boundary diffusion of Tb on the same magnet (using TbF$_3$ powder or Tb-containing alloy sheets as the diffusion source), and compare the magnetic property differences between the two-step procedure and direct Tb diffusion.

\section{Conclusions}
Al$_{80}$Zn$_{20}$ alloy sheets were used as the diffusion source and placed on both end faces of small cylindrical sintered NdFeB magnets. Vacuum diffusion treatments were performed at 900$^\circ$C and 700$^\circ$C, respectively, with a holding time of 7 hours. By comparing the magnetic properties, microstructure, and elemental distribution of the magnets before and after diffusion, the diffusion behavior of Al and Zn and their influence mechanisms on magnetic properties were analyzed. The following main conclusions were drawn.

Al-Zn grain boundary diffusion can effectively enhance the coercivity of sintered NdFeB magnets, and the enhancement effect is positively correlated with the diffusion temperature. After diffusion at 900$^\circ$C for 7 hours, the coercivity increased from the original value of 951.5 kA/m (11.96 kOe) to 1158.2 kA/m (14.55 kOe), an enhancement of approximately 206.7 kA/m. After diffusion at 700$^\circ$C for 7 hours, the coercivity increased to 1039.6 kA/m (13.06 kOe), an enhancement of approximately 88 kA/m. The enhancement at 900$^\circ$C is approximately 2.35 times that at 700$^\circ$C. The temperature dependence of the coercivity enhancement can be attributed to two reasons: first, the atomic diffusion coefficient increases significantly at higher temperatures (the diffusion coefficient at 900$^\circ$C is approximately 15-20 times that at 700$^\circ$C); second, the viscosity of the grain boundary phase decreases and its fluidity increases at higher temperatures, facilitating rapid atomic migration.

Higher diffusion temperatures lead to more pronounced decreases in remanence. After diffusion at 900$^\circ$C, the remanence decreased from the original value of 1.282 T (12.82 kGs) to 1.256 T (12.56 kGs), a decrease of 0.026 T (0.26 kGs). After diffusion at 700$^\circ$C, the remanence decreased to 1.276 T (12.76 kGs), a decrease of 0.006 T (0.06 kGs). The decrease in remanence is attributed to the entry of Al atoms into the lattice of the main phase Nd$_2$Fe$_{14}$B, which dilutes the concentration of magnetic atoms and may disrupt the magnetic exchange coupling between Fe atoms \cite{liuxiang_2008}. The greater decrease in remanence for the 900$^\circ$C sample indicates that more Al atoms diffused into the main-phase grains at the higher temperature.

The distribution patterns of Al and Zn in the magnet are different: Al tends to enter the main-phase lattice, while Zn mainly remains in the grain boundary phase. EDS analysis results show that in the 900$^\circ$C diffused sample, the Al content at the edge (shell layer) of the main-phase grains reaches 1.24wt.\%, while that at the grain center is 0.31wt.\%, indicating that Al mainly aggregates in the surface layer of the grains. The Zn content in the grain boundary phase is 0.42wt.\%, at the grain edge is 0.15wt.\%, and at the grain center is almost undetectable (0.02wt.\%), indicating that Zn essentially does not enter the main-phase lattice \cite{qu_2025}. The reason for the different distribution patterns lies in the fact that the atomic radius of Al (143 pm) is relatively close to that of Fe (124 pm), making it easy for Al to substitute for Fe and enter the main phase; although the atomic radius of Zn (134 pm) is also close to that of Fe, the mixing enthalpy of Zn and Fe is positive (they repel each other), and Zn has a high vapor pressure, making it more volatile at high temperatures rather than dissolving into the main phase.

Al-Zn diffusion optimizes the morphology and distribution of the grain boundary phase. SEM observations show that in the original magnet, the grain boundary phase mainly exists in blocky aggregates at triple junctions, with many main-phase grains having no obvious grain boundaries between them. After diffusion at 900$^\circ$C, the grain boundary phase becomes thin and continuous, uniformly enveloping each main-phase grain and forming a shell-layer structure with a thickness of approximately 1-2 $\mu$m. Although the 700$^\circ$C diffused sample shows some improvement, the continuity and uniformity of the grain boundary phase remain suboptimal. The uniform and continuous thin grain boundary layer can effectively isolate the main-phase grains and weaken the magnetic exchange coupling between them, which is one of the important reasons for the enhancement of coercivity.

Al and Zn exhibit synergistic effects during diffusion. Based on the experimental results, a synergistic model of Al and Zn can be proposed: Zn is mainly distributed in the grain boundary phase, and when it volatilizes at high temperature, it generates local disturbances in the grain boundary liquid phase, preventing the formation of local stagnant regions and keeping the diffusion channels unobstructed; in addition to being present in the grain boundary phase, Al further diffuses into the main-phase grains, entering the surface layer of the grains and substituting for Fe atoms, creating conditions for the subsequent entry of heavy rare-earth atoms. The synergistic effects of the two can be specifically summarized as follows: Zn enhances the diffusion rate, while Al enhances the diffusion depth and solid solution efficiency.

The experimental results demonstrate that Al-Zn alloy wafer grain boundary diffusion modifies the microstructure and magnetic properties of the sintered Nd-Fe-B magnets. The 900$^\circ$C treatment produces a greater coercivity enhancement than the 700$^\circ$C treatment, accompanied by Al enrichment near the grain edges and modification of the grain-boundary region. The EDS results indicate that Al is preferentially enriched at the grain-edge region and extends into the adjacent main phase, whereas Zn is more strongly concentrated in the grain-boundary region. These observations are consistent with a contribution from grain-boundary modification and improved magnetic isolation of the Nd-Fe-B grains. However, the present results do not provide direct evidence that Al substitution produces a high-anisotropy shell, and this mechanism is therefore not used as an independent basis for explaining the coercivity enhancement. The observed Al and Zn distributions are interpreted as experimental observations rather than definitive evidence for a specific atomic-scale diffusion mechanism. The XRD results show no significant peak displacement beyond the experimental uncertainty and therefore do not establish a measurable lattice expansion associated with Al substitution.

\section{Declaration of competing interest}
The authors declare that there are no competing interests related to this research.
\bibliographystyle{elsarticle-num}
\bibliography{ref}

@article{ZHONG2023,
title = {Grain boundary diffusion efficiency of sintered Nd-Fe-B magnets by Al, Sn and Zn addition},
journal = {Journal of Alloys and Compounds},
volume = {967},
pages = {171676},
year = {2023},
issn = {0925-8388},
doi = {10.1016/j.jallcom.2023.171676},
url = {https://www.sciencedirect.com/science/article/pii/S0925838823029791},
author = {Shuwei Zhong and Munan Yang and Hang Wang and Sajjad Ur Rehman and Xi Yu and Sangen Luo and Longgui Li and Chao Li and Shuhua Xiong and Bin Yang}}

@article{MATSURA2006,
title = {Recent development of Nd–Fe–B sintered magnets and their applications},
journal = {Journal of Magnetism and Magnetic Materials},
volume = {303},
number = {2},
pages = {344-347},
year = {2006},
note = {The 6th International Symposium on Physics of Magnetic Materials},
issn = {0304-8853},
doi = {10.1016/j.jmmm.2006.01.171},
url = {https://www.sciencedirect.com/science/article/pii/S0304885306000424},
author = {Yutaka Matsura}}

@article{strnat_1970,
author = {Karl Strnat},
title = {The recent development of permanent magnet materials containing rare earth metals},
journal = {journal of magnetism and magnetic materials},
year = {1970},
volume={6},
number={2},
pages={182-190},
doi={10.1109/TMAG.1970.1066743}}

@article{ojima_1977,
author={Ojima, T. and Tomizawa, S. and Yoneyama, T. and Hori, T.},
journal={IEEE Transactions on Magnetics}, 
title={Magnetic properties of a new type of rare-earth cobalt magnets Sm2(Co, Cu, Fe, M)17}, 
year={1977},
volume={13},
number={5},
pages={1317-1319},
doi={10.1109/TMAG.1977.1059703}}

@article{sagawa_1984,
author = {Sagawa, M. and Fujimura, S. and Togawa, N. and Yamamoto, H. and Matsuura, Y.},
title = {New material for permanent magnets on a base of Nd and Fe},
journal = {Journal of Applied Physics},
year = {1984},
volume = {55},
number = {6},
pages = {2083-2087},
issn = {0021-8979},
doi = {10.1063/1.333572},
url = {https://doi.org/10.1063/1.333572}}

@article{zhou_2011,
author={Zhou, Shouzeng and Dong, Qingfei},
title={Sintered Nd-Fe-B Rare-Earth Permanent Magnet Materials and Technology},
journal={Beijing: Metallurgical Industry Press},
url={https://metalworld.ustb.edu.cn/cn/article/id/JSSJ201303004},
year={2011}}

@article{herbst_1991_int,
title = {${\mathrm{R}}_{2}{\mathrm{Fe}}_{14}$B materials: Intrinsic properties and technological aspects},
author = {Herbst, J. F.},
journal = {Rev. Mod. Phys.},
year = {1991},
publisher = {American Physical Society},
volume = {63},
issue = {4},
pages = {819-898},
numpages = {0},
doi = {10.1103/RevModPhys.63.819},
url = {https://link.aps.org/doi/10.1103/RevModPhys.63.819}}

@article{HONO2012,
title = {Strategy for high-coercivity Nd-Fe-B magnets},
journal = {Scripta Materialia},
year = {2012},
note = {Viewpoint Set No. 51: Magnetic Materials for Energy},
volume = {67},
number = {6},
pages = {530-535},
year = {2012},
issn = {1359-6462},
doi = {10.1016/j.scriptamat.2012.06.038},
url = {https://www.sciencedirect.com/science/article/pii/S1359646212004320},
author = {K. Hono and H. Sepehri-Amin}}

@article{Kronmuler_1988,
title={Micromagnetic analysis of the magnetic hardening mechanisms in RE-Fe-B magnets},
journal = {Journal of Magnetism and Magnetic Materials},
volume = {74},
number = {3},
pages = {291-302},
year = {1988},
issn = {0304-8853},
doi = {10.1016/0304-8853(88)90202-8},
url = {https://www.sciencedirect.com/science/article/pii/0304885388902028},
author = {H. Kronmüller and K.-D. Durst and M. Sagawa}}

@article{li2020,
title = {On the temperature-dependent coercivities of anisotropic Nd-Fe-B magnet},
journal = {Acta Materialia},
volume = {199},
pages = {288-296},
year = {2020},
issn = {1359-6454},
doi = {10.1016/j.actamat.2020.08.040},
url = {https://www.sciencedirect.com/science/article/pii/S1359645420306376},
author = {J. Li and Xin Tang and H. Sepehri-Amin and T. Ohkubo and K. Hioki and A. Hattori and K. Hono}}

@article{HIROSAWA_1985,
author = {HIROSAWA, S and MATSUURA, Y and YAMAMOTO, H and FUJIMURA, S},
title = {Single crystal measurements of anisotropy constants of R2Fe14B R=Y, Ce, Pr, Nd, Gd, Tb, Dy and Ho},
journal = {Japanese journal of applied physics},
year = {1985},
volume = {24},
number = {10A},
pages = {L803},
doi = {10.1143/JJAP.24.L803}}

@article{jiang_2025,
author = {\begin{CJK*}{UTF8}{bsmi}姜琛 and 崔紅兵 and 張茂彩\end{CJK*} and
Jiang, Chen and Cui, Hongbing and Zhang, Maocai},
title = {\begin{CJK*}{UTF8}{bsmi}磁控溅射Dy-Cu合金晶界扩散钕铁硼磁体磁性能研究\end{CJK*} Study on Magnetic Properties of Nd-Fe-B Magnets via Magnetron Sputtering Dy-Cu Alloy Grain Boundary Diffusion},
journal = {\begin{CJK*}{UTF8}{bsmi}中國稀土學報\end{CJK*} Journal of the Chinese Society of Rare Earths},
year = {2025}}

@article{zhang_1998,
author = {\begin{CJK*}{UTF8}{bsmi}張建成 and 李衛 and 李岫梅\end{CJK*} and
Zhang, Jiancheng and Li, Wei and Li, Xiumei},
title = {\begin{CJK*}{UTF8}{bsmi}晶粒尺寸及取向對釹鐵硼磁體磁性能的影響\end{CJK*} Effect of Grain Size and Orientation on the Magnetic Properties of Nd-Fe-B Magnets},
journal = {\begin{CJK*}{UTF8}{bsmi}金屬功能材料\end{CJK*} Metallic Functional Materials},
year = {1998}}

@article{zhu_2025,
author = {\begin{CJK*}{UTF8}{bsmi}朱琳 and 李志杰 and 張超超\end{CJK*} 
and Zhu, Lin and Li, Zhijie and Zhang, Chaochao},
title = {\begin{CJK*}{UTF8}{bsmi}納米粉晶界擴散對燒結釹鐵硼性能的影響\end{CJK*} Effect of Al-Zn Nano-Powder Grain Boundary Diffusion on the Properties of Sintered Nd-Fe-B Magnets},
journal = {\begin{CJK*}{UTF8}{bsmi}遼寧化工\end{CJK*} Liaoning Chemical Industry},
year = {2025}}

@article{HE2024_boost,
title = {Boosting efficiency and oxidation resistance of Pr-Tb-Cu diffusion source for Nd-Fe-B magnets by synergistic modification of Al or Ni},
journal = {Corrosion Science},
volume = {237},
pages = {112300},
year = {2024},
issn = {0010-938X},
doi = {10.1016/j.corsci.2024.112300},
url = {https://www.sciencedirect.com/science/article/pii/S0010938X24004955},
author = {Jiayi He and Chaochao Zeng and Lizhong Zhao and Jiali Cao and Xiaolian Liu}}

@article{cheng_2013,
author = {\begin{CJK*}{UTF8}{bsmi}程星華 and 李建 and 喻曉軍\end{CJK*}
and Cheng, Xinghua and Li, Jian and Yu, Xiaojun},
title = {\begin{CJK*}{UTF8}{bsmi}釹鐵硼磁體鍍層結合力分析\end{CJK*} Analysis of Coating Adhesion on Nd-Fe-B Magnets},
journal = {\begin{CJK*}{UTF8}{bsmi}金屬功能材料\end{CJK*} Metallic Functional Materials},
year = {2013}}

@article{zhou_2020_prep,
author = {\begin{CJK*}{UTF8}{bsmi}周巧英 and 陳仁杰 and 郭帥\end{CJK*}
and Zhou, Qiaoying and Chen, Renjie and Guo, Shuai},
title = {\begin{CJK*}{UTF8}{bsmi}釹鐵硼磁體表面高耐蝕性鋅鋁塗層的製備及性能研究\end{CJK*} Preparation and Performance Study of Highly Corrosion-Resistant Zinc-Aluminum Coating on Nd-Fe-B Magnet Surfaces},
journal = {\begin{CJK*}{UTF8}{bsmi}電鍍與塗飾\end{CJK*} Electroplating \& Finishing},
year = {2020}}

@article{wangfang_2023,
author = {\begin{CJK*}{UTF8}{bsmi}王芳\end{CJK*} and Wang Fang},
title = {\begin{CJK*}{UTF8}{bsmi}多元素晶界擴散燒結釹鐵硼磁體的結構與性能調控\end{CJK*} Structure and Property Regulation of Multi-Element Grain Boundary Diffused Sintered Nd-Fe-B Magnets},
journal = {\begin{CJK*}{UTF8}{bsmi}內蒙古科技大學包頭\end{CJK*} Inner Mongolia University of Science and Technology Baotou},
year = {2023}}

@article{gaowei_2025,
author = {\begin{CJK*}{UTF8}{bsmi}高偉波 and 謝志斌 and 岑志波\end{CJK*}
and Gao, Weibo and Xie, Zhibin and Cen, Zhibo},
title = {\begin{CJK*}{UTF8}{bsmi}利用x射綫螢光譜研究釹鐵硼磁鋼晶界擴散後鋱的分佈\end{CJK*} Study on the Distribution of Terbium in Nd-Fe-B Permanent Magnets After Grain Boundary Diffusion Using X-Ray Fluorescence Spectrometry},
journal = {\begin{CJK*}{UTF8}{bsmi}中國稀土學報\end{CJK*} Journal of the Chinese Society of Rare Earths},
year = {2025}}

@article{ZENG2025,
title = {A galvanic corrosion mechanism of the selected area grain boundary diffusion in sintered Nd-Fe-B},
journal = {Materials Letters},
volume = {384},
pages = {138070},
year = {2025},
issn = {0167-577X},
doi = {10.1016/j.matlet.2025.138070},
url = {https://www.sciencedirect.com/science/article/pii/S0167577X25000990},
author = {Jia Zeng and Xiaojun Sun and Ruzhi Wang and Yang Luo and Dunbo Yu and Hui Yan}}

@article{wangzhan_2026,
author = {Wang, Zhanjia and Li, Yuqing and Liu, Weiqiang and Wu, Haihui and Ji, Ming},
title = {Performance and diffusion optimization mechanism in Al-assisted grain boundary diffused Nd–Fe–B magnets},
journal = {Journal of Materials Chemistry A},
volume = {14},
number = {3},
pages = {1806-1820},
year = {2026},
doi = {10.1039/d5ta07777j},
url = {https://doi.org/10.1039/d5ta07777j}}

@article{fangzhu_2025,
author = {\begin{CJK*}{UTF8}{bsmi}方祝福\end{CJK*} and Fang, Zhufu},
title = {\begin{CJK*}{UTF8}{bsmi}釹鐵硼磁體表面處理技術研究\end{CJK*} Research on Surface Treatment Technology of Nd-Fe-B Magnets},
journal = {\begin{CJK*}{UTF8}{bsmi}冶金與材料\end{CJK*} Metallurgy and Materials},
year = {2025}}

@article{wangjing_2021,
author = {\begin{CJK*}{UTF8}{bsmi}王晶 and 張懷武\end{CJK*}
and Wang, Jing and Zhang, Huaiwu},
title = {\begin{CJK*}{UTF8}{bsmi}燒結釹鐵硼永磁材料的晶界擴散技術研究進展\end{CJK*} Research Progress on Grain Boundary Diffusion Technology of Sintered Nd-Fe-B Permanent Magnet Materials},
journal = {\begin{CJK*}{UTF8}{bsmi}功能材料\end{CJK*} Journal of Functional Materials},
year = {2021}}

@article{huangmin_2025,
author = {\begin{CJK*}{UTF8}{bsmi}黃敏 and 丁勇 and 王春國\end{CJK*} and Huang, Min and Ding, Yong and Wang, Chunguo},
title = {\begin{CJK*}{UTF8}{bsmi}熱處理工藝對晶界擴散釹鐵硼磁體顯微組織和矯頑力的影響研究\end{CJK*} Effect of Heat Treatment Process on Microstructure and Coercivity of Grain Boundary Diffused Nd-Fe-B Magnets},
journal = {\begin{CJK*}{UTF8}{bsmi}稀土\end{CJK*} Chinese Rare Earths},
year = {2025}}

@article{xujia_2025,
author = {\begin{CJK*}{UTF8}{bsmi}徐嘉誠\end{CJK*} and Xu, Jiacheng},
title = {\begin{CJK*}{UTF8}{bsmi}釹鐵硼磁體晶界擴散優化研究\end{CJK*} Research on Grain Boundary Diffusion Optimization of Nd-Fe-B Magnets},
journal = {\begin{CJK*}{UTF8}{bsmi}冶金與材料\end{CJK*} Metallurgy and Materials},
year = {2025}}

@article{liuxiang_2008,
author = {\begin{CJK*}{UTF8}{bsmi}劉湘漣 and 李永利\end{CJK*}
and Liu, Xianglian and Li, Yongli},
title = {\begin{CJK*}{UTF8}{bsmi}Al元素對燒結NdFeB磁體磁性能和微觀結構的影響\end{CJK*} Effect of Al Element on Magnetic Properties and Microstructure of Sintered NdFeB Magnets},
journal = {\begin{CJK*}{UTF8}{bsmi}稀有金屬材料與工程\end{CJK*} Rare Metal Materials and Engineering},
year = {2008}}

@article{zhangshou_2000,
author = {\begin{CJK*}{UTF8}{bsmi}張守民 and 歐陽砥 and 周永治\end{CJK*} and Zhang, Shoumin and Ouyang, Di and Zhou, Yongzhi},
title = {\begin{CJK*}{UTF8}{bsmi}釹鐵硼磁體的有機溶液電鍍鋁研究\end{CJK*} Study on Aluminum Electroplating from Organic Solution for NdFeB Magnets},
journal = {\begin{CJK*}{UTF8}{bsmi}材料工程\end{CJK*} Journal of Materials Engineering},
year = {2000}}

@article{qu_2025,
author = {\begin{CJK*}{UTF8}{bsmi}瞿朝雲 and 劉正堂 and 張定軍\end{CJK*} and Qu, Chaoyun and Liu, Zhengtang and Zhang, Dingjun},
title = {\begin{CJK*}{UTF8}{bsmi}釹鐵硼磁體表面複合材料的防腐性能分析\end{CJK*} Analysis of Anti-corrosion Performance of Composite Materials on NdFeB Magnet Surface},
journal = {\begin{CJK*}{UTF8}{bsmi}冶金與材料\end{CJK*} Metallurgy and Materials},
year = {2025}}

@article{zhang_2021_pred_mag,
author = {Yun Zhang and Xiaojie Xu},
title = {Predicting Magnetic Remanence of NdFeB Magnets from Composition},
journal = {Journal of Superconductivity and Novel Magnetism},
volume  = {34},
number  = {11},
pages   = {2711-2715},
issn    = {1557-1947},
year = {2021},
doi = {10.1007/s10948-021-05901-2}}

@article{kim_1996_high,
author = {Kim, A. S. and Camp, F. E.},
title = {High performance NdFeB magnets},
journal = {Journal of Applied Physics},
year = {1996},
volume = {79},
number = {8},
pages = {5035-5039},
issn = {0021-8979},
doi = {10.1063/1.361566}}

@article{brown2002develop,
title = {Developments in the processing and properties of NdFeb-type permanent magnets},
journal = {Journal of Magnetism and Magnetic Materials},
year = {2002},
volume = {248},
number = {3},
pages = {432-440},
issn = {0304-8853},
doi = {10.1016/S0304-8853(02)00334-7},
url = {https://www.sciencedirect.com/science/article/pii/S0304885302003347},
author = {David Brown and Bao-Min Ma and Zhongmin Chen}}

@ARTICLE{fan2016form,
author={Fan, M. and Liu, Y. and Jha, Rajesh and Dulikravich, George S. and Schwartz, J. and Koch, C. C.},
journal={IEEE Transactions on Magnetics}, 
title={On the Formation and Evolution of Cu-Ni-Rich Bridges of Alnico Alloys With Thermomagnetic Treatment},
year={2016},
volume={52},
number={8},
pages={1-10},
doi={10.1109/TMAG.2016.2555956}}

@ARTICLE{hirosawa1990coercivity,
author={Hirosawa, S. and Tomizawa, H. and Mino, S. and Hamamura, A.},
journal={IEEE Transactions on Magnetics}, 
title={High-coercivity Nd-Fe-B-type permanent magnets with less dysprosium}, 
year={1990},
volume={26},
number={5},
pages={1960-1962},
doi={10.1109/20.104582}}

@ARTICLE{sagawa1990high,
author={Sagawa, M. and Tenaud, P. and Vial, F. and Hiraga, K.},
journal={IEEE Transactions on Magnetics}, 
title={High coercivity Nd-Fe-B sintered magnet containing vanadium with new microstructure}, 
year={1990},
volume={26},
number={5},
pages={1957-1959},
doi={10.1109/20.104581}}

@ARTICLE{kim1992corrosion,
author={Kim, A.S. and Camp, F.E.},
journal={IEEE Transactions on Magnetics}, 
title={A high performance NdFeB magnet with improved corrosion resistance}, 
year={1992},
volume={28},
number={5},
pages={2151-2153},
doi={10.1109/20.179426}}

@article{jurczyk1995powder,
title = {Magnets produced by hot pressing Nd2(Fe,Co,Zr)14B-$\alpha$-Fe and Nd(Fe,Mo)12Nx-$\alpha$-Fe powders},
journal = {Journal of Alloys and Compounds},
volume = {230},
number = {1},
pages = {L1-L3},
year = {1995},
issn = {0925-8388},
doi = {10.1016/0925-8388(95)01979-0},
author = {M. Jurczyk and P.B. Gwan}}

@article{cheng2014ml,
author = {Cheng, WenDe},
title = {Magnetic Remanence Prediction of NdFeB Magnets Based on a Novel Machine Learning Intelligence Approach Using a Particle Swarm Optimization Support Vector Regression},
year = {2014},
volume = {6},
number = {4},
issn = {1942-9045},
pages = {72–81},
numpages = {10},
doi = {10.4018/IJSSCI.2014100105},
journal = {International Journal of Software Science and Computational Intelligence}}

@Inbook{Jha2017alloy,
author="Jha, R.
and Dulikravich, G. S.
and Cola{\c{c}}o, M. J.
and Fan, M.
and Schwartz, J.
and Koch, C. C.",
title="Magnetic Alloys Design Using Multi-objective Optimization",
bookTitle="Properties and Characterization of Modern Materials  ",
year="2017",
pages="261-284",
publisher="Springer Singapore",
doi="10.1007/978-981-10-1602-8_22"}

@article{camp1991micro,
author = {Camp, F. E. and Kim, A. S.},
title = {Effect of microstructure on the corrosion behavior of NdFeB and NdFeCoAlB magnets},
journal = {Journal of Applied Physics},
year = {1991},
volume = {70},
number = {10},
pages = {6348-6350},
issn = {0021-8979},
doi = {10.1063/1.349938}}

@article{bae2015oxygen,
author = {Bae, Kyoung-Hoon and Lee, Seong-Rae and Kim, Hyo-Jun and Lee, Min-Woo and Jang, Tae-Suk},
title = {Effect of oxygen content of Nd-Fe-B sintered magnet on grain boundary diffusion process of DyH2 dip-coating},
journal = {Journal of Applied Physics},
year = {2015},
volume = {118},
number = {20},
pages = {203902},
issn = {0021-8979},
doi = {10.1063/1.4936172}}

@article{Bessais2006nano,
author = {Bessais, L. and Djéga-Mariadassou, C. and Beaunier, P.},
title = {Effect of nanocrystallization on the structure and the magnetic properties of Nd–Fe–Co–Al–B glassy alloy},
journal = {Journal of Applied Physics},
year = {2006},
volume = {99},
number = {9},
pages = {093906},
issn = {0021-8979},
doi = {10.1063/1.2195367}}

@article{Kobayashi2019magnetostriction,
author = {Kobayashi, Shintaro and Martín-Cid, Andrés and Toyoki, Kentaro and Okazaki, Hiroyuki and Hirosawa, Satoshi and Nakamura, Tetsuya},
title = {Influence of magnetostriction on the lattice constants of the secondary phases in Nd-Fe-B sintered magnets studied by synchrotron X-ray diffraction},
journal = {AIP Advances},
year = {2019},
volume = {9},
number = {12},
pages = {125154},
issn = {2158-3226},
doi = {10.1063/1.5129849}}
\end{document}